\documentclass[10pt,a4paper]{article}
 \usepackage{benstyle}

\usepackage[normalem]{ulem}
\usepackage{times}
\usepackage{caption}
\usepackage{overpic}
\usepackage[margin=2ex]{subcaption}

\DefineNamedColor{named}{Purple}{cmyk}{0.45,0.86,0,0}

\DefineNamedColor{named}{Gray}{cmyk}{0.3,0.3,0.3,0.3}

\DefineNamedColor{named}{orange} {rgb}{1,0.55,0}

\newcommand{\res}[1]{\ensuremath{#1\!\times\!#1}}

\DeclareGraphicsExtensions{.png, .jpg, .pdf}
\graphicspath{%
{./}%
{tmp/}%
{Diagrams/}%
{Diagrams/bogdan/}%
{Diagrams/Coherence/}%
{Diagrams/fliptests/cells/}%
{Diagrams/fliptests/geoeye/}%
{Diagrams/fliptests/glpu/}%
{Diagrams/fliptests/glpu_radial/}%
{Diagrams/fliptests/radioi/}%
{Diagrams/fliptests/watch/}%
{Diagrams/fixed-n-glpu/}%
{Diagrams/fixed-n-glpu-noise/}%
{Diagrams/fixed-n-watch/}%
{Diagrams/fixed-p-cells/}%
{Diagrams/fixed-p-cells-bern/}%
{Diagrams/fixed-p-glpu/}%
{Diagrams/fixed-p-glpu-bern/}%
{Diagrams/universality/}%
{Diagrams/frames-watch/}%
{Diagrams/modelcs/}%
{Diagrams/12p/}%
{Diagrams/structure/}%
{Diagrams/infdim/}%
{Diagrams/boats/}%
}

\begin{document}
\title{Breaking the coherence barrier: A new theory for\\ compressed sensing}

\author{B. Adcock\\ \small Purdue Univ.
\and
A. C. Hansen\\ \small
Univ. of
Cambridge \and
C. Poon\\ \small
Univ. of Cambridge
\and
B. Roman\\ \small
Univ. of Cambridge
}
\date{}
\maketitle


\section{Introduction}\label{s:introduction}

This paper provides an extension of compressed sensing which 
bridges a substantial gap between existing theory and its current use in 
real-world applications.

Compressed sensing (CS), introduced by Cand\`es, Romberg \& Tao 
\cite{CandesRombergTao} and Donoho \cite{donohoCS},  has been one of the major 
developments in applied mathematics in the last decade 
\cite{candesCSMag,DonohoTannerCounting,DonohoTannerNeighbor,EldarDuarteCSReview,EldarKutyniokCSBook,FornasierRauhutCS,FoucartRauhutBook}.
  Subject to appropriate conditions, it allows one to circumvent the 
traditional barriers of sampling theory (e.g.\ the Nyquist rate), and thereby 
recover signals from far fewer measurements than is classically 
considered possible.  This has important implications in many practical 
applications, and for this reason CS has, and continues to be, 
very intensively researched.  

The theory of CS is based on three fundamental concepts: {\it 
sparsity}, {\it incoherence} and {\it uniform random subsampling}. Whilst 
there are examples where these apply, in many  applications one 
or more of these principles may be lacking.  This includes virtually all 
of medical imaging -- Magnetic Resonance Imaging (MRI), Computerized 
Tomography (CT) and other versions of tomography such as Thermoacoustic, 
Photoacoustic or Electrical Impedance Tomography -- most of electron 
microscopy, as well as seismic tomography, fluorescence microscopy, Hadamard 
spectroscopy and radio interferometry.  In many of these problems, it is the 
principle of incoherence that is lacking, rendering the standard theory 
inapplicable.  Despite this issue, compressed sensing has been, and continues 
to be, used with great success in many of these areas.  Yet, to do so it is 
typically implemented with sampling patterns that differ substantially from 
the uniform subsampling strategies suggested by the theory.  In fact, in many 
cases uniform random subsampling yields highly suboptimal numerical results. 

The standard mathematical theory of CS has now reached a 
mature state.  However, as this discussion attests, there is a substantial, 
and arguably widening gap between the theoretical and applied sides of the 
field.  New developments and sampling strategies are increasingly based on 
empirical evidence lacking mathematical justification.  Furthermore, in the 
above applications one also witnesses a number of intriguing phenomena that 
are not explained by the standard theory.  For example, in such problems, the 
optimal sampling strategy depends not just on the overall sparsity of the 
signal, but also on its structure, as will be documented 
thoroughly in this paper. This phenomenon is in direct contradiction with 
the usual sparsity-based theory of CS.  Theorems that explain 
this observation -- i.e. that reflect how the optimal subsampling strategy 
depends on the structure of the signal -- do not currently exist.



The purpose of this paper is to provide a bridge across this divide.  It 
does so by generalizing the three traditional pillars of CS  
to three new concepts:  {\it asymptotic sparsity},  {\it asymptotic 
incoherence} and {\it multilevel random subsampling}.  This new theory 
shows that CS is also possible, and reveals several 
advantages, under these substantially more general conditions. Critically, 
it also addresses the important issue raised above: the 
dependence of the subsampling strategy on the structure of the signal.

The importance of this generalization is threefold. First, as will be explained, real-world inverse problems are typically 
not incoherent and sparse, but asymptotically incoherent and asymptotically 
sparse. This paper provides the first comprehensive mathematical 
explanation for a range of empirical usages of CS 
in applications such as those listed above.  Second, in showing that incoherence is not a requirement for CS, but 
instead that asymptotic incoherence suffices, the new theory offers 
markedly greater flexibility in the design of sensing mechanisms. In the 
future, sensors need only satisfy this significantly more relaxed condition.  Third, by using asymptotic incoherence and multilevel sampling to exploit not just sparsity, but also structure, i.e.\ asymptotic sparsity, the 
new theory paves the way for an improved CS paradigm that achieve better reconstructions in practice from fewer measurements. 

A critical aspect of many practical problems such as those listed above is that 
they do not offer the freedom to design or choose the sensing 
operator, but instead impose it (e.g.\ Fourier sampling in MRI). As such, 
much of the existing CS work, which relies on random or custom-designed sensing matrices, typically to provide universality, is not 
applicable.  This paper shows that in many such applications the imposed sensing 
operators are highly non-universal and coherent with popular sparsifying 
bases.  Yet they are asymptotically incoherent, and thus fall within the remit of the new theory.  Spurred by this observation, this paper also raises the question of whether universality is actually desirable in practice, even in applications where there is flexibility to design sensing operators with this property (e.g.\ in compressive imaging).  The new theory shows that asymptotically incoherent sensing and multilevel sampling allow one to exploit structure, not just sparsity.  Doing so leads to notable advantages over universal operators, even for problems where the latter are applicable.  Moreover, and crucially, this can be done in a computationally efficient manner using fast Fourier or Hadamard transforms (see \S \ref{RIP?}). 
 
This aside, another outcome of this work is that the Restricted Isometry 
Property (RIP), although a popular tool in CS theory, is of 
little relevance in many practical inverse problems.  As confirmed later via the 
so-called \textit{flip test}, the RIP does not hold in such applications. 

Before we commence with the remainder of this paper, let us make several further remarks. 
First, many of the problems listed above are analog, i.e. they are 
modelled with continuous transforms, such as the Fourier or Radon transforms.  
Conversely, the standard theory of CS is based on a 
finite-dimensional model.  Such \textit{mismatch} can lead to critical errors 
when applied to real data arising from continuous models, or inverse crimes 
when the data is inappropriately simulated 
\cite{CalderbankEtAlBasisMismatch,GLPU}.  To overcome this issue, a theory of 
CS in infinite dimensions was recently introduced in 
\cite{BAACHGSCS}.  This paper fundamentally extends \cite{BAACHGSCS} by 
presenting new theory in both the finite- and infinite-dimensional 
settings, the infinite-dimensional analysis also being instrumental for 
obtaining the Fourier 
and wavelets estimates in \S \ref{ss:Fourier_wavelets}.

Second, this is primarily a mathematical paper.  However, as one may expect in light of the above discussion, there are a range of practical implications.  We therefore encourage the reader to consult the paper \cite{Roman} for further discussions on the practical aspects and more extensive numerical experiments.

\section{The need for a new theory}
Let us ask the following question: 
does the standard theory of CS explain its empirical success 
in the aforementioned applications?  We now argue that the answer is no.  
Specifically, even in well-known applications such as MRI (recall that MRI 
was one of the first applications of CS, due to the pioneering 
work of Lustig et al.\ \cite{Lustig2,LustigThesis,Lustig3,Lustig}), there is a 
significant gap between theory and practice. 

\subsection{Compressed sensing}\label{ss:fin_dim_CS}
Let us commence with a short review of finite-dimensional CS
theory -- infinite-dimensional CS will be considered in \S 
\ref{s:main_thmsII}.  A typical setup, and one which we shall follow in part 
of this paper, is as follows.  Let $\{ \psi_j \}^{N}_{j=1}$ and $\{ \varphi_j 
\}^{N}_{j=1}$ be two orthonormal bases of $\bbC^N$, the \textit{sampling} and 
\textit{sparsity} bases respectively, and write
$
U = \left ( u_{ij}\right )^{N}_{i,j=1} \in \bbC^{N \times N},
$
$u_{ij} = 
\ip{\varphi_j}{\psi_i}.
$
Note that $U$ is an isometry, i.e.\ $U^* U = I$.

\defn{
Let $U = (u_{ij})^{N}_{i,j=1} \in \bbC^{N \times N}$ be an isometry.  The 
coherence of $U$ is precisely
\be{
\label{coherence_def}
\mu(U) = \max_{i,j=1,\ldots,N} | u_{ij} |^2 \in [N^{-1},1].
} 
We say that $U$ is perfectly incoherent if $\mu(U) = N^{-1}$. 
}

A signal $f \in \bbC^N$ is said to be $s$-sparse in the orthonormal basis $\{ 
\varphi_j \}^{N}_{j=1}$ if at most $s$ of its coefficients in this basis are 
nonzero.
In other words, $f = \sum^{N}_{j=1} x_j \varphi_j$, and the vector $x \in 
\bbC^N$ satisfies $| \mathrm{supp}(x) | \leq s$, where
$
\mathrm{supp}(x) = \{ j : x_j \neq 0 \}.
$
Let $f \in \bbC^N$ be $s$-sparse in $\{ \varphi_j \}^{N}_{j=1}$, and suppose 
we have access to the samples
$
\hat{f}_j = \ip{f}{\psi_j},$ 
$j=1,\ldots,N.$
Let $\Omega \subseteq \{ 1,\ldots,N \}$ be of cardinality $m$ and chosen 
uniformly at random.  According to a result of Cand\`es \& Plan 
\cite{Candes_Plan} and Adcock \& Hansen \cite{BAACHGSCS}, $f$ can be recovered 
exactly with probability exceeding $1-\epsilon$ from the subset of measurements
$
\{ \hat{f}_j : j \in \Omega \},
$
provided
\be{
\label{m_est_Candes_Plan}
m \gtrsim  \mu(U) \cdot N \cdot s \cdot \left (1+\log (\epsilon^{-1}) \right ) 
\cdot \log (N),
}
(here and elsewhere in this paper we shall use the notation $a \gtrsim b$ to 
mean that there exists a constant $C > 0$ independent of all relevant 
parameters such that $a \geq C b$).  In practice, recovery is achieved by 
solving the following convex optimization problem:
\be{
\label{fin_dim_l1}
\min_{\eta \in \bbC^N} \| \eta \|_{l^1}\ \mbox{subject to $P_{\Omega} U \eta = 
P_{\Omega} \hat{f}$},
}
where $\hat{f} = (\hat{f}_1,\ldots,\hat{f}_N)^{\top}$ and $P_{\Omega} \in 
\bbC^{N \times N}$ is the diagonal projection matrix with $j^{\rth}$ entry $1$ 
if $j \in \Omega$ and zero otherwise.
The key estimate \R{m_est_Candes_Plan} shows that the number of measurements 
$m$ required is, up to a log factor, on the order of the sparsity $s$, 
provided the coherence $\mu(U) = \ord{N^{-1}}$.  This is the case, for 
example, when $U$ is the DFT matrix; a problem which was studied in some of 
the first papers on CS \cite{CandesRombergTao}.

\subsection{Incoherence is rare in practice}\label{ss:inc_rare}
\begin{figure}
\begin{center}
\includegraphics[width=0.24\linewidth]{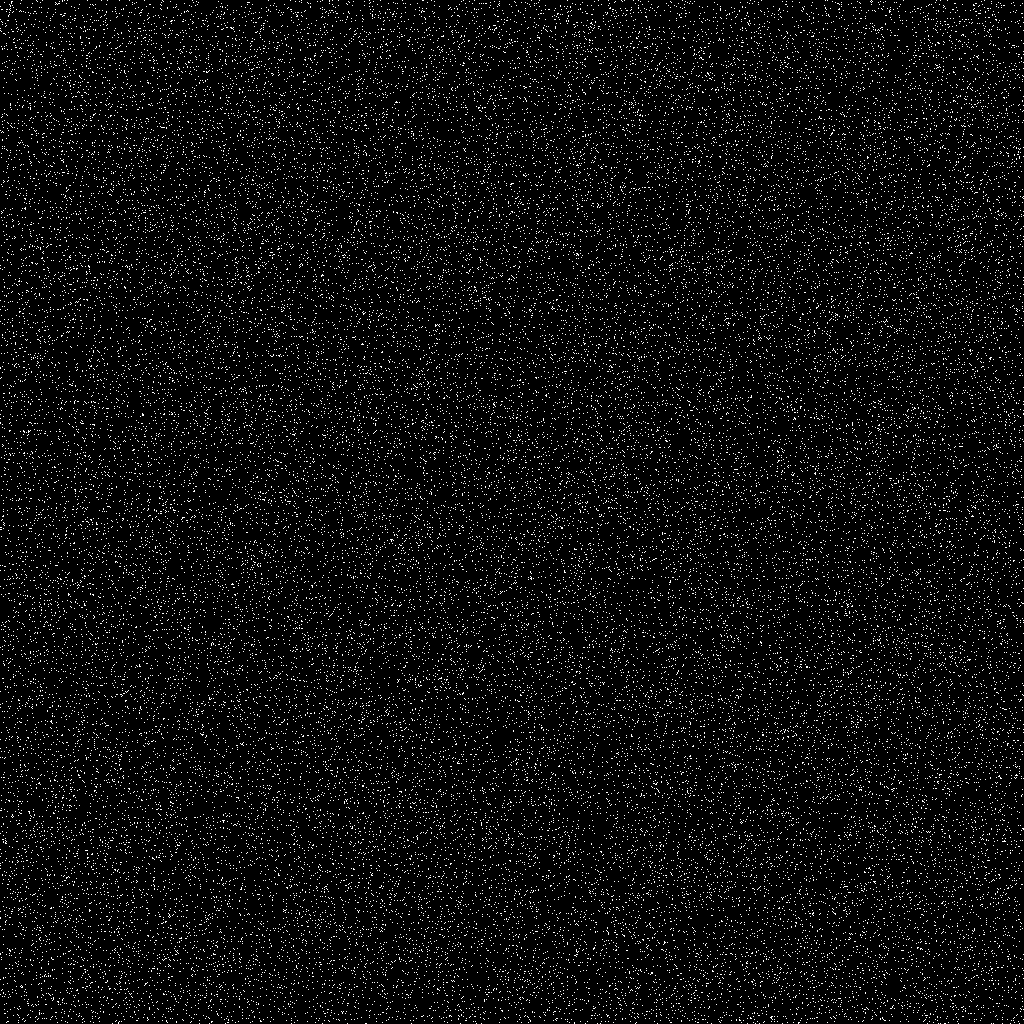}
\includegraphics[width=0.24\textwidth]{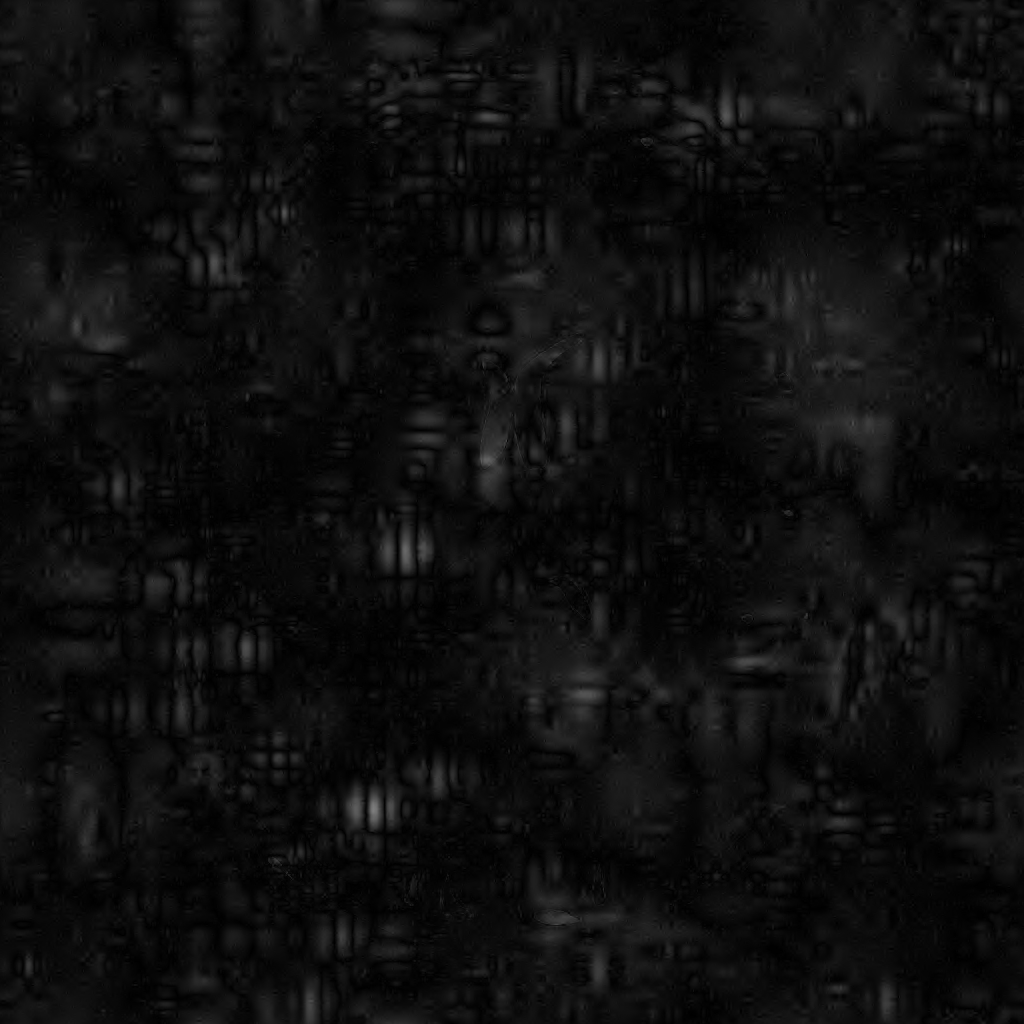}
 ~
\includegraphics[width=0.24\textwidth]{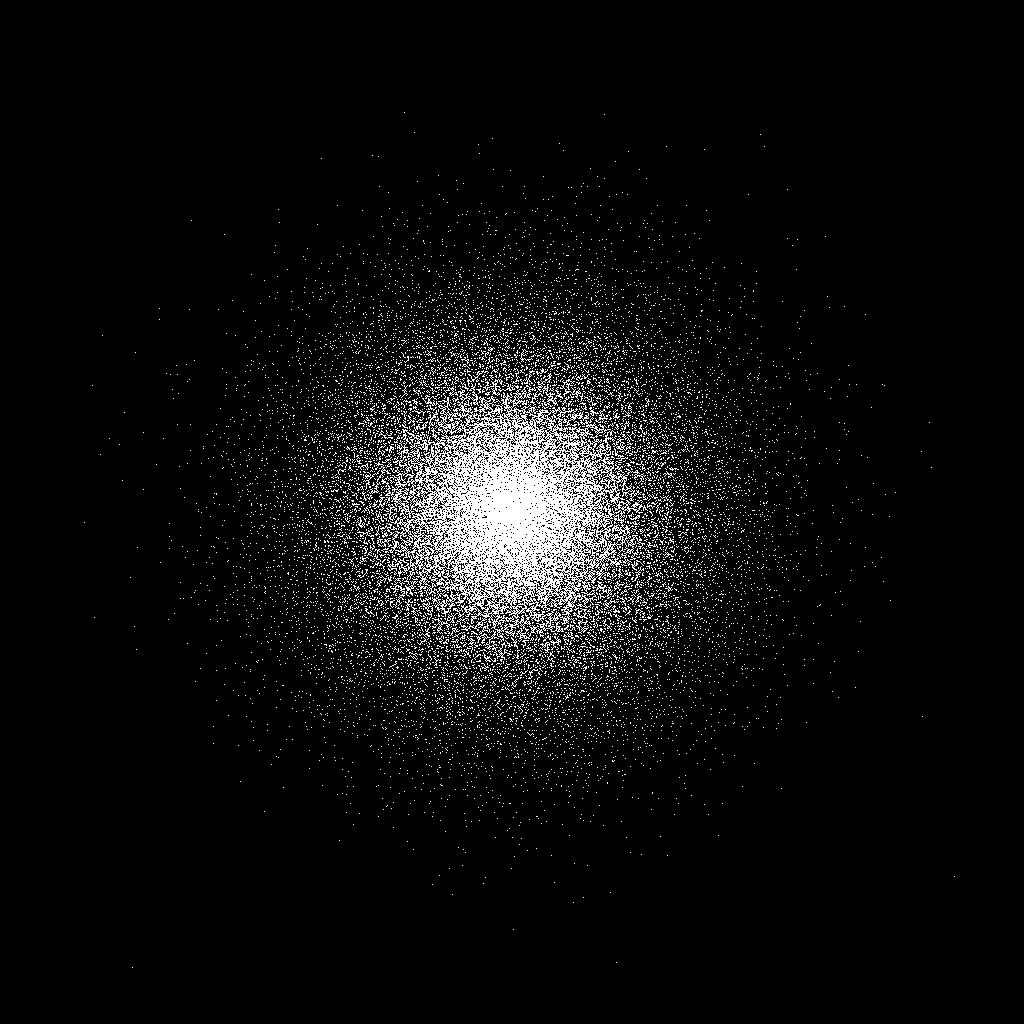}
\includegraphics[width=0.24\textwidth]{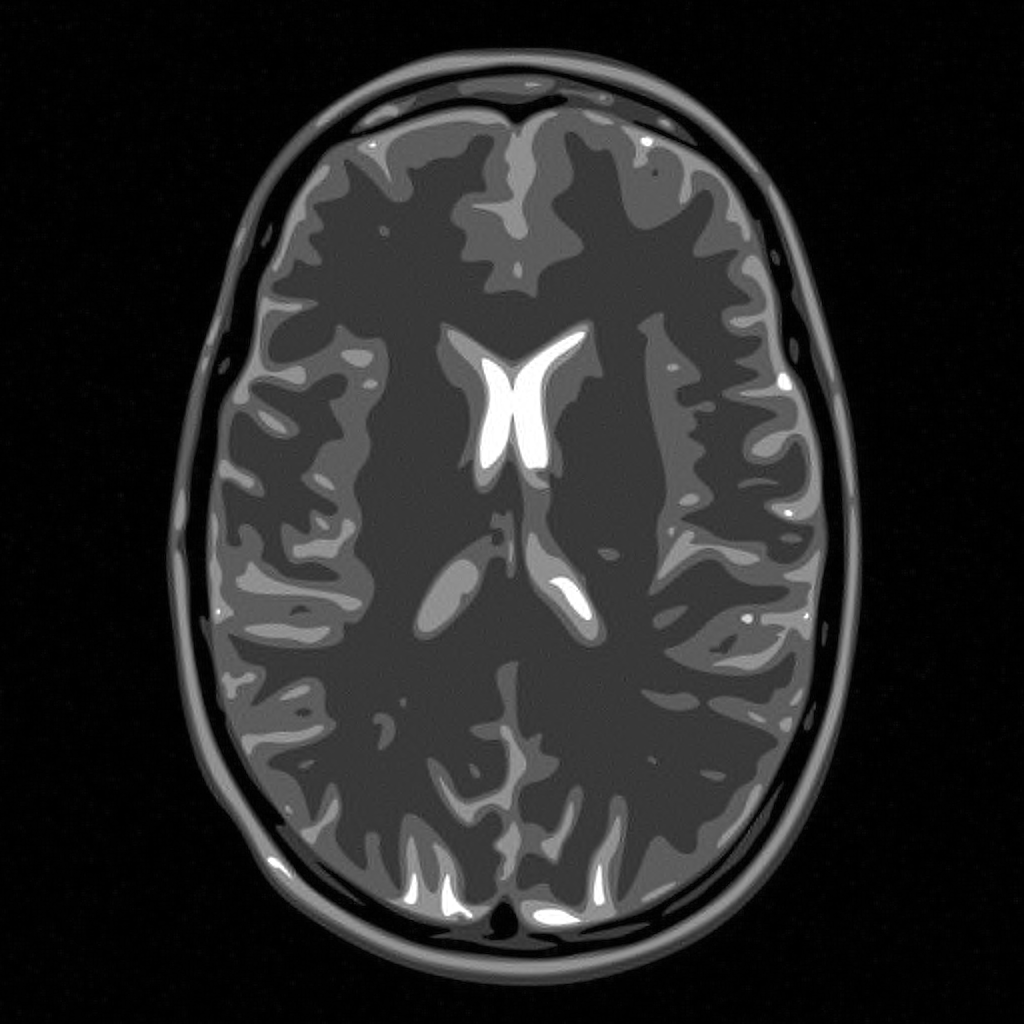}
\end{center}
\caption{Left to right: (i) $5\%$ uniform random subsampling scheme, (ii) CS 
reconstruction from uniform subsampling, (iii) $5\%$ multilevel subsampling 
scheme, (iv) CS reconstruction from multilevel subsampling. }
\label{f:CS_Evol}
\end{figure}

To test the practicality of the incoherence condition, let us consider a typical 
CS problem. In a 
number of important applications, not least MRI, the sampling is carried out 
in the Fourier domain.  Since images are sparse in wavelets, the usual CS 
setup is to form the a matrix $U_N =  U_{\mathrm{df}} V^{-1}_{\mathrm{dw}} 
\in \mathbb{C}^{N \times N}$, where $U_{\mathrm{df}}$ and $V_{\mathrm{dw}}$ 
represent the discrete Fourier and wavelet transforms respectively.  However, in the case the coherence satisfies $\mu(U_N) = \ord{1}$ as $N \rightarrow \infty$, for any wavelet basis.  Thus, this problem has the 
worst possible coherence, and the standard CS estimate \R{m_est_Candes_Plan}
 states that $m = N$ samples are 
needed in this case (i.e.\ full sampling), even though the object to recover 
is typically highly sparse.  Note that this is not an insufficiency of 
the theory.  If uniform random subsampling is employed, then the lack of 
incoherence does indeed lead to a very poor reconstruction.  This can be seen in 
Figure \ref{f:CS_Evol}.  

The underlying reason for this lack of incoherence can be traced to the fact 
that this finite-dimensional problem is a discretization of an 
infinite-dimensional problem.  Specifically, 
\be{
\label{WOT}
\underset{N \rightarrow \infty}{\text{WOT-lim \,}}
 U_{\mathrm{df}} V^{-1}_{\mathrm{dw}} = U,
}
where $U : l^2(\bbN) \rightarrow l^2(\bbN)$ is the operator represented as the 
infinite matrix
\be{\label{Umatrix}
U =
  \left(\begin{array}{ccc} \left < \varphi_1 , \psi_1 \right >   & \left < 
  \varphi_2 , \psi_1 \right > & \cdots \\
\left < \varphi_1 , \psi_2 \right >   & \left < \varphi_2 , \psi_2\right > & 
\cdots \\
\vdots  & \vdots  & \ddots   \end{array}\right),
}
and the functions $\varphi_j$ are the wavelets used, the $\psi_j$'s are the 
standard complex exponentials and WOT denotes the weak operator topology.  Since the 
coherence of the infinite matrix $U$ -- i.e.\ the supremum of its entries in 
absolute value -- is a fixed number independent of $N$, we cannot expect incoherence of the 
discretization $U_N$ for large $N$.  At some point, one will always encounter 
the so-called \textit{coherence barrier}.  
Such an issue is not isolated to this example.  Heuristically, any problem that 
arises as a discretization of an infinite-dimensional problem will suffer from 
the same phenomenon.  The list of applications of this type is long, and 
includes for example, MRI, CT, microscopy and seismology.  

To mitigate this 
problem, one may naturally try to change $\{\varphi_j\}$ or $\{\psi_j\}$.  However, this will deliver only marginal benefits, since (\ref{WOT}) demonstrates that the 
coherence barrier will always occur for large enough $N$. 

In view of this, one may wonder how it is possible that CS is applied so successfully to many such problems.
The key is so-called \textit{asymptotic} incoherence (see \S \ref{asymp_inco}) and the 
use of a variable density/multilevel subsampling strategy. The success of such 
subsampling is confirmed numerically in Figure  \ref{f:CS_Evol}. However, it 
is important to note that this is an empirical solution to the problem.  None 
of the usual theory explains the effectiveness of CS when 
implemented in this way.

\subsection{Sparsity and the flip test}\label{flip}
\begin{figure}
\begin{center}
\newcommand{\abrcap}[2]{\raisebox{80pt}{\begin{minipage}[t]{65pt}{\footnotesize
 #1x#1\\[15pt]Error:\\[2pt]}#2\%\end{minipage}}}%
\newcommand{\abrdesc}[5]{\raisebox{90pt}{\begin{minipage}[t]{60pt}{\footnotesize
  \res{#1}\\#2\%\\[7pt]#3$\cdot$#4\\[7pt]{#5}}\end{minipage}}}%
\newcommand{\abrflip}[1]{\includegraphics[width=0.28\textwidth]{#1}}%
\vspace{-10pt}
\begin{tabular}{@{\hspace{0pt}}c@{\hspace{0pt}}c@{\hspace{3pt}}c@{\hspace{3pt}}c@{\hspace{0pt}}}
& CS reconstruction & CS reconstruction w/ flip & Subsampling pattern 
used\\[5pt]
  \abrdesc{512}{10}{$U_{\mathrm{Had}}$}{$V_{\mathrm{dwt}}^{-1}$}{Fluorescence\\[-2pt]Microscopy}
   &\abrflip{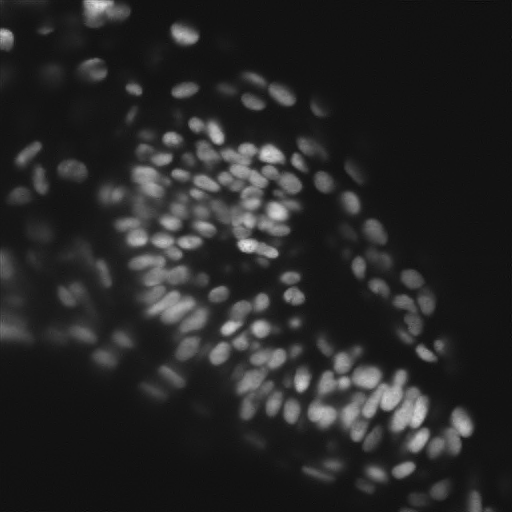}
  &\abrflip{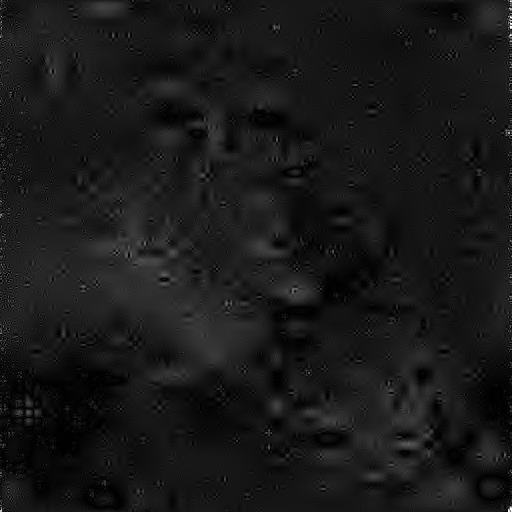}
  &\abrflip{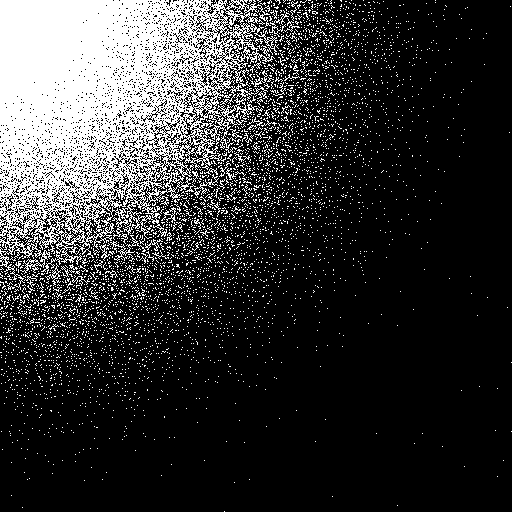}
 \\
   \abrdesc{512}{15}{$U_{\mathrm{Had}}$}{$V_{\mathrm{dwt}}^{-1}$}{Compressive 
   Imaging, Hadamard Spectroscopy}
  &\abrflip{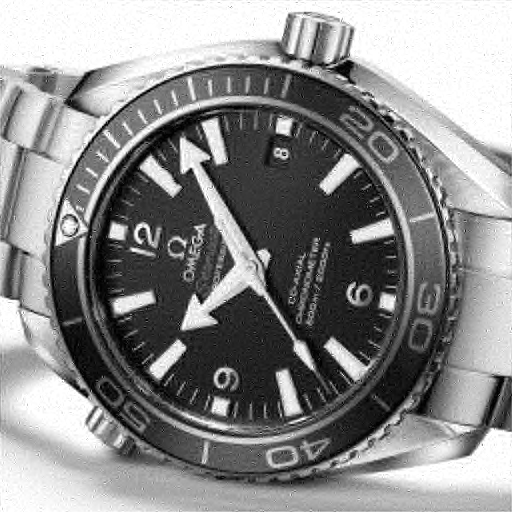}
  &\abrflip{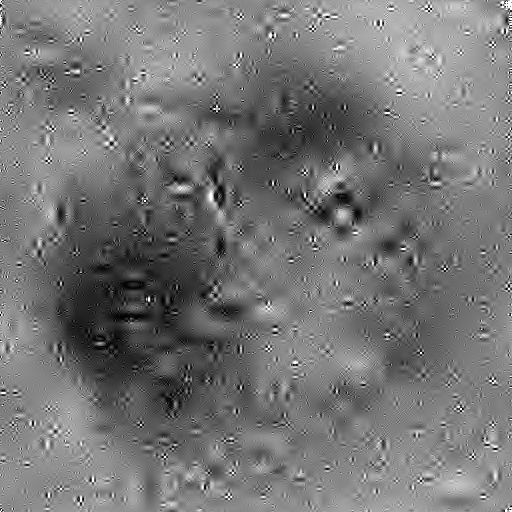}
  &\abrflip{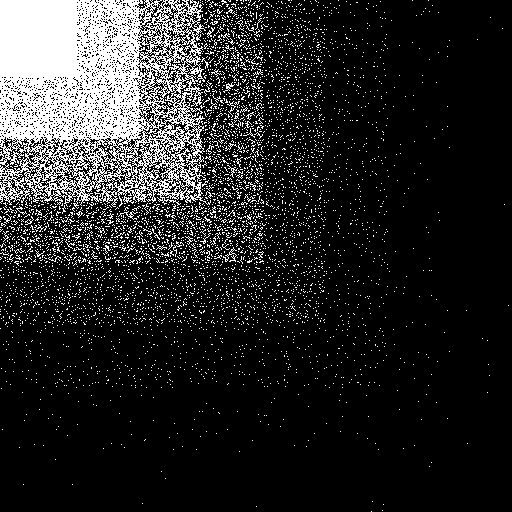}
 \\
  \abrdesc{1024}{20}{$U_{\mathrm{dft}}$}{$V_{\mathrm{dwt}}^{-1}$}{Magnetic\\[-2pt]Resonance\\[-2pt]Imaging}
  &\abrflip{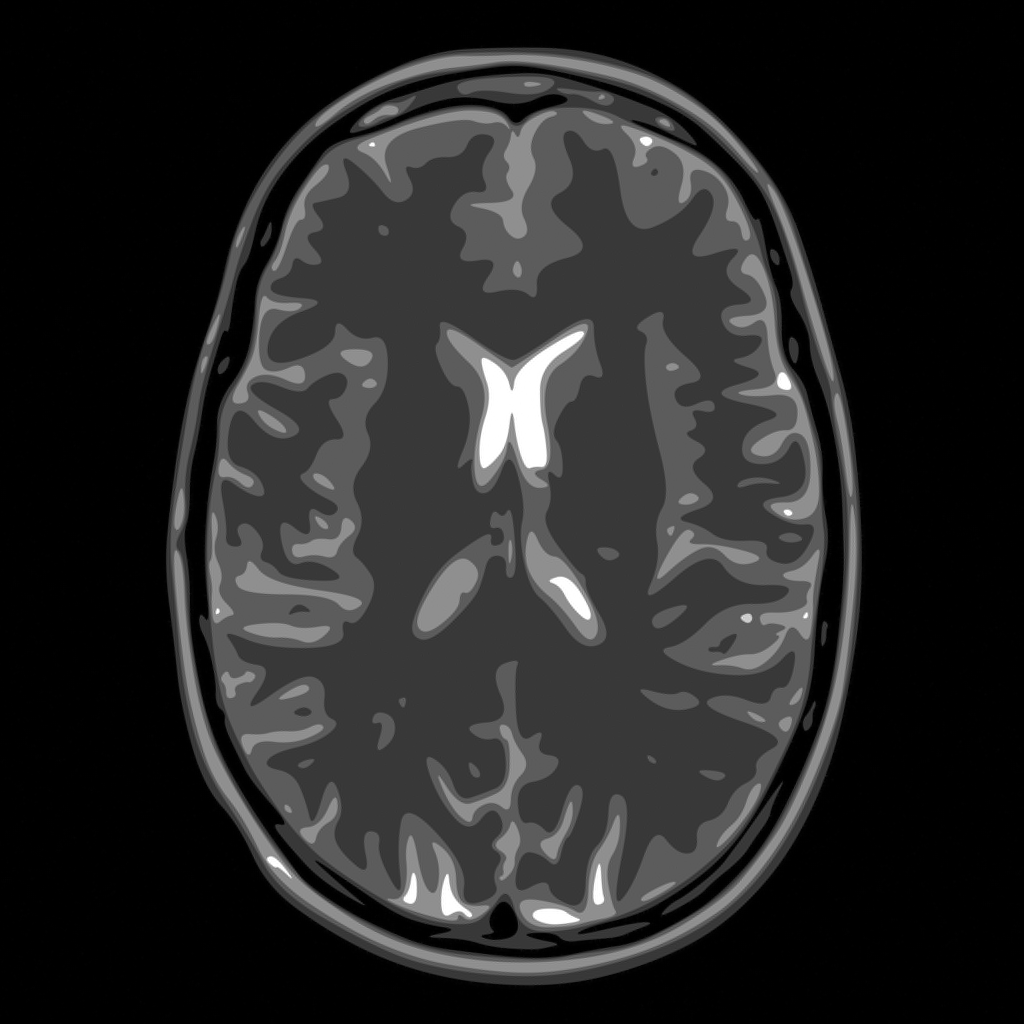}
  &\abrflip{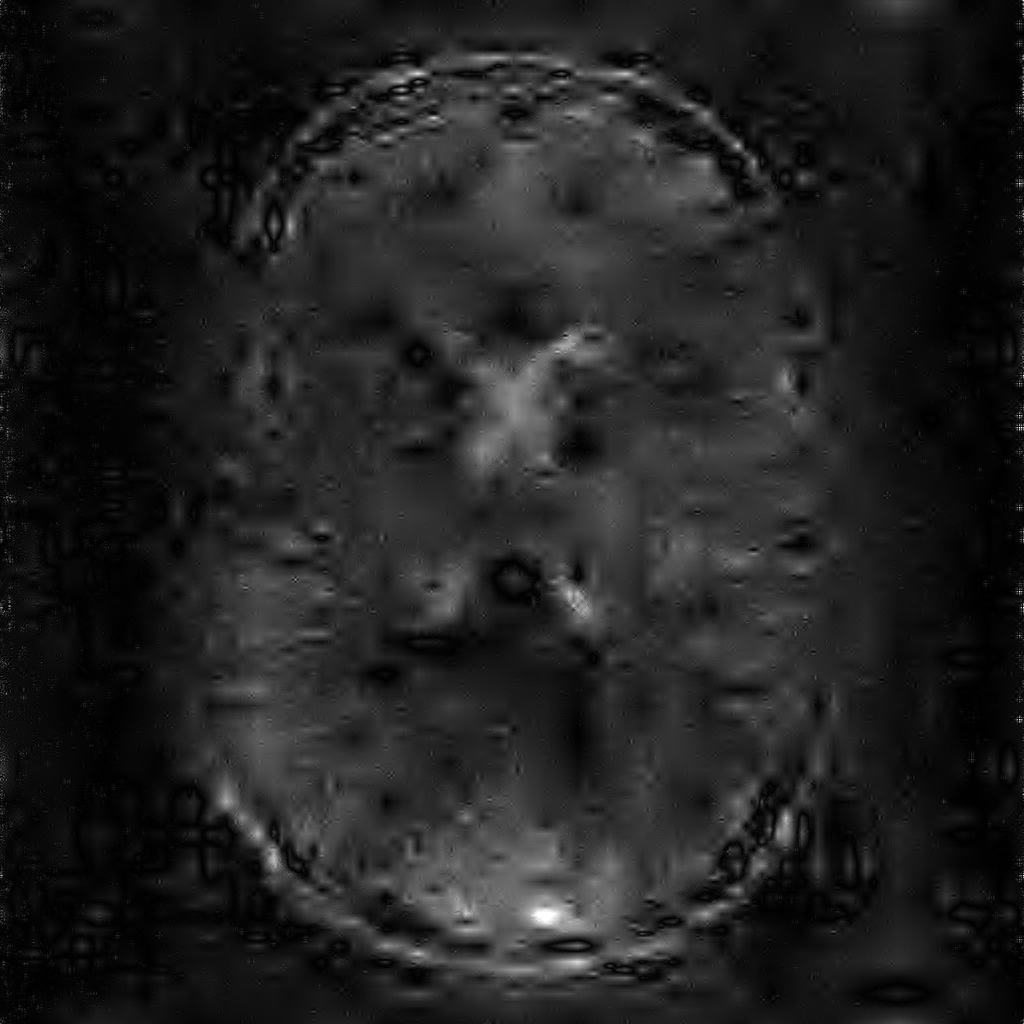}
  &\abrflip{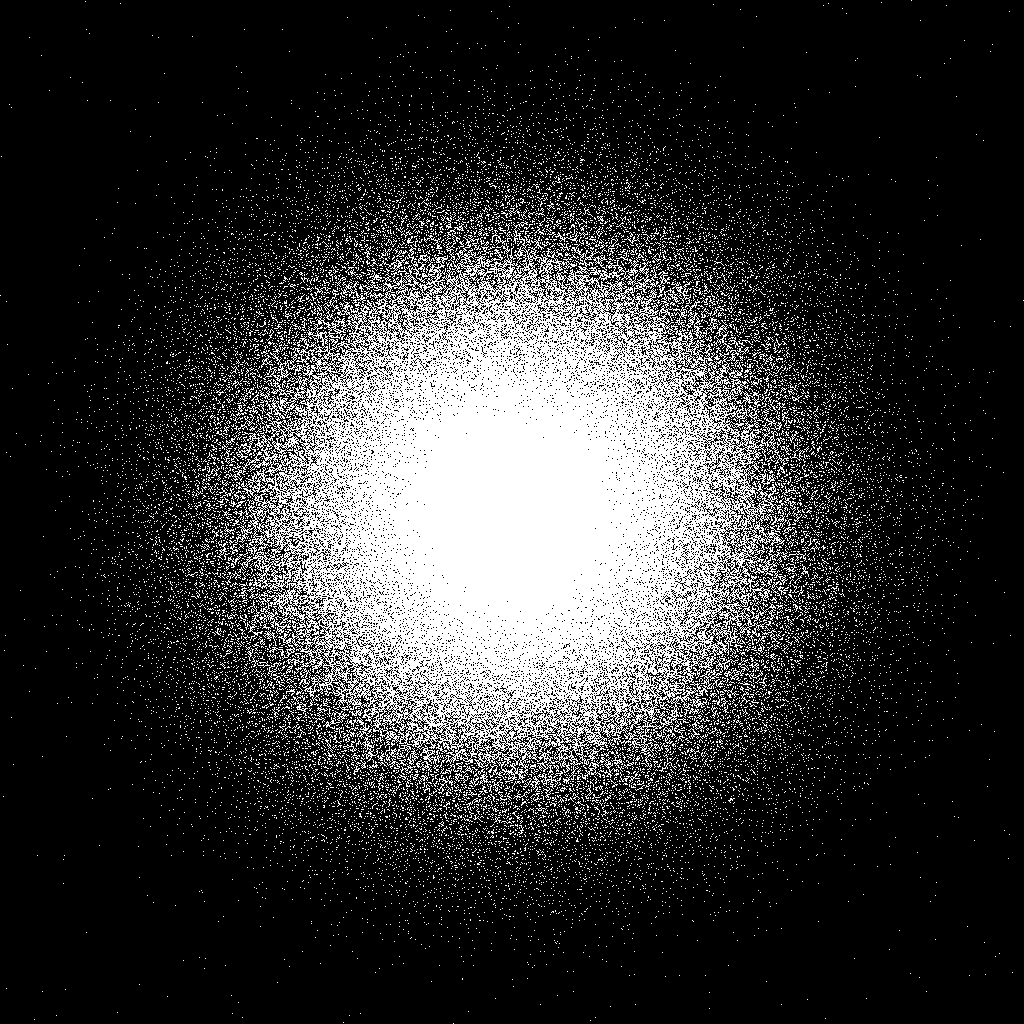}
 \\
  \abrdesc{512}{12}{$U_{\mathrm{dft}}$}{$V_{\mathrm{dwt}}^{-1}$}{Tomography,\\[0pt]Electron\\[-2pt]Microscopy}
  &\abrflip{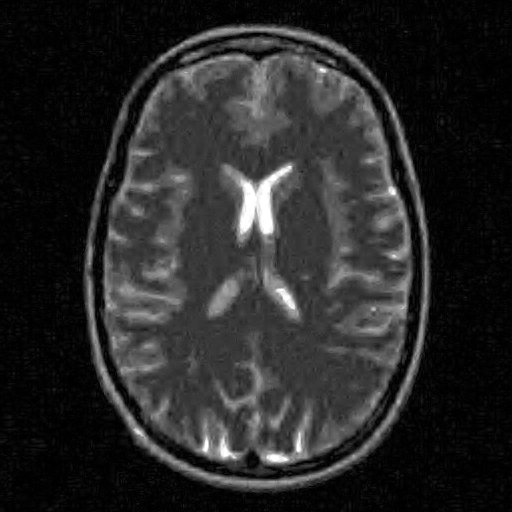}
  &\abrflip{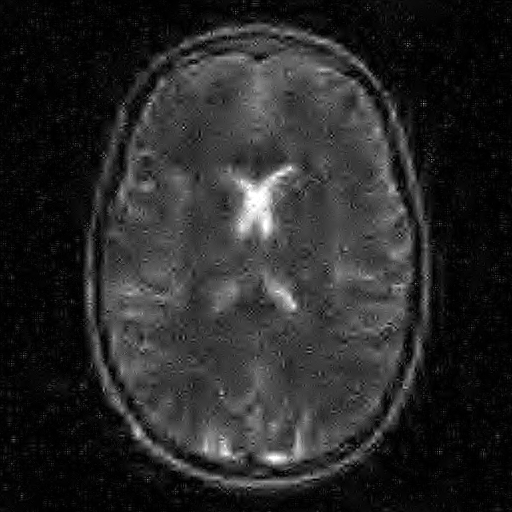}
  &\abrflip{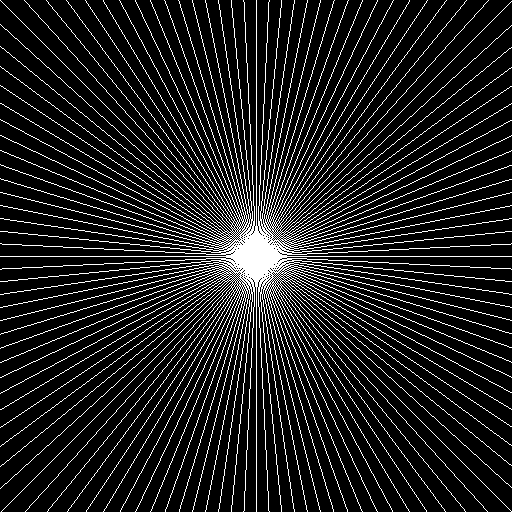}
 \\
  \abrdesc{512}{15}{$U_{\mathrm{dft}}$}{$V_{\mathrm{dwt}}^{-1}$}{Radio\\[-2pt]interferometry}
  &\abrflip{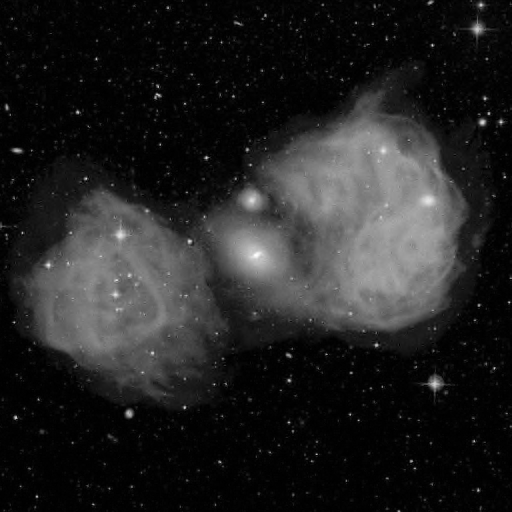}
  &\abrflip{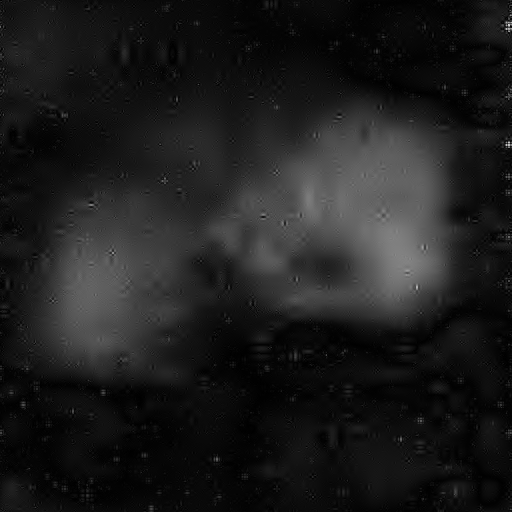}
  &\abrflip{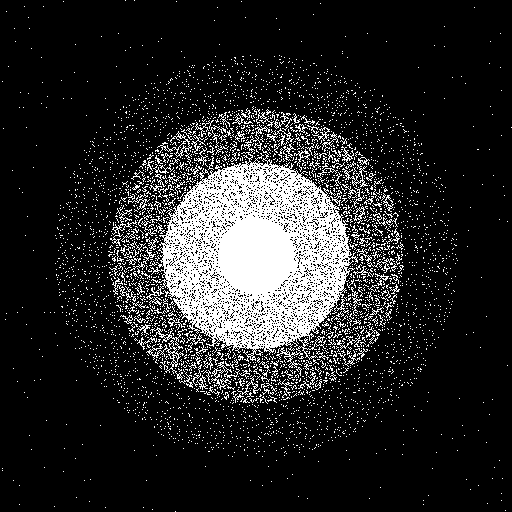}
 \end{tabular}
\caption{Reconstructions via CS (left column) and the flipped wavelet coefficients (middle 
column). The right column shows the subsampling map used. The percentage shown 
is the fraction of Fourier or Hadamard coefficients that 
were sampled. The reconstruction basis was DB4 for the Fluorescence microscopy 
example, and DB6 for the rest. }
\label{f:flip-tests}
\end{center}
\end{figure}
The previous discussion demonstrates that we must dispense with the principles 
of incoherence and uniform random subsampling in order to develop a new theory 
of CS.  We now claim that sparsity must also be replaced with 
a more general concept.  This may come as a surprise to the reader, since 
sparsity is a central pillar of not just CS, but much of 
modern signal processing.  However, this can be confirmed by a simple experiment we refer to as the \textit{flip test}.

Sparsity asserts that an unknown vector $x$ has $s$ important 
coefficients, where the locations can be arbitrary.  CS 
establishes that all $s$-sparse vectors can be recovered from the same sampling strategy.  In 
particular, the sampling strategy is completely independent of the location of 
these coefficients.  The flip test, described next, allows one to evaluate 
whether this holds in a given application. Let $x \in \bbC^N$ and $U \in \bbC^{N \times N}$.  Next we take samples 
according to some appropriate subset $\Omega \subseteq \{1,\hdots,N\}$ with 
$|\Omega| = m$, and solve:
\begin{equation}\label{l1_test}
\min_{z \in \bbC^N}  \|z\|_1 \ \mbox{subject to $P_{\Omega}Uz = P_{\Omega}Ux$}.
\end{equation}
This gives a reconstruction $z = z_1$.  Now we flip $x$ 
through the operation
$
x \mapsto x^{\mathrm{fp}} \in \mathbb{C}^N,
$ $ x^{\mathrm{fp}}_1 = x_N,$ $ 
x^{\mathrm{fp}}_2 = x_{N-1}, \hdots, x^{\mathrm{fp}}_N = x_1,
$
giving a new vector $x^{\mathrm{fp}}$ with reversed entries. We next apply the 
same CS reconstruction to $x^{\mathrm{fp}}$, using the same 
matrix $U$ and the same subset $\Omega$.  That is we solve
\begin{equation}\label{l1_test2}
\min_{z \in \bbC^N} \|z\|_1 \  \mbox{subject to $P_{\Omega}Uz = P_{\Omega}Ux^{\mathrm{fp}}$}.
\end{equation}
Let $z$ be a solution of \R{l1_test2}.  In order to get a reconstruction of 
the original vector $x$, we perform the flipping operation once more and form 
the final reconstruction $z_2 = z^{\mathrm{fp}}$.

Suppose now that $\Omega$ is a good sampling pattern for recovering $x$ using 
the solution $z_1$ of \R{l1_test}.  If sparsity is the key structure that 
determines such reconstruction quality, then we expect exactly the same 
quality in the approximation $z_2$ obtained via \R{l1_test2}, since 
$x^{\mathrm{fp}}$ is merely a permutation of $x$. To investigate whether or not this is true, we consider several examples arising from the following applications: fluorescence microscopy, compressive imaging, MRI, CT, electron 
microscopy and radio interferometry. These examples are based on the matrix $U 
= U_{\mathrm{dft}}V_{\mathrm{dwt}}^{-1}$ or $U = 
U_{\mathrm{Had}}V_{\mathrm{dwt}}^{-1}$, where $U_{\mathrm{dft}}$ is the 
discrete Fourier transform,  $U_{\mathrm{Had}}$ is a Hadamard matrix and 
$V_{\mathrm{dwt}}$ is the discrete wavelet transform.

The results of this experiment are shown in Figure \ref{f:flip-tests}. As is 
evident, in all cases the flipped reconstructions $z_2$ are substantially 
worse than their unflipped counterparts $z_1$.  Hence, we conclude that sparsity alone does not 
govern the reconstruction quality, and consequently the success in the 
unflipped case must also be due in part to the structure of the signal. 
 In other words:
\bes{
\mbox{\textit{The optimal subsampling strategy depends on the 
signal structure}.}
}
Note that the flip test reveals another interesting phenomenon:
\bes{
\mbox{\textit{There is no Restricted Isometry Property (RIP)}.}
}
Suppose the matrix $P_{\Omega} U$ satisfied an RIP for realistic parameter 
values (i.e. problem size $N$, subsampling percentage $m$, and sparsity $s$) 
found in applications.  Then this would imply recovery of all approximately 
sparse vectors with the same error.  This is in direct contradiction with the 
results of the flip test.  

Note that in all the examples in Figure \ref{f:flip-tests}, uniform random 
subsampling would have given nonsensical results, analogously to what was 
shown in Figure \ref{f:CS_Evol}.

\subsection{Signals and images are asymptotically sparse in -lets}
\label{Real_world_sparse}

Given that structure is key, we now ask the question: what, if any, structure 
is characteristic of such applications?  
Let us consider a wavelet basis $\{ \varphi_n \}_{n \in \bbN}$.  Recall that 
associated to such a basis, there is a natural 
decomposition of $\bbN$ into finite subsets according to different scales, 
i.e.
$
\bbN = \bigcup_{k \in \bbN} \{ M_{k-1}+1,\ldots,M_k \},
$
where $0 = M_0 < M_1 < M_2 < \ldots$ and $\{ M_{k-1}+1,\ldots,M_k \}$ is the 
set of indices corresponding to the $k^{\rth}$ scale. 
Let $x \in l^2(\bbN)$ be the coefficients of a function $f$ in this basis.  
Suppose that $\epsilon \in (0,1]$ is 
given, and define
\begin{equation}\label{sk_def1}
s_k = s_k(\epsilon) = 
\min\Big\{K:\Big\|\sum_{i=1}^K x_{\pi(i)}\varphi_{\pi(i)}\Big\| \geq 
\epsilon\, \Big\| \sum_{i=M_{k-1}+1}^{M_k} x_j\varphi_j  \Big\|\,\Big\},
\end{equation}
where $\pi: \{1,\hdots, M_k-M_{k-1}\} \rightarrow  \{M_{k-1}+1,\hdots, M_k\}$ 
is a bijection such that $|x_{\pi(i)}| \geq |x_{\pi(i+1)}|$ for 
$i=1,\ldots,M_{k}-M_{k-1}-1$.  In order words, the quantity $s_k$ is the 
effective sparsity
of the wavelet coefficients of $f$ at the $k^{\rth}$ scale. 

\begin{figure}[!t]
\begin{center}
\begin{tabular}{@{\hspace{0pt}}c@{\hspace{0.02\textwidth}}c@{\hspace{0pt}}}
\qquad\includegraphics[width=0.26\linewidth]{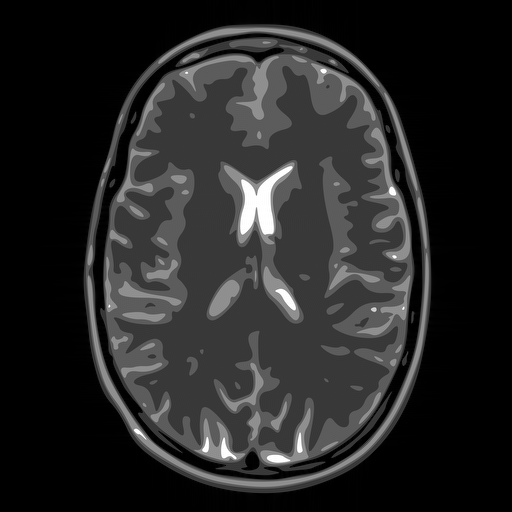}&
\qquad\includegraphics[width=0.26\textwidth]{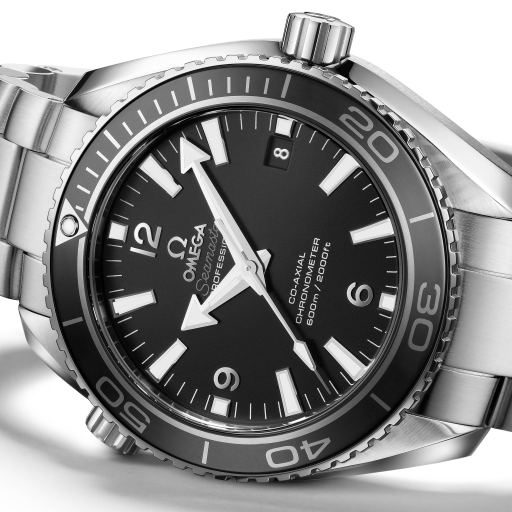}\\[3pt]
\includegraphics[width=0.42\textwidth]{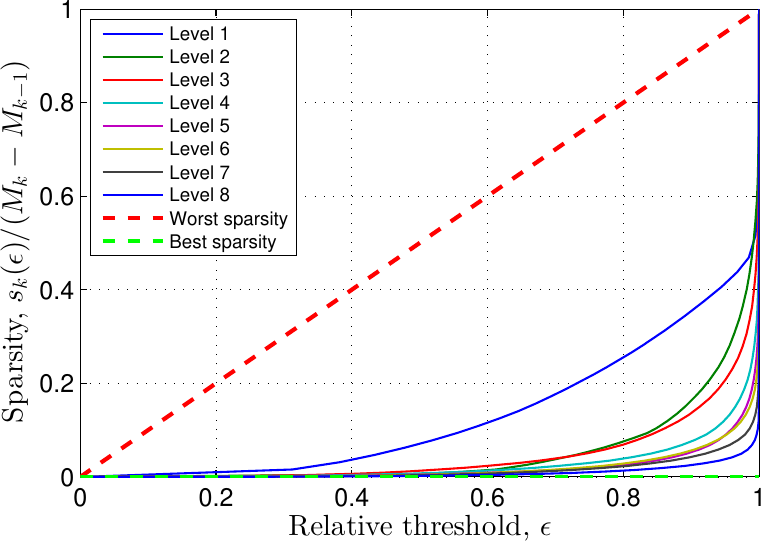}&
\includegraphics[width=0.42\textwidth]{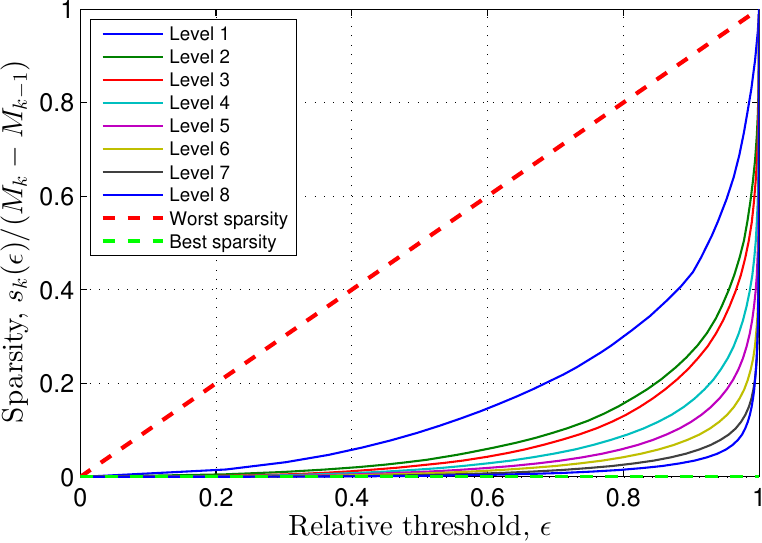}\\[3pt]
\end{tabular}
\caption{Relative sparsity of the Daubechies-8 wavelet coefficients of two 
images.  Here the levels correspond to wavelet scales and $s_k(\epsilon)$ is 
given by \R{sk_def1}.  Each curve shows the relative sparsity at level $k$ as 
a function of $\epsilon$.  The decreasing nature of the curves for increasing 
$k$ confirms \R{fine_decay}.
}
\label{f:CS_LevelsSparsity}
\end{center}
\end{figure}

Sparsity of $f$ in a wavelet basis means that for a given maximal scale $r 
\in\bbN$, the ratio $s / M_r \ll 1$, where $M = M_r$ and $s = s_1+\ldots+s_r$ 
is the total effective sparsity of $f$.  The observation that typical images 
and signals are approximately sparse in wavelet bases is one of the key 
results in nonlinear approximation \cite{DeVoreNLACTA,mallat09wavelet}.  
However, such objects exhibit far more than sparsity alone.  In fact, the 
ratios
\be{
\label{fine_decay}
s_k / (M_k - M_{k-1}) \rightarrow 0,
}
rapidly as $k \rightarrow \infty$, for every fixed $\epsilon \in (0,1]$.  Thus 
typical signals and images have a distinct sparsity \textit{structure}.  They 
are much more sparse at fine scales (large 
$k$) than at coarse scales (small $k$). 
This is confirmed in Figure \ref{f:CS_LevelsSparsity}. 
Note that this conclusion does not 
change if one replaces wavelets by other related approximation 
systems, such as curvelets \cite{Cand,candes2004new}, contourlets 
\cite{Vetterli,Do} or shearlets \cite{Gitta,Gitta2,Gitta3}.

\section{New principles}\label{s:new_concepts}
Having argued for their need, we now introduce the main new concepts of the 
paper: namely, asymptotic incoherence, asymptotic sparsity and multilevel 
sampling.  

\subsection{Asymptotic incoherence}\label{asymp_inco}
Recall from \S \ref{ss:inc_rare} that the case of Fourier sampling with 
wavelets as the sparsity basis is a standard example of a coherent problem.  
Similarly, Fourier sampling with Legendre polynomials is also coherent, as is the case of Hadamard 
sampling with wavelets.  In Figure \ref{f:Haar_Asy_Inc} we plot the absolute 
values of the entries of the matrix $U$ for these three examples.  As is 
evident, whilst $U$ does indeed have large entries in all three case (since it 
is coherent), these are isolated to a leading submatrix (note that we 
enumerate over $\bbZ$ for the Fourier sampling basis and $\bbN$ for the 
wavelet/Legendre sparsity bases).  As one moves away from this region the 
values get progressively smaller.  That is, the matrix $U$ is incoherent aside 
from a leading coherent submatrix. This motivates the following definition:
\begin{figure}
\centering
\includegraphics[width=0.3\textwidth]{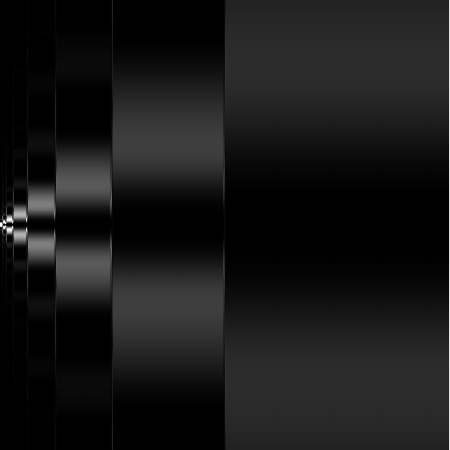}~~
\includegraphics[width=0.3\textwidth]{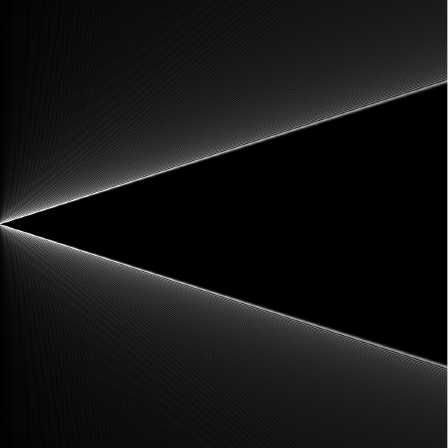}~~
\includegraphics[width=0.3\textwidth]{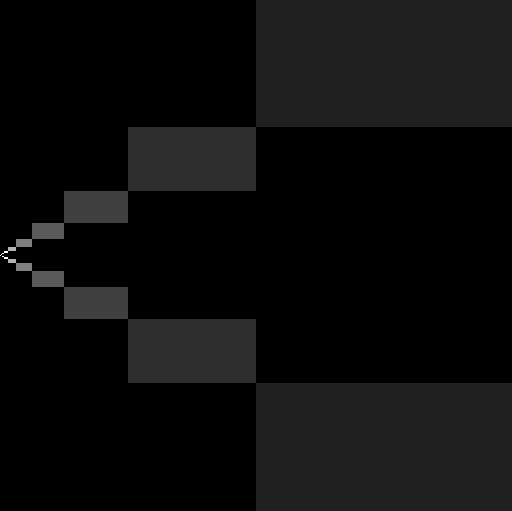}
\caption{The absolute values of the matrix $U$ in (\ref{Umatrix}): ({\it 
left}): DB2 wavelets with Fourier sampling. ({\it middle}): Legendre 
polynomials with Fourier sampling. ({\it right}): The absolute values of 
$U_{\mathrm{Had}} V^{-1}_{\mathrm{dwt}},$ where $U_{\mathrm{Had}}$ is a 
Hadamard matrix and $V^{-1}_{\mathrm{dwt}}$ is the discrete Haar transform. 
Light regions correspond to large values and dark regions to small values. }
\label{f:Haar_Asy_Inc}
\end{figure}

\defn{[Asymptotic incoherence] Let be $\{U_N\}$  be a sequence of isometries 
with $U_N \in \bbC^N$ or let $U \in \cB(l^2(\bbN))$ be an isometry. Then
\begin{itemize}
\item[(i)] $\{U_N\}$ is asymptotically incoherent if
$
\mu(P^{\perp}_K U_N),\ \mu(U_N P^{\perp}_K) \rightarrow 0, 
$
when $K 
\rightarrow \infty, 
$ with $N/K = c,$ for all $c \geq 1$.
\item[(ii)] $U$ is asymptotically incoherent if
$
\mu(P^{\perp}_K U),\ \mu(U P^{\perp}_K) \rightarrow 0,$
when $K \rightarrow 
\infty.$
\end{itemize}
Here $P_K$ is the projection onto $\spn \{e_j : j = 1,...,K \}$, where $\{ e_j 
\}$ is the canonical basis of either $\bbC^N$ or $l^2(\bbN)$, and 
$P^{\perp}_K$ is its orthogonal complement.
}

In other words, $U$ is asymptotically incoherent if the coherences of the 
matrices formed by replacing either the first $K$ rows or columns of $U$ are 
small.  As it transpires, the Fourier/wavelets, Fourier/Legendre and 
Hadamard/wavelets problems are asymptotically incoherent.  In particular, 
$\mu(P^{\perp}_K U),\ \mu(U P^{\perp}_K) = \ord{K^{-1}}$ as $K \rightarrow 
\infty$ for the former (see \S \ref{ss:Fourier_wavelets}).

\subsection{Multi-level sampling}\label{multilevel}
Asymptotic incoherence suggests a different subsampling strategy should be 
used instead of uniform random sampling.  High coherence in the first few rows 
of $U$ means that important information about the signal to be recovered may 
well be contained in its corresponding measurements.  Hence to ensure good 
recovery we should fully sample these rows.  Conversely, once outside of this 
region, when the coherence starts to decrease, we can begin to subsample.   
Let $N_1 , N,m \in \bbN$ be given.  This now leads us to consider an index set 
$\Omega$ of the form $\Omega = \Omega_1 \cup \Omega_2$, where $\Omega_{1} = 
\{1,\ldots,N_1 \}$, and $\Omega_2 \subseteq \{ N_1+1,\ldots,N \}$ is chosen 
uniformly at random with $| \Omega_2 | = m$.  We refer to this as a 
\textit{two-level} sampling scheme.  As we shall prove later, the amount of 
subsampling possible (i.e.\ the parameter $m$) in the region corresponding to 
$\Omega_2$ will depend solely on the sparsity of the signal and coherence 
$\mu(P^{\perp}_{N_1} U)$.  

The two-level scheme represents the simplest type of nonuniform density 
sampling.  There is no reason, however, to restrict our attention to just two 
levels, full and subsampled.  In general, we shall consider 
\textit{multilevel} schemes, defined as follows:

\defn{[Multilevel random sampling]
\label{multi_level_dfn}
Let $r \in \bbN$, $\mathbf{N} = (N_1,\ldots,N_r) \in \bbN^r$ with $1 \leq N_1 
< \ldots < N_r$, $\mathbf{m} = (m_1,\ldots,m_r) \in \bbN^r$, with $m_k \leq 
N_k-N_{k-1}$, $k=1,\ldots,r$, and suppose that
$
\Omega_k \subseteq \{ N_{k-1}+1,\ldots,N_{k} \},\quad | \Omega_k | = m_k,\quad 
k=1,\ldots,r,
$
are chosen uniformly at random, where $N_0 = 0$.  We refer to the set
$
\Omega = \Omega_{\mathbf{N},\mathbf{m}} = \Omega_1 \cup \ldots \cup \Omega_r.
$
as an $(\mathbf{N},\mathbf{m})$-multilevel sampling scheme.
}

Note that the idea of sampling the low-order coefficients of an image differently goes 
back to the early days of CS.  In particular, Donoho considers 
a two-level approach for recovering wavelet coefficients in his seminal paper 
\cite{donohoCS}, based on acquiring the coarse scale coefficients directly.  
This was later extended by Tsaig \& Donoho to so-called `multiscale CS' in \cite{DonohoCSext}, where distinct subbands were sensed 
separately.  See also Romberg's work \cite{RombergCompImg}, and as well as 
Cand\`es \& Romberg \cite{Candes_Romberg}.

We also remark that, although motivated by wavelets, our 
definition is completely general, as are the theorems we present in \S 
\ref{s:main_thms_I} and \S \ref{s:main_thmsII}.  Moreover, and critically, we 
do not assume separation of the coefficients into distinct levels before 
sampling (as done above), which is often infeasible in practice (in particular, any 
application based on Fourier or Hadamard sampling).
Note also that in MRI similar sampling strategies to what we introduce 
here are found in most implementations of CS 
\cite{Lustig3,Lustig,VanderEtAlSpreadSpectrum,VanderEtAlVariable}.  
Additionally, a so-called ``half-half'' scheme (an example of a two-level 
strategy) was used by \cite{Candes_PNAS} in application of CS
in fluorescence microscopy, albeit without theoretical recovery guarantees.

\subsection{Asymptotic sparsity in levels}

The flip test, the discussion in \S \ref{Real_world_sparse} and Figure 
\ref{f:CS_LevelsSparsity} suggest that we need a different concept to 
sparsity. Given the structure of modern function systems such as wavelets and 
their generalizations, we propose the notion of sparsity in levels:
\defn{[Sparsity in levels]
\label{d:Asy_Sparse}
Let $x$ be an element of either $\bbC^N$ or $l^2(\bbN)$. For $r \in \bbN$ let 
$\mathbf{M} = (M_1,\ldots,M_r) \in \bbN^r$ with $1 \leq M_1 < \ldots < M_r$ 
and $\mathbf{s} = (s_1,\ldots,s_r) \in \bbN^r$, with $s_k \leq M_k - M_{k-1}$, 
$k=1,\ldots,r$, where $M_0 = 0$.  We say that $x$ is 
$(\mathbf{s},\mathbf{M})$-sparse if, for each $k=1,\ldots,r$,
$
\Delta_k : = \mathrm{supp}(x) \cap \{ M_{k-1}+1,\ldots,M_{k} \},
$
satisfies $| \Delta_k | \leq s_k$.  We denote the set of 
$(\mathbf{s},\mathbf{M})$-sparse vectors by $\Sigma_{\mathbf{s},\mathbf{M}}$.
}

\defn{[$(\mathbf{s},\mathbf{M}$)-term approximation]
Let $f = \sum_{j} x_j \varphi_j$, where $\{\varphi_j\}$ is some orthonormal basis of a Hilbert space and $x = (x_j )$ is 
an element of either $\bbC^N$ or $l^2(\bbN)$. 
We define the ($\mathbf{s},\mathbf{M}$)-term approximation
\be{
\label{sigma_s_m}
\sigma_{\mathbf{s},\mathbf{M}}(f) = \min_{\eta \in 
\Sigma_{\mathbf{s},\mathbf{M}} } \| x - \eta \|_{l^1}.
}
}

Typically, it is the case that $s_k / (M_k - M_{k-1}) \rightarrow 0$ as $k 
\rightarrow \infty$, in which case we say that $x$ is \textit{asymptotically 
sparse in levels}.  


\section{Main theorems I: the finite-dimensional case}\label{s:main_thms_I}
We now present the main theorems in the finite-dimensional setting.  In \S 
\ref{s:main_thmsII} we address the infinite-dimensional case. To avoid 
pathological examples we will assume  throughout that the total sparsity $s = 
s_1+\ldots+s_r \geq 3$. This is simply to ensure that $\log(s) \geq 1$, which 
is convenient in the proofs.

\subsection{Two-level sampling schemes}
We commence with the case of two-level sampling schemes.  Recall that in 
practice, signals are never exactly sparse (or sparse in levels), and their 
measurements are always contaminated by noise.  Let $f = \sum_{j} x_j 
\varphi_j$ be a fixed signal, and write 
$
y = P_{\Omega} \hat{f} + z = P_{\Omega} U x + z,
$
for its noisy measurements, where $z \in \mathrm{ran}(P_{\Omega})$ is a noise 
vector satisfying $\| z \| \leq \delta$ for some $\delta \geq 0$.  If $\delta$ 
is known, we now consider the following problem: 
\begin{equation}\label{eq:problem12_noise}
\min_{\eta \in \bbC^N} \|\eta\|_{l^1}\ \mbox{subject to $\|P_{\Omega}U\eta - y 
\| \leq \delta$.}
\end{equation}
Our aim now is to recover $x$ up to an error proportional to 
$\delta$ and the best approximation error 
$\sigma_{\mathbf{s},\mathbf{M}}(f)$. 

Before stating our theorem, it is useful 
to make the following definition. For $K \in \bbN$, we write $\mu_{K} = \mu(P^{\perp}_K U)$.  We now have the following:  

\thm{
\label{main_two_level_fin_dim2}
Let $U \in \bbC^{N \times N}$ be an isometry and $x \in \bbC^N$.  Suppose that 
$\Omega = \Omega_{\mathbf{N},\mathbf{m}}$ is a two-level sampling scheme, 
where $\mathbf{N} = (N_1,N_2)$, $N_2 = N$, and $\mathbf{m} = (N_1,m_2)$.  Let 
$(\mathbf{s},\mathbf{M})$, where $\mathbf{M} = (M_1,M_2) \in \bbN^2$, $M_1 < 
M_2$, $M_2 = N$, and $\mathbf{s} = (M_1,s_2) \in \bbN^2$, $s_2 \leq M_2-M_1$, 
be any pair such that the following holds: 
\begin{enumerate}
\item[(i)] we have 
\be{
\label{for_free}
\| P^{\perp}_{N_1} U P_{M_1} \| \leq \frac{\gamma}{ \sqrt{M_1}}
}
and $\gamma \leq s_2 \sqrt{\mu_{N_1}}$ for some $\gamma \in (0,2/5]$; 
\item[(ii)] for $\epsilon \in (0,e^{-1}]$, let
\bes{
m_2 \gtrsim (N-N_1) \cdot \log(\epsilon^{-1})   \cdot  \mu_{N_1}  \cdot s_2 
\cdot \log\left(N\right).
}
\end{enumerate}
Suppose that $\xi \in \mathbb{C}^N$ is a minimizer of 
(\ref{eq:problem12_noise}) with $\delta = \tilde \delta \sqrt{K^{-1}}$ and $K = 
(N_2-N_1)/m_2$.  Then, with probability exceeding $1-s\epsilon$, 
we have
\be{
\label{need_more_coffee} 
\| \xi - x \| \leq C  \cdot \left(\tilde \delta \cdot  \left(1+L 
\cdot\sqrt{s}\right) + \sigma_{\mathbf{s},\mathbf{M}}(f) \right),
}
for some constant $C$, where $\sigma_{\mathbf{s},\mathbf{M}}(f)$ is as in 
\R{sigma_s_m}, $L= 1+ 
\frac{\sqrt{\log_2\left(6\epsilon^{-1}\right)}}{\log_2(4KM\sqrt{s})}$.  If $m_2 = N-N_1$ then this holds with probability $1$.
}

To interpret Theorem \ref{main_two_level_fin_dim2}, and in particular, 
show how it overcomes the coherence barrier, we note the following:

\begin{enumerate}
\item[(i)] The condition $\|P_{N_1}^{\perp} U P_{M_1}\| \leq  \frac{2}{5 
\sqrt{M_1}}$ (which is always satisfied for some $N_1$) implies that fully 
sampling the first $N_1$ measurements allows one to recover the first $M_1$ 
coefficients of $f$.
\item[(ii)] To recover the remaining $s_2$ coefficients we require, up to log 
factors, an additional
$
m_2 \gtrsim (N-N_1) \cdot \mu_{N_1} \cdot s_2,
$
measurements, taken randomly from the range $M_1+1,\ldots,M_2$.  In 
particular, if $N_1$ is a fixed fraction of $N$, and if $\mu_{N_1} = 
\ord{N^{-1}_1}$, such as for wavelets with Fourier measurements (Theorem 
\ref{t:Wavelet_Asy_Inc}), then one requires only $m_2 \gtrsim s_2$ additional 
measurements to recover the sparse part of the signal.
\end{enumerate}
Thus, in the case where $x$ is asymptotically sparse, we require a fixed 
number $N_1$ measurements to recover the nonsparse part of $x$, and then a 
number $m_2$ depending on $s_2$ and the asymptotic coherence $\mu_{N_1}$ to 
recover the sparse part.

\rem{
\label{sparsitylevels}
It is not necessary to know the sparsity structure, i.e.\ the values 
$\mathbf{s}$ and $\mathbf{M}$, of the signal $f$ in order to implement the 
two-level sampling technique (the same also applies to the multilevel 
technique discussed in the next section).  Given a two-level scheme $\Omega = 
\Omega_{\mathbf{N},\mathbf{m}}$, Theorem \ref{main_two_level_fin_dim2} 
demonstrates that $f$ will be recovered exactly up to an error on the order of 
$\sigma_{\mathbf{s},\mathbf{M}}(f)$, where $\mathbf{s}$ and $\mathbf{M}$ are 
determined implicitly by $\mathbf{N}$, $\mathbf{m}$ and the conditions (i) and 
(ii) of the theorem.  Of course, some \textit{a priori} knowledge of 
$\mathbf{s}$ and $\mathbf{M}$ will greatly assist in selecting the parameters 
$\mathbf{N}$ and $\mathbf{m}$ so as to get the best recovery results.  
However, this is not strictly necessary for implementation.
}

\subsection{Multilevel sampling schemes}\label{ss:multilevel_thms}
We now consider the case of multilevel sampling schemes.  Before presenting 
this case, we need several definitions.  The first is key concept in this 
paper: namely, \emph{the local coherence}.

\defn{[Local coherence]\label{loc_coherence}
Let $U$ be an isometry of either $\bbC^{N}$ or $l^2(\bbN)$.
If $\mathbf{N} = (N_1,\ldots,N_r) \in \bbN^r$ and $\mathbf{M} = 
(M_1,\ldots,M_r) \in \bbN^r$ with $1 \leq N_1 < \ldots N_r $ and $1 \leq M_1 < 
\ldots < M_r $ the $(k,l)^{\rth}$ local coherence of $U$ with respect to 
$\mathbf{N}$ and $\mathbf{M}$ is given by
\eas{
\mu_{\mathbf{N},\mathbf{M}}(k,l) &= 
\sqrt{\mu(P^{N_{k-1}}_{N_{k}}UP^{M_{l-1}}_{M_{l}}) \cdot  
\mu(P^{N_{k-1}}_{N_{k}}U)},\quad k,l=1,\ldots,r,
}
where $N_0 = M_0 = 0$ and $P^{a}_{b}$ denotes the projection matrix 
corresponding to indices $\{a+1,\hdots, b\}$.  In the case where $U \in 
\cB(l^2(\bbN))$ (i.e.\ $U$ belongs to the space of bounded operators on 
$l^2(\bbN)$), we also define
\eas{
\mu_{\mathbf{N},\mathbf{M}}(k,\infty) &= 
\sqrt{\mu(P^{N_{k-1}}_{N_{k}}UP_{M_{r-1}}^\perp) \cdot  
\mu(P^{N_{k-1}}_{N_{k}}U)},\quad k=1,\ldots,r.
}}

Besides the local sparsities $s_k$, we shall also require the notion of a 
relative sparsity:
\defn{[Relative sparsity]
\label{S}
Let $U$ be an isometry of either $\mathbb{C}^{N}$ or $l^2(\bbN)$.  For 
$\mathbf{N} = (N_1,\ldots,N_r) \in \bbN^r$, $\mathbf{M} = (M_1,\ldots,M_r) \in 
\bbN^r$ with $1 \leq N_1 < \ldots < N_r$ and $1 \leq M_1 < \ldots < M_r$, 
$\mathbf{s} = (s_1,\ldots,s_r) \in \bbN^r$ and $1 \leq k \leq r$, the 
$k^{\rth}$ relative sparsity is given by
$
S_k = S_k(\mathbf{N},\mathbf{M},\mathbf{s}) =  \max_{\eta \in 
\Theta}\|P_{N_k}^{N_{k-1}}U\eta\|^2,
$
where $N_0 = M_0 = 0$ and $\Theta$ is the set
\bes{
\Theta = \{\eta : \|\eta\|_{l^{\infty}} \leq 1, 
|\mathrm{supp}(P_{M_l}^{M_{l-1}}\eta)| = s_l, \, l=1,\hdots, r\}.
}
}

We can now present our main theorem:

\thm{
\label{main_full_fin_noise2}
Let $U \in \mathbb{C}^{N \times N}$ be an isometry and $x \in 
\mathbb{C}^{N}$.  Suppose that $\Omega = \Omega_{\mathbf{N},\mathbf{m}}$ is a 
multilevel sampling scheme, where $\mathbf{N} = (N_1,\ldots,N_r) \in \bbN^r$, 
$N_r = N$, and $\mathbf{m} = (m_1,\ldots,m_r) \in \bbN^r$.  Let 
$(\mathbf{s},\mathbf{M})$, where $\mathbf{M} = (M_1,\ldots,M_r) \in \bbN^r$,  
$M_r = N$, and $\mathbf{s} = (s_1,\ldots,s_r) \in \bbN^r$, be any pair such 
that the following holds: for $\epsilon \in (0,e^{-1}]$ and $1 \leq k \leq r$,
\be{
\label{conditions31_levels}
1 \gtrsim \frac{N_k-N_{k-1}}{m_k} \cdot \log(\epsilon^{-1}) \cdot \left(
\sum_{l=1}^r \mu_{\mathbf{N},\mathbf{M}}(k,l) \cdot s_l\right) \cdot 
\log\left(N\right),
 }
 where
$
m_k \gtrsim \hat m_k \cdot  \log(\epsilon^{-1})  \cdot \log\left(N\right),
$
and $\hat{m}_k$ is such that
\be{
\label{conditions_on_hatm}
 1 \gtrsim \sum_{k=1}^r \left(\frac{N_k-N_{k-1}}{\hat m_k} - 1\right) \cdot 
 \mu_{\mathbf{N},\mathbf{M}}(k,l)\cdot \tilde s_k,
 }
for all $l=1,\ldots,r$ and all $\tilde{s}_1,\ldots,\tilde{s}_r \in (0,\infty)$ 
satisfying
\bes{
\tilde s_{1}+ \hdots + \tilde s_{r}  \leq s_1+ \hdots + s_r, \qquad \tilde s_k 
\leq S_k(\mathbf{N},\mathbf{M},\mathbf{s}).
}
Suppose that $\xi \in \bbC^N$ is a minimizer of (\ref{eq:problem12_noise}) with $\delta = \tilde \delta \sqrt{K^{-1}}$ and $K = \max_{1 \leq k \leq r} \{ (N_{k}-N_{k-1})/m_k \}$.  
Then, with probability exceeding $1-s\epsilon$,  where $s = s_1+\ldots+s_r$, 
we have that 
\bes{
\|\xi - x\|\leq  C \cdot \left(\tilde \delta \cdot \left(1 +L 
\cdot\sqrt{s}\right) +\sigma_{\mathbf{s},\mathbf{M}}(f) \right),}
for some constant $C$, where $\sigma_{\mathbf{s},\mathbf{M}}(f)$ is as in 
\R{sigma_s_m}, $
L= 1+ \frac{\sqrt{\log_2\left(6\epsilon^{-1}\right)}}{\log_2(4KM\sqrt{s})}$.  If $m_k = 
N_{k}-N_{k-1}$, $1 \leq k \leq r$, then this holds with probability $1$.
}

The key component of this theorem is the bounds \R{conditions31_levels} and 
\R{conditions_on_hatm}.  Whereas the standard CS estimate 
\R{m_est_Candes_Plan} relates the total number of samples $m$ to the global 
coherence and the global sparsity, these bounds now relate the local sampling 
$m_k$ to the local coherences $\mu_{\mathbf{N},\mathbf{M}}(k,l)$ and local and 
relative sparsities $s_k$ and $S_k$. In particular, by relating these local 
quantities this theorem conforms with the conclusions of the flip test in \S 
\ref{flip}: namely, that the optimal sampling strategy must depend on the signal 
structure.  This is exactly what is described in \R{conditions31_levels} 
and \R{conditions_on_hatm}.

On the face of it, the bounds  \R{conditions31_levels} and 
\R{conditions_on_hatm} may appear somewhat complicated, not least because they 
involve the relative sparsities $S_k$.  As we next show, however, they are 
indeed sharp in the sense that they reduce to the correct 
information-theoretic limits in several important cases.  Furthermore, in the 
important case of wavelet sparsity with Fourier sampling, they can be used to 
provide near-optimal recovery guarantees.  We discuss this in \S 
\ref{ss:Fourier_wavelets}. Note, however, that to do this it is first 
necessary to generalize Theorem \ref{main_full_fin_noise2} to the 
infinite-dimensional setting, which we do in \S \ref{s:main_thmsII}.

\subsubsection{Sharpness of the estimates -- the block-diagonal case}
Suppose that $\Omega = \Omega_{\mathbf{N},\mathbf{m}}$ 
is a multilevel sampling scheme, where $\mathbf{N} = (N_1,\ldots,N_r) \in 
\bbN^r$ and $\mathbf{m} = (m_1,\ldots,m_r) \in \bbN^r$.  Let 
$(\mathbf{s},\mathbf{M})$, where $\mathbf{M} = (M_1,\ldots,M_r) \in \bbN^r$,
 and suppose for simplicity that $\mathbf{M} = \mathbf{N}$.
Consider the block-diagonal matrix
$$
A = A_1 \oplus \hdots \oplus A_r \in \bbC^{N \times N}, \quad 
A_k \in 
\mathbb{C}^{(N_k - N_{k-1}) \times (N_k - N_{k-1})}, 
\quad
A_k^*A_k = I,
$$
where $N_0 = 0$.  Note that in this setting we have
$
S_k = s_k,
$ 
$
\mu_{\mathbf{N},\mathbf{M}}(k,l) = 0, 
$
$k \neq l.
$
Also, since 
$\mu(\mathbf{N},\mathbf{M})(k,k) = \mu(A_k)$, equations 
\R{conditions31_levels} and \R{conditions_on_hatm} reduce to
\bes{
1 \gtrsim \frac{N_k - N_{k-1}}{m_k} \cdot  \log (\epsilon^{-1}) \cdot \mu(A_k) 
\cdot s_k \cdot \log(N), \quad 1 \gtrsim \left(\frac{N_k-N_{k-1}}{\hat m_k} - 
1\right)\cdot \mu(A_k) 
\cdot s_k .
}
In particular, it suffices to take
\be{
\label{block_samples}
m_k \gtrsim (N_k - N_{k-1} ) \cdot \log (\epsilon^{-1})  
\cdot\mu(A_k) \cdot s_k \cdot \log(N),\quad 1 \leq k \leq r.
}
This is exactly as one expects: the number of measurements in the $k^{\rth}$ 
level depends on the size of the level multiplied by the local 
coherence and the sparsity in that level. Note that this result recovers 
the standard one-level results in finite dimensions 
\cite{BAACHGSCS,Candes_Plan} up to a slight deterioration in the probability bound to $1-s \epsilon$.  Specifically, 
the usual bound would be $1-\epsilon$.  The question as to 
whether or not this $s$ can be removed in the multilevel setting is open, 
although such a result
would be more of a cosmetic improvement.

\subsubsection{Sharpness of the estimates -- the non-block diagonal case}
The previous argument demonstrated that Theorem \ref{main_full_fin_noise2} is 
sharp, up to the probability term, in the sense that it reduces to the usual 
estimate \R{block_samples} for block-diagonal matrices, i.e. $S_k = s_k$.  
 This is not true in the general 
setting.  Clearly, 
$
S_k \leq s = s_1+\ldots+s_r.
$
However in general there is usually \textit{interference} between different 
sparsity levels, which means that $S_k$ need not have anything to do with 
$s_k$, or can indeed be proportional to the total sparsity $s$.  
This may seem an undesirable aspect of the theorems, since 
$S_k$ may be significantly larger than $s_k$, and thus the estimate on the 
number of measurements $m_k$ required in the $k^{\rth}$ level may also be much 
larger than the corresponding sparsity $s_k$.  Could it therefore be that the 
$S_k$s are an unfortunate artefact of the proof?  As we now show by example, 
this is not the case.

Let $N = r n$ for some $n \in 
\bbN$ and $\mathbf{N} = \mathbf{M} = 
(n,2n,\ldots,r n)$.  Let $W \in \bbC^{n \times n}$ and $V \in \bbC^{r\times 
r}$ be isometries and consider the matrix
$$
A = V \otimes W,
$$
where $\otimes$ is the usual Kronecker product.  Note that $A \in \bbC^{N 
\times N}$ is also an isometry.  Now suppose that $x = (x_1,\ldots,x_r) \in 
\bbC^N$ is an $(\mathbf{s},\mathbf{M})$-sparse vector, where each $x_k\in 
\bbC^n$ is $s_k$-sparse.  Then
$
A x= y,\quad y = (y_1,\ldots,y_r),\ y_k = W z_k,\  z_k = \sum^{r}_{l=1} 
v_{kl} x_l.
$
Hence the problem of recovering $x$ from measurements $y$ with an 
$(\mathbf{N},\mathbf{m})$-multilevel strategy decouples into $r$ problems of 
recovering the vector $z_k$ from the measurements $y_k = W z_k$, 
$k=1,\ldots,r$.  Let $\hat{s}_k$ denote the sparsity of $z_k$.  Since the 
coherence provides an information-theoretic limit \cite{Candes_Plan}, one 
requires at least
\be{
\label{inf_limit}
m_k \gtrsim n \cdot \mu(W) \cdot \hat{s}_k \cdot \log(n),\quad 1 \leq k \leq r.
}
measurements at level $k$ in order to recover each $z_k$, and therefore 
recover $x$, regardless of the reconstruction method used.  We now consider 
two examples of this setup:

\examp{
Let $\pi : \{1,\ldots,r\} \rightarrow \{1,\ldots,r\}$ be a permutation and let 
$V$ be the matrix with entries $v_{kl} = \delta_{l,\pi(k)}$.  Since $z_k = 
x_{\pi(k)}$ in this case,  the lower bound \R{inf_limit} reads
\be{
\label{inf_limit_perm}
m_k \gtrsim n \cdot \mu(W) \cdot s_{\pi(k)} \cdot \log(n),\quad 1 \leq k \leq r.
}
Now consider Theorem \ref{main_full_fin_noise2} for this matrix.  First, we 
note that $S_k = s_{\pi(k)}$.  In particular, $S_k$ is completely unrelated to 
$s_k$.  Substituting this into Theorem \ref{main_full_fin_noise2} and noting 
that $\mu_{\mathbf{N},\mathbf{M}}(k,l) = \mu(W) \delta_{l,\pi(k)}$ in this 
case, we arrive at the condition
$
m_k \gtrsim n  \cdot \mu(W)\cdot s_{\pi(k)} \cdot \left ( \log (\epsilon^{-1}) 
+ 1 \right )  \cdot \log(n r),
$
which is equivalent to \R{inf_limit_perm} provided $r \lesssim n$.
}

\examp{
Now suppose that $V$ is the $r \times r$ DFT matrix.  Suppose also that $s 
\leq n / r$ and that the $x_k$'s have disjoint support sets, i.e.\ 
$\mathrm{supp}(x_k) \cap \mathrm{supp}(x_l) = \emptyset$, $k \neq l$.  Then by 
construction, each $z_k$ is $s$-sparse, and therefore the lower bound 
\R{inf_limit} reads
$
m_k \gtrsim n \cdot \mu(W) \cdot s \cdot \log n,$
for $1 \leq k \leq r.$
After a short argument, one finds that $s/r \leq S_k \leq s$ in this case.  
Hence, $S_k$ is typically much larger than $s_k$.  Moreover, after noting that 
$\mu_{\mathbf{N},\mathbf{M}}(k,l) = \frac{1}{r} \mu(W)$, we find that Theorem 
\ref{main_full_fin_noise2} gives the condition
$
m_k \gtrsim n \cdot  \mu(W) \cdot s \cdot \left ( \log (\epsilon^{-1}) + 1 
\right ) \cdot \log(n r).
$
Thus, Theorem \ref{main_full_fin_noise2} obtains the lower bound in this case 
as well. 
}

\subsubsection{Sparsity leads to pessimistic reconstruction 
guarantees}\label{sss:sparsity_pessimistic}
The flip test demonstrates that any sparsity-based theory of CS cannot describe the quality of the reconstructions seen in practice.  To 
conclude this section, we now use the block-diagonal case to further emphasize 
the need for theorems that go beyond sparsity, such as Theorems 
\ref{main_two_level_fin_dim2} and \ref{main_full_fin_noise2}. 
To see this, consider the block-diagonal matrix
$$
U = U_1 \oplus \hdots \oplus U_r, \qquad U_k \in \bbC^{(N_k-N_{k-1}) \times (N_k - 
N_{k-1})},
$$
where each $U_k$ is perfectly incoherent, i.e.\ $\mu(U_k) = (N_k - 
N_{k-1})^{-1}$, and suppose we take $m_k$ measurements within each block 
$U_k$.  Let $x \in \bbC^N$ be the signal we wish to recover, where $N = N_r$.  
The question is, how many samples $m = m_1+\ldots+m_r$ do we require?

 Suppose we assume that $x$ is $s$-sparse, where $s \leq \min_{k=1,\ldots,r} 
 \{ N_k - N_{k-1} \}$.  Given no further information about the sparsity 
 structure, it is necessary to take $m_k \gtrsim s \log(N)$ measurements in 
 each block, giving $m \gtrsim r s \log(N)$ in total.  However, suppose now 
 that $x$ is known to be $s_k$-sparse within each level, i.e.\ 
 $| 
 \mathrm{supp}(x) \cap \{ N_{k-1}+1,\ldots,N_k \} | = s_k.
 $  Then we now 
 require only $m_k \gtrsim s_k \log(N)$, and therefore $m \gtrsim s \log(N)$ 
 total measurements.  Thus, structured sparsity leads to a significant saving 
 by a factor of $r$ in the total number of measurements required.  
 
  Although this may appear insignificant on the face of it, this factor represents a substantial saving in practice.  Given that a $512 \times 512$ 
image corresponds to $r=9$ wavelet scales, any sparsity-based theorem will 
lead to a nine-fold overestimate in the number of measurements required.  
Since $m \approx 5-10\%$ are typically necessary in applications (see, for 
example, Figure \ref{f:flip-tests}), such an overestimate, i.e.\ $m \approx 45- 90\%$, is therefore of little or no practical use. Although this argument is based on a simplified model, the block-diagonal structure described above is a good approximation to the Fourier/wavelets recovery problem, which we discuss in detail in \S \ref{ss:Fourier_wavelets}.

\section{Main theorems II: the infinite-dimensional case}\label{s:main_thmsII}
Finite-dimensional CS is suitable in many cases.  However, 
there are some important problems where it can lead to significant 
problems, since the underlying problem is continuous/analog.  Discretization 
of the problem in order to produce a finite-dimensional, vector-space model 
can lead to substantial errors \cite{BAACHGSCS, Adcock_Hansen_Book, 
CalderbankEtAlBasisMismatch,StrohmerMeasure}, due to the phenomenon of model 
mismatch.

To address this issue, a theory of infinite-dimensional CS 
was introduced by Adcock \& Hansen in \cite{BAACHGSCS}, based on a new 
approach to classical sampling known as \textit{generalized sampling} 
\cite{BAACHShannon,BAACHAccRecov,AHHTillposed,BAACHOptimality,AHPWavelet,hrycakIPRM}.  We 
describe this theory next.  Note that this infinite-dimensional CS model has also been advocated for and implemented in MRI by 
Guerquin--Kern, H\"aberlin, Pruessmann \& Unser \cite{Unser}.  Note also that sampling theories such as generalized sampling and finite rate of innovation \cite{VetterliFRI} are infinite-dimensional, and hence it is most natural that CS has an infinite-dimensional theory as well.

\subsection{Infinite-dimensional CS}\label{ss:inf_dim_CS}
Suppose 
that $\cH$ is a separable Hilbert space over $\bbC$, and let $\{ \psi_j \}_{j 
\in \bbN}$ be an orthonormal basis on $\cH$ (the sampling basis).  Let $\{ 
\varphi_j \}_{j \in \bbN}$ be an orthonormal system in $\cH$ (the sparsity 
system), and suppose that
\be{
\label{U_inf_dim}
U = (u_{ij})_{i,j \in \bbN},\qquad u_{ij} = \ip{\varphi_j}{\psi_i},
}
is an infinite matrix.  We may consider $U$ as an element of $\cB(l^2(\bbN))$; 
the space of bounded operators on $l^2(\bbN)$.  As in 
the finite-dimensional case, $U$ is an isometry, and we may define its 
coherence $\mu(U) \in (0,1]$ analogously to \R{coherence_def}.  
We want to recover $f = \sum_{j\in \mathbb{N}} x_j \varphi_j \in \cH$  from a 
small number of the measurements
$
\hat f = \{\hat{f}_j\}_{j \in \mathbb{N}}, 
$
where $\hat{f}_j = \ip{f}{\psi_j}$.
To do this, we introduce a second parameter $N \in \bbN$, and let $\Omega$ be 
a randomly-chosen subset of indices $1,\ldots,N$ of size $m$.  
Unlike in finite dimensions, we now consider two cases.  Suppose first that 
$P^{\perp}_M x = 0$, i.e.\ $x$ has no tail.  Then we solve 
\be{
\label{GSCS_l1_semi}
\inf_{\eta \in l^1(\bbN) } \| \eta \|_{l^1}\ \mbox{subject to $\| P_{\Omega} 
UP_M \eta - y \| \leq \delta$},
}
where $y = P_{\Omega} \hat{f} + z$ and $z \in \mathrm{ran}(P_{\Omega})$ is a noise vector satisfying $\| 
z \| \leq \delta$, and $P_{\Omega}$ is the projection operator corresponding 
to the index set $\Omega$.  In \cite{BAACHGSCS} it was proved that any 
solution to \R{GSCS_l1_semi} recovers $f$ exactly up to an error determined by 
$\sigma_{s,M}(f)$, provided $N$ and $m$ satisfy the so-called \textit{weak 
balancing property} with respect to $M$ and $s$ (see Definition 
\ref{balancing_property}, as well as Remark \ref{r:balancing_role} for a 
discussion), and provided
\be{
\label{inf_CS_bound}
m \gtrsim  \mu(U)\cdot  N \cdot s \cdot \left(1+\log (\epsilon^{-1}) \right ) 
\cdot \log \left ( m^{-1} M N \sqrt{s} \right ).
}
As in the finite-dimensional case, which turns out to be a corollary of this 
result, we find that $m$ is on the order of the sparsity $s$ whenever $\mu(U)$ 
is sufficiently small.

In practice, the condition $P^{\perp}_M x = 0$ is unrealistic.  In the more 
general case, $P^{\perp}_M x \neq 0$, we solve the following problem:
\be{
\label{GSCS_l1}
\inf_{\eta \in l^1(\bbN) } \| \eta \|_{l^1}\ \mbox{subject to $\| P_{\Omega} 
U\eta - y \| \leq \delta$}.
}
In \cite{BAACHGSCS} it was shown that any solution of \R{GSCS_l1} recovers $f$ 
exactly up to an error determined by $\sigma_{s,M}(f)$, provided $N$ and $m$ 
satisfy the so-called \textit{strong balancing property} with respect to $M$ 
and $s$ (see Definition \ref{balancing_property}), and provided a bound 
similar to \R{inf_CS_bound} holds, where the $M$ is replaced by a slightly 
larger constant (we give the details in the next section in the more general 
setting of multilevel sampling).  Note that \R{GSCS_l1} cannot be solved 
numerically, since it is infinite-dimensional.  Therefore in practice we 
replace \R{GSCS_l1} by 
\be{
\label{GSCS_l1_trunc}
\inf_{\eta \in l^1(\bbN) } \| \eta \|_{l^1}\ \mbox{subject to $\| P_{\Omega} U 
P_R\eta - y \| \leq \delta$},
}
where $R$ is taken sufficiently large.  See \cite{BAACHGSCS} for more information.

\subsection{Main theorems}
We first require the definition of the so-called \textit{balancing property} 
\cite{BAACHGSCS}:
\begin{definition}[Balancing property]\label{balancing_property}
Let $U \in \mathcal{B}(l^2(\mathbb{N}))$ be an isometry.  Then $N \in \bbN$ 
and $K \geq 1$ satisfy the weak balancing property with respect to 
$U,$ $M \in \bbN$ and $s \in \bbN$ if 
\begin{equation}\label{conditions1}
\begin{split}
\|P_{M}U^* P_NUP_{M} -P_{M}\|_{l^{\infty} \rightarrow l^{\infty}} \leq  
\frac{1}{8}\left(\log_2^{1/2}\left(4 \sqrt{s}KM\right)\right)^{-1},
\end{split}
\end{equation}
where $\nm{\cdot}_{l^\infty \rightarrow l^{\infty}}$ is the norm on 
$\cB(l^{\infty}(\bbN))$. 
We say that $N$ and $K$ satisfy the strong balancing property with respect to 
$U,$ $M$ and $s$ if 
(\ref{conditions1}) holds, as well as
\begin{equation}\label{conditions34}
\begin{split}
\|P_M^{\perp}U^*P_NUP_M\|_{l^{\infty} \rightarrow l^{\infty}}
\leq \frac{1}{8}.
\end{split}
\end{equation}
\end{definition}

As in the previous section, we commence with the two-level case.  Furthermore, 
to illustrate the differences between the weak/strong balancing property, we 
first consider the setting of \R{GSCS_l1_semi}:

\thm{\label{main_two_level_weak}
Let $U \in \mathcal{B}(l^2(\mathbb{N}))$ be an isometry and $x \in 
l^1(\bbN)$.  Suppose that $\Omega = \Omega_{\mathbf{N},\mathbf{m}}$ is a 
two-level sampling scheme, where $\mathbf{N} = (N_1,N_2)$ and $\mathbf{m} = 
(N_1,m_2)$.  Let $(\mathbf{s},\mathbf{M})$, where $\mathbf{M} = (M_1,M_2) \in 
\bbN^2$, $M_1 < M_2$, and $\mathbf{s} = (M_1,s_2) \in \bbN^2$, be any pair 
such that the following holds: 
\begin{enumerate}
\item[(i)] we have $\| P^{\perp}_{N_1} U P_{M_1} \| \leq \frac{\gamma}{ 
\sqrt{M_1}}$ and $\gamma \leq s_2 \sqrt{\mu_{N_1}}$ for some $\gamma \in 
(0,2/5]$; 
\item[(ii)] the parameters
$
N=N_2,  K = (N_2-N_1)/m_2
$
satisfy the weak balancing property with respect to $U$, $M:= M_2$ and $s : = 
M_1 + s_2$;
\item[(iii)] for $\epsilon \in (0,e^{-1}]$, let
\bes{
m_2 \gtrsim (N-N_1) \cdot \log(\epsilon^{-1})  \cdot  \mu_{N_1}  \cdot s_2 
\cdot \log\left(K M \sqrt{s}\right).
}
\end{enumerate}
Suppose that $P^{\perp}_{M_2} x = 0$ and let $\xi \in l^1(\bbN)$ be a 
minimizer of \R{GSCS_l1_semi} with $\delta = \tilde \delta \sqrt{K^{-1}}$.  Then, with probability exceeding 
$1-s\epsilon$, we have
\be{
\label{need_more_coffee2} 
\| \xi - x \| \leq C  \cdot \left(\tilde \delta \cdot \left(1+L 
\cdot\sqrt{s}\right) + \sigma_{\mathbf{s},\mathbf{M}}(f) \right),
}
for some constant $C$, where $\sigma_{\mathbf{s},\mathbf{M}}(f)$ is as in 
\R{sigma_s_m}, and $L= 1+ 
\frac{\sqrt{\log_2\left(6\epsilon^{-1}\right)}}{\log_2(4KM\sqrt{s})}$.  If 
$m_2 = N-N_1$ then this holds with probability $1$.
}

We next state a result for multilevel sampling in the more general setting of 
\R{GSCS_l1}.  For this, we require the following notation:
$
\tilde{M} = \min\{i\in\bbN: \max_{k\geq i}\| P_N U e_k \| \leq 
1/(32K\sqrt{s})\},
$
where $N$, $s$ and $K$ are as defined below.

\thm{
\label{main_full_inf_noise2}
Let $U \in \mathcal{B}(l^2(\mathbb{N}))$ be an isometry and $x \in 
l^1(\bbN)$.  Suppose that $\Omega = \Omega_{\mathbf{N},\mathbf{m}}$ is a 
multilevel sampling scheme, where $\mathbf{N} = (N_1,\ldots,N_r) \in \bbN^r$ 
and $\mathbf{m} = (m_1,\ldots,m_r) \in \bbN^r$.  Let 
$(\mathbf{s},\mathbf{M})$, where $\mathbf{M} = (M_1,\ldots,M_r) \in \bbN^r$, 
$M_1 < \ldots < M_r$, and $\mathbf{s} = (s_1,\ldots,s_r) \in \bbN^r$, be any 
pair such that the following holds: 
\begin{enumerate}
\item[(i)] the parameters
$
N=N_r, K = \max_{k=1,\ldots,r} \left \{ \frac{N_{k}-N_{k-1}}{m_k} 
\right \},
$
satisfy the strong balancing property with respect to $U$, $M:= M_r$ and $s : 
= s_1+\ldots + s_r$;
\item[(ii)] for $\epsilon \in (0,e^{-1}]$ and $1 \leq k \leq r$,
\bes{
1 \gtrsim \frac{N_k-N_{k-1}}{m_k} \cdot \log(\epsilon^{-1})  \cdot \left(
\sum_{l=1}^r \mu_{\mathbf{N},\mathbf{M}}(k,l) \cdot s_l\right) \cdot 
\log\left(K \tilde M \sqrt{s}\right),
 }
 (with $\mu_{\mathbf{N},\mathbf{M}}(k,r)$ replaced by 
 $\mu_{\mathbf{N},\mathbf{M}}(k,\infty)$)
 and 
$
m_k \gtrsim \hat m_k \cdot  \log(\epsilon^{-1}) \cdot \log\left(K \tilde M 
\sqrt{s}\right),
$
where $\hat m_k$ satisfies (\ref{conditions_on_hatm}).
\end{enumerate}
Suppose that $\xi \in l^1(\bbN)$ is a minimizer of 
(\ref{GSCS_l1}) with $\delta = \tilde \delta \sqrt{K^{-1}}$.  Then, with probability exceeding $1-s\epsilon$, 
\bes{
\|\xi - x\|\leq  C  \cdot \left(\tilde \delta \cdot \left(1+L 
\cdot\sqrt{s}\right) +\sigma_{\mathbf{s},\mathbf{M}}(f) \right),
}
for some constant $C$, where $\sigma_{\mathbf{s},\mathbf{M}}(f)$ is as in 
\R{sigma_s_m}, and $
L= C \cdot \left(1+ 
\frac{\sqrt{\log_2\left(6\epsilon^{-1}\right)}}{\log_2(4KM\sqrt{s})}\right).
$
 If $m_k = N_{k}-N_{k-1}$ for $1 \leq k \leq r$ then this holds with 
 probability $1$.
}

This theorem removes the condition 
in Theorem \ref{main_two_level_weak} that $x$ has zero tail. Note that the 
price to pay is the $\tilde M$ in the logarithmic term rather than $M$ ($\tilde M \geq M$ because of the balancing property).  Observe 
that $\tilde M$ is finite, and in the case of 
Fourier sampling with wavelets, we have that $\tilde M = \ord{K N}$ (see \S 
\ref{ss:Fourier_wavelets}).
Note that Theorem \ref{main_two_level_weak} has a strong form 
analogous to Theorem \ref{main_full_inf_noise2} which removes the tail 
condition. The only difference is the requirement of the strong, as opposed to 
the weak, balancing property, and the replacement of $M$ by $\tilde{M}$ in the 
log factor. Similarly, Theorem \ref{main_full_inf_noise2} has a weak form 
involving a tail condition.  For succinctness we do not state these.

\rem{\label{r:balancing_role}
The balancing property is the main difference between the finite- and 
infinite-dimensional theorems.    Its role is to ensure that the truncated 
matrix $P_N U P_M$ is close to an isometry.  In reconstruction problems, the 
presence of an isometry ensures stability in the mapping between measurements 
and coefficients \cite{BAACHShannon}, which explains the need for a such a 
property in our theorems.  As explained in \cite{BAACHGSCS}, without the 
balancing property the lack of stability in this mapping leads to 
numerically useless reconstructions.  Note that the balancing property is 
usually not satisfied for $N=M$.  In general, one requires $N > M$ for the balancing property to 
hold.  However, there is always a finite $N$ for which it is satisfied, since 
the infinite matrix $U$ is an isometry.  For details we refer to 
\cite{BAACHGSCS}.  We will provide specific estimates in \S 
\ref{ss:Fourier_wavelets} for the required magnitude of $N$ in the case of 
Fourier sampling with wavelet sparsity.
}

\subsection{The need for infinite-dimensional CS}\label{inf_dim_exp}
\begin{figure}[t]
\begin{center}
\small
\begin{tabular}{@{}c@{\hspace{0.005\linewidth}}c@{\hspace{0.005\linewidth}}c@{\hspace{0.005\linewidth}}c@{}}
\includegraphics[width=0.24\linewidth]{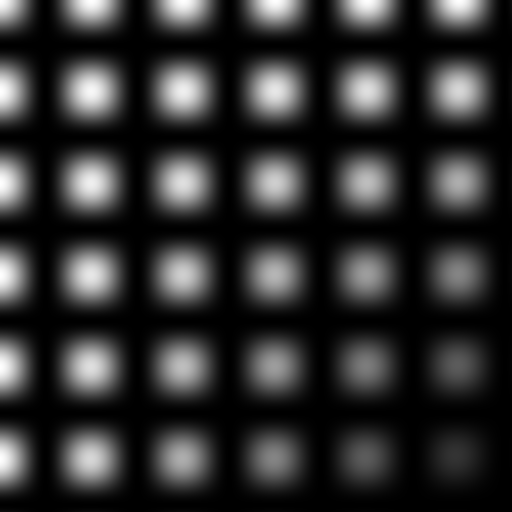}&
\includegraphics[width=0.24\linewidth]{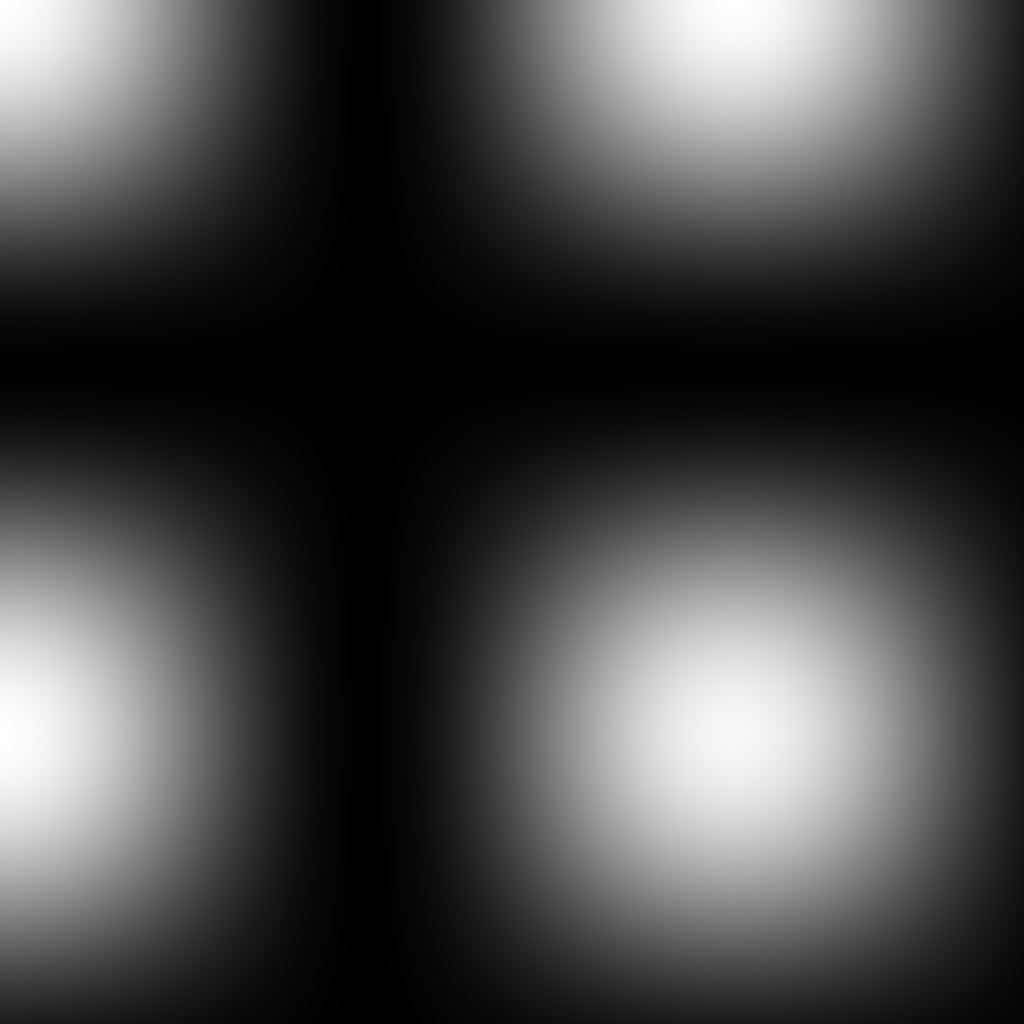}&
\includegraphics[width=0.24\linewidth]{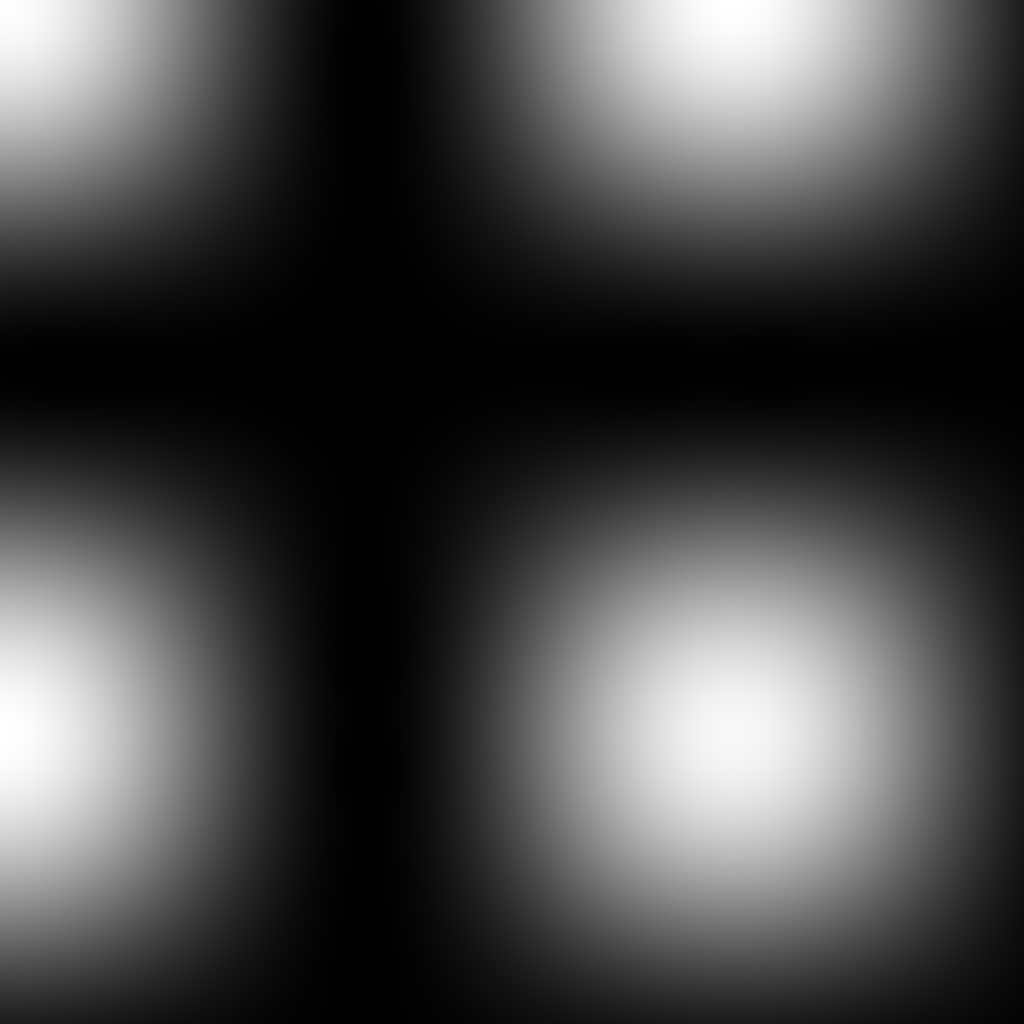}&
\includegraphics[width=0.24\linewidth]{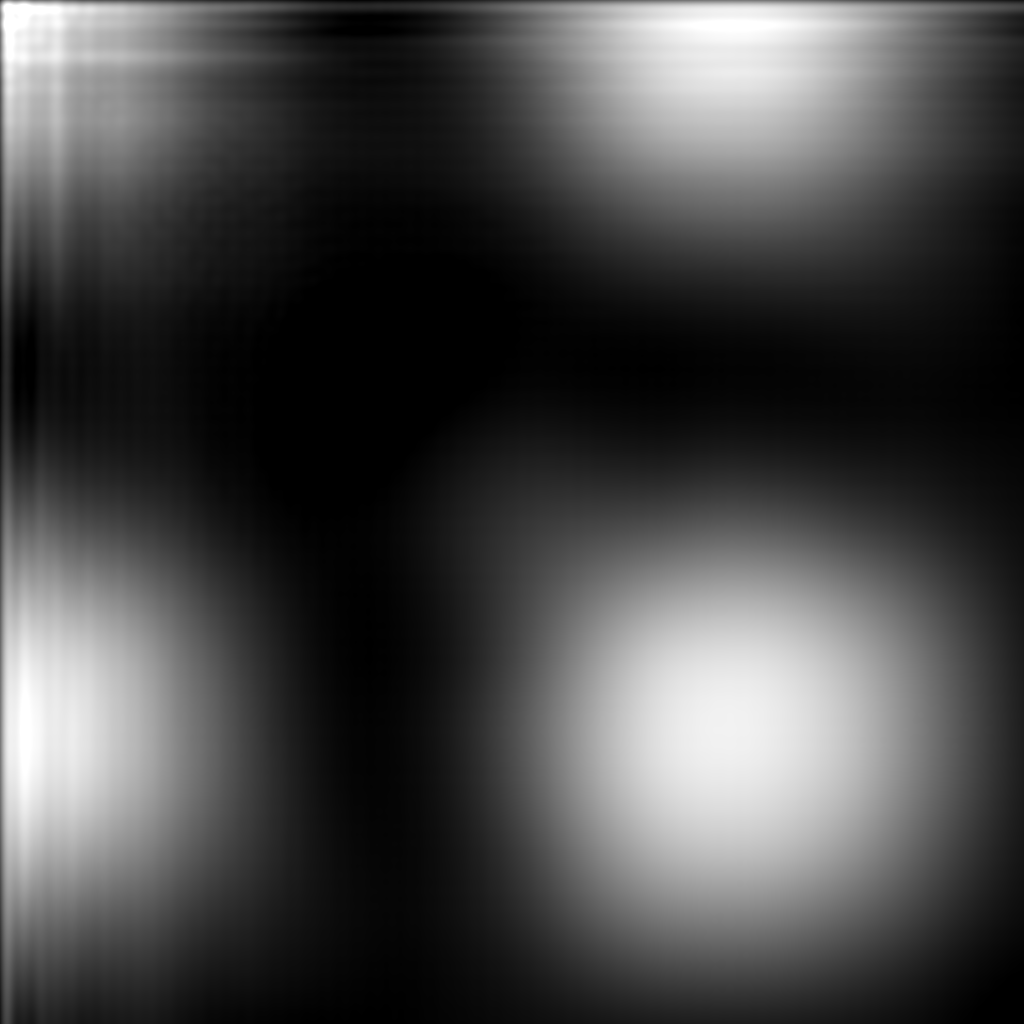}\\
Original & Original (zoomed) & Infinite-dim. CS (zoomed) & Finite-dim. CS (zoomed) \\
 & & Err 0.6\% & Err 12.7\% 
\end{tabular}
\caption{Subsampling 6.15\%. Both reconstructions are based on identical sampling information. 
}
\label{inf_dim_CS}
\end{center}
\end{figure}

As mentioned, infinite-dimensional CS is necessary to avoid the artefacts that are introduced when one applies finite-dimensional CS techniques to analog problems.  To illustrate this, we consider the problem of recovering a smooth phantom, i.e.\ a $C^{\infty}$ bivariate function, from its Fourier data.  Note that this arises in both electron microscopy and spectroscopy.  The test function is $f(x,y) = \cos^2(17\pi x/2)\cos^2(17\pi y/2)\exp(-x-y)$.  In Figure \ref{inf_dim_CS}, we compare finite-dimensional CS, based on solving (\ref{eq:problem12_noise}) with $U = U_{\mathrm{dft}}V^{-1}_{\mathrm{dwt}}$ (discrete Fourier and wavelet transform respectively) with infinite-dimensional CS, which solves \R{GSCS_l1_trunc} with the Fourier basis $\{\psi_j\}_{j\in \mathbb{N}}$ and boundary wavelet basis $\{\varphi_j\}_{j\in \mathbb{N}}$.  The improvement one gets is due to that fact that that the error in infinite-dimensional case is dominated by the wavelet approximation error,  whereas in the finite-dimensional case (due mismatch between the continuous Fourier samples and the discrete Fourier transform) the error is dominated by the Fourier approximation error.  As is well known \cite{mallat09wavelet}, wavelet approximation is superior to Fourier approximation and depends on the number of vanishing moments of the wavelet used (DB4 in this case).


\section{Recovery of wavelet coefficients from Fourier samples}\label{ss:Fourier_wavelets}
 As noted, Fourier sampling with wavelet sparsity is a 
 important reconstruction problem in CS, with numerous 
 applications ranging from medical imaging to seismology and interferometry. Here we consider the Fourier sampling basis $\{ \psi_j \}_{j 
\in \bbN}$ and wavelet reconstruction basis $\{ 
\varphi_j \}_{j \in \bbN}$ (see \S \ref{Setup} for a formal definition) with the infinite matrix $U$ as in (\ref{U_inf_dim}).
 The incoherence properties can be described 
 as follows.
\thm{
\label{t:Wavelet_Asy_Inc}
Let $U \in \cB(l^2(\bbN))$ be the matrix from (\ref{inf_U}) corresponding to the Fourier/wavelets 
system described in \S \ref{ss:Four_wave}.  Then $\mu(U) \geq \omega$, where $\omega$ is the sampling density, and $\mu(P^{\perp}_N U) , \mu(U 
P^{\perp}_N ) = \ord{N^{-1}}$. 
}

Thus, Fourier sampling with wavelet sparsity is indeed globally coherent, yet 
asymptotically incoherent. This result holds for essentially any 
wavelet basis in one dimension (see \cite{Jones2013Incoherence} for the multidimensional case). 
To recover wavelet coefficients, we seek to appl a multilevel 
sampling strategy, which raises the question: how do we design this strategy, 
and how many measurements are required? If the levels $\mathbf{M} = 
(M_1,\ldots,M_r)$ correspond to the wavelet scales, and 
$\mathbf{s}=(s_1,\ldots,s_r)$ to the sparsities within them, then the best one 
could hope to achieve is that the number of measurements $m_k$ in the 
$k^{\rth}$ sampling level is proportional to the sparsity $s_k$ in the 
corresponding sparsity level.  Our main theorem below shows that multilevel 
sampling can achieve this, up to an exponentially-localized factor and the 
usual log terms.

\begin{theorem} 
\label{Four_to_wave} 

Consider an orthonormal basis of compactly supported wavelets with a 
multiresolution analysis (MRA).  Let $\Phi$ and $\Psi$ denote the scaling 
function and mother wavelet respectively satisfying (\ref{wavelet_decay}) with $\alpha \geq 1$. Suppose that $\Psi$ has $v\geq 1$ vanishing moments, that the Fourier sampling density $\omega$ 
satisfies \R{eq:sampling_den_assump} and that the wavelets $\{\varphi_j\}$ are ordered according to (\ref{ordering}). 
Let 
$
f = \sum_{j=1}^{\infty} x_j \varphi_j.
$
Suppose that $\mathbf{M} = 
(M_1,\ldots,M_r)$ corresponds to wavelet scales with $M_k = \ord{2^{R_k}}$ with $R_k \in \mathbb{N}$, $R_{k+1} = a + R_k,$ $a \geq 1$, $k=1,\ldots,r$ and $\mathbf{s} = 
(s_1,\ldots,s_r)$ corresponds to the sparsities within them. Let  $\epsilon \in (0,e^{-1}]$ and let $\Omega = \Omega_{\mathbf{N},\mathbf{m}}$ be a multilevel sampling 
scheme such that the following holds:
 
\begin{enumerate}
\item[(i)] The parameters
$
N=N_r,
$ $K = \max_{k=1,\ldots,r},
$ $\{(N_{k}-N_{k-1})/m_k\},$ $M=M_r,$ $s =s_1+\ldots + s_r$
satisfy $N \gtrsim M^{1+1/(2\alpha -1)} \cdot 
\left(\log_2(4MK\sqrt{s})\right)^{1/(2\alpha -1)} $.  Alternatively, if $\Phi$ and $\Psi$ satisfy the slightly stronger Fourier decay property (\ref{eq:cond2_fdecay_mthm}), then $N\gtrsim
M \cdot \left(\log_2(4K M \sqrt{s})\right)^{1/(4\alpha-2)}$.
\item[(ii)]  For each $k=1,\ldots,r-1,$ $N_k = 2^{R_k} \omega^{-1}$ and for each $k=1,\ldots,r,$
\be{\label{eq:cond_wav1}
\begin{split}
m_k \gtrsim \log(\epsilon^{-1})  \cdot & \log(\tilde N) \cdot \frac{N_k - N_{k-1}}{N_{k-1}} \cdot \left( \hat s_k +\sum_{l=1}^{k-2}  s_l \cdot   2^{-(\alpha-1/2)A_{k,l}} +   \sum_{l=k+2}^r s_l \cdot 2^{-v B_{k,l}} \right),
\end{split}
 } 
where $A_{k,l} = R_{k-1} - 
R_{l}$, $B_{k,l} = R_{l-1}-R_k$,  $\tilde N = (K\sqrt{s})^{1+ 1/v} 
N$ and $\hat s_k = \max\{s_{k-1},s_k,s_{k+1}\}$ (see Remark \ref{notation_abuse}).
\end{enumerate}
  Then, with probability exceeding $1- s\epsilon$, any minimizer $\xi \in 
  l^1(\bbN)$ of (\ref{GSCS_l1}) with $\delta = \tilde \delta \sqrt{K^{-1}}$ satisfies
\bes{
\|\xi - x\|\leq  C  \cdot \left(\tilde \delta \cdot \left(1+ L 
\cdot\sqrt{s}\right) +\sigma_{\mathbf{s},\mathbf{M}}(f) \right),
}
for some constant $C$, where $\sigma_{\mathbf{s},\mathbf{M}}(f)$ is as in 
\R{sigma_s_m}, and $
L= C \cdot \left(1+ 
\frac{\sqrt{\log_2\left(6\epsilon^{-1}\right)}}{\log_2(4KM\sqrt{s})}\right).
$
 If $m_k = N_k - N_{k-1}$ for $1 \leq k \leq r$ then this holds with 
 probability $1$.
\end{theorem}

\begin{remark}\label{notation_abuse}
To avoid cluttered notation we have abused notation slightly in (ii) of Theorem \ref{Four_to_wave}. In particular, we interpret $s_0 = 0$, $\frac{N_k - N_{k-1}}{N_{k-1}} = N_1$ for $k=1$, and $\sum_{l=1}^{k-2}  s_l \cdot   2^{-(\alpha-1/2)A_{k,l}} = 0$ when $k \leq 2$.
\end{remark}

This theorem provides the first 
comprehensive explanation for the observed success of CS in applications based on the Fourier/wavelets model.  To see why, note that the key estimate \R{eq:cond_wav1} shows that $m_k$ need only scale as a linear combination of the local sparsities $s_l$, $1 \leq l \leq r$, and critically, the 
dependence of the sparsities $s_l$ for $l \neq k$ is exponentially diminishing 
in $|k-l|$.  Note that the presence of the off-diagonal terms 
is due to the previously-discussed phenomenon of interference, which occurs 
since the Fourier/wavelets system is not exactly block diagonal.  Nonetheless, 
the system is nearly block-diagonal, and this results in the near-optimality 
seen in \R{eq:cond_wav1}. 

Observe that \R{eq:cond_wav1} is in agreement with the flip test: if the local sparsities $s_k$ change, then the subsampling factors $m_k$ must also change to ensure the same quality reconstruction.  Having said that, it is straightforward to deduce from \R{eq:cond_wav1} the following global sparsity bound:
$$
m  \gtrsim s \cdot \log(\epsilon^{-1}) \cdot   \log(\tilde N), 
$$
where $m = m_1+\ldots+m_r$ is the total number of measurements and $s = s_1+\ldots+s_r$ is the total sparsity.  Note in particular the optimal exponent in the log factor.

\rem{
The Fourier/wavelets recovery problem was studied by Cand\`es \& Romberg in 
\cite{Candes_Romberg}.  Their result shows that if, in an ideal setting, an image can be first 
separated into separate wavelet subbands before sampling, then it can be 
recovered using approximately $s_k$ measurements (up to a log factor) in each 
sampling band.  Unfortunately, such separation into separate wavelet subbands before sampling is 
infeasible in most practical situations. Theorem \ref{Four_to_wave} improves 
on this result by removing this substantial restriction, with the sole penalty being the 
slightly worse bound \R{eq:cond_wav1}. 

Note also that a recovery result for bivariate Haar wavelets, as well as the related 
technique of TV minimization, was given in \cite{KrahmerWardCSImaging}. 
Similarly \cite{BigotBlockCS} analyzes block sampling strategies with 
application to MRI.  However, these results are based on sparsity, and 
therefore they do not explain how the sampling strategy will depend on the 
signal structure.
}

\begin{figure}[t]
\begin{center}
\small
\begin{tabular}{@{}c@{\hspace{0.003\linewidth}}c@{\hspace{0.003\linewidth}}c@{\hspace{0.003\linewidth}}c@{}}
\includegraphics[width=0.247\linewidth]{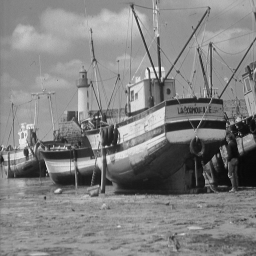}&
\includegraphics[width=0.247\linewidth]{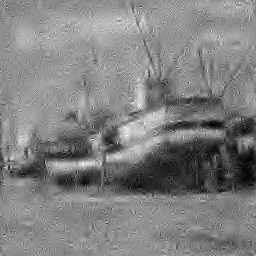}&
\includegraphics[width=0.247\linewidth]{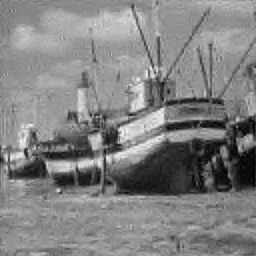}&
\includegraphics[width=0.247\linewidth]{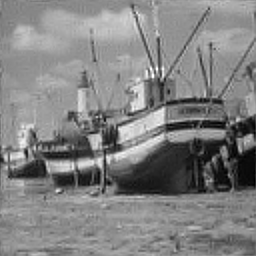}\\
Original image & Random Bernoulli & Multilevel Hadamard & Multilevel Fourier \\
 & Err = 15.7\% & Err = 9.6\% & Err 8.7\% 
\end{tabular}
\caption{12.5\% subsampling at \res{256} resolution using DB4 wavelets and various different measurements.
}
\label{structure}
\end{center}
\end{figure}


\subsection{Universality and RIP or structure?}\label{RIP?}
Theorem \ref{Four_to_wave} explains the success of CS when one is constrained to acquire Fourier measurements.  Yet, due primarily to the their high global coherence with wavelets, Fourier measurements are often viewed as suboptimal for CS.  If one had complete freedom to choose the measurements, and no physical constraints (such as are always present in MRI, for example), then standard CS intuition would suggest random Gaussian or Bernoulli measurements, since they are universal and satisfy the RIP.

However, in reality such measurements are actually highly suboptimal in the presence of structured sparsity.  This is demonstrated in Figure \ref{structure}, where an image is recovered from $m=8192$ measurements taken either as random Bernoulli or multilevel Hadamard or Fourier.  As is evident, the latter gives an error that is almost 50\% smaller.  The reason for this improvement is that whilst Fourier or Hadamard measurements are highly coherent with wavelets, they are asymptotically incoherent, and this can be exploited through multilevel random subsampling to recover asymptotically sparse wavelet coefficients.  Random Gaussian/Bernoulli measurements on the other hand cannot take advantage of this structure since they satisfy an RIP.

This observation is an important consequence of our theory.  In conclusion, whenever structured sparsity is present (such is the case in the majority of imaging applications, for example) there are substantial improvements to be gained by designing the measurements according to not just the sparsity, but also the additional structure.  For a more comprehensive discussion see \cite{Roman}, see also \cite{CalderbankCommInspCS, CalderbankNIPS}.

\section{Proofs}\label{s:proofs}
The proofs rely on some key propositions from which one can deduce the main 
theorems. The main work is to prove these proposition, and that will be done 
subsequently.

\subsection{Key results}

\begin{proposition}\label{sufficient}
Let  $U \in  \mathcal{B}(l^2(\mathbb{N}))$  and suppose that $\Delta$ and 
$\Omega = \Omega_1 \cup \hdots \cup \Omega_r$ (where the union is disjoint) 
are subsets of $\mathbb{N}.$ Let $x_0 \in \mathcal{H}$ and 
$z\in\mathrm{ran}(P_\Omega U)$ be such that $\| z\| \leq \delta$ for $\delta 
\geq 0$. Let $M \in \mathbb{N}$ and 
$y=P_\Omega Ux_0+z$ and $y_M=P_\Omega UP_Mx_0+z$.
Suppose that $\xi \in \mathcal{H}$ and $\xi_M \in \mathcal{H}$ satisfiy
\begin{equation}\label{eq:problem}
\|\xi\|_{l^1} = \inf_{\eta \in \mathcal{H}}
\{\|\eta\|_{l^1}: \| P_{\Omega}U\eta - y \| \leq \delta \}.
\end{equation}
\begin{equation}\label{eq:problemM}
\|\xi_M\|_{l^1} = \inf_{\eta \in \bbC^M}
\{\|\eta\|_{l^1}: \| P_{\Omega}UP_M\eta - y_M \| \leq \delta \}.
\end{equation}
If there exists a vector $\rho= U^* P_\Omega w$ such that 
\begin{enumerate}
\item[(i)]$\|P_{\Delta} U^*\left(q_1^{-1} P_{\Omega_1} \oplus \ldots \oplus 
q_r^{-1} P_{\Omega_r}\right)U P_{\Delta}- I_\Delta\| \leq \frac{1}{4}$
\item[(ii)] $\max_{i\in\Delta^c}\|\left(q_1^{-1/2} P_{\Omega_1} \oplus \ldots 
\oplus q_r^{-1/2} P_{\Omega_r}\right) U e_i \| \leq \sqrt{\frac{5}{4}}$
\item[(iii)]$\|P_\Delta \rho - \mathrm{sgn}(P_{\Delta}x_0)\|\leq \frac{q}{8}.$
\item[(iv)] $\|P_\Delta^\perp \rho\|_{l^\infty} \leq \frac{1}{2}$
\item[(v)] $\|w\| \leq L\cdot \sqrt{|\Delta|} $
\end{enumerate}
for some $L > 0$ and $0 < q_k \leq 1$, $k=1,\ldots,r$,
then we have that 
\bes{
\|\xi - x_0\|\leq C \cdot \left(\delta \cdot \left(\frac{1}{\sqrt{q}} + L 
\sqrt{s} \right) + \| P_{\Delta}^\perp x_0\|_{l^1}\right),
}
for some constant $C$, where $s = | \Delta |$ and $q= \min \{q_k\}_{k=1}^r$.  
Also, if (ii) is replaced by
 $$
 \max_{i\in\{1,\ldots,M\}\cap \Delta^c}\|\left(q_1^{-1/2} P_{\Omega_1} \oplus 
 \ldots \oplus q_r^{-1/2} P_{\Omega_r}\right) U e_i \| \leq \sqrt{\frac{5}{4}}
 $$ and (iv) is replaced by  $\|P_MP_\Delta^\perp \rho\|_{l^\infty} \leq 
 \frac{1}{2}$ then 
\be{\label{eq:error2}
\|\xi_M - x_0\|\leq C \cdot \left(\delta \cdot \left(\frac{1}{\sqrt{q}} + L 
\sqrt{s} \right) + \| P_M P_{\Delta}^\perp x_0\|_{l^1}\right).
}
\end{proposition}
\begin{proof}
First observe that (i) implies that $(P_{\Delta} U^*\left(q_1^{-1} 
P_{\Omega_1} \oplus \ldots \oplus q_r^{-1} P_{\Omega_r}\right)U 
P_{\Delta}|_{P_{\Delta}(\cH)})^{-1}$ exists and
\be{\label{eq:inv_exists}
\|(P_{\Delta} U^*\left(q_1^{-1} P_{\Omega_1} \oplus \ldots \oplus q_r^{-1} 
P_{\Omega_r}\right)U P_{\Delta}|_{P_{\Delta}(\cH)})^{-1}\| \leq \frac{4}{3}.
}
Also, (i) implies that
\begin{align}\label{first_est}
\|\left(q_1^{-1/2} P_{\Omega_1} \oplus \ldots \oplus q_r^{-1/2} 
P_{\Omega_r}\right) U P_\Delta\|^2 = \|P_\Delta U^*\left(q_1^{-1} P_{\Omega_1} 
\oplus \ldots \oplus q_r^{-1} P_{\Omega_r}\right) U P_\Delta\| \leq  
\frac{5}{4},
\end{align}
and 
\begin{equation}\label{sec_est}
\begin{split}
&\|P_\Delta U^*\left(q_1^{-1} P_{\Omega_1} \oplus \ldots \oplus q_r^{-1} 
P_{\Omega_r}\right)\|^2 = \|\left(q_1^{-1} P_{\Omega_1} \oplus \ldots \oplus 
q_r^{-1} P_{\Omega_r}\right)U P_\Delta\|^2  \\
&= \sup_{\|{\eta}\|=1} \|\left(q_1^{-1} P_{\Omega_1} \oplus \ldots \oplus 
q_r^{-1} P_{\Omega_r}\right)U P_\Delta \eta \|^2 \\
&=  \sup_{\|{\eta}\|=1} \sum_{k=1}^r \|q_k^{-1} P_{\Omega_k} U P_\Delta \eta 
\|^2 \leq \frac{1}{q} \sup_{\|{\eta}\|=1} \sum_{k=1}^r q_k^{-1}  
\|P_{\Omega_k} U P_\Delta \eta \|^2, \quad \frac{1}{q} = \max_{1\leq k \leq r} 
\{\frac{1}{q_k}\}\\
&=  \frac{1}{q} \sup_{\|{\eta}\|=1} \ip{P_\Delta U^* \left(\sum_{k=1}^r 
q_k^{-1} P_{\Omega_k}\right) U P_\Delta \eta}{\eta}\leq \frac{1}{q} \| 
P_\Delta U^* \left(q_1^{-1} P_{\Omega_1} \oplus \ldots \oplus q_r^{-1} 
P_{\Omega_r}\right) U P_\Delta\|.
\end{split}
\end{equation}
Thus, (\ref{first_est}) and (\ref{sec_est}) imply 
\be{\label{eq:side_est}
\|P_\Delta U^*\left(q_1^{-1} P_{\Omega_1} \oplus \ldots \oplus q_r^{-1} 
P_{\Omega_r}\right)\| \leq \sqrt{\frac{5}{4q}}.
}
Suppose that there exists a vector $\rho$, constructed with $y_0 = P_\Delta 
x_0$, satisfying  (iii)-(v). Let $\xi$ be a solution to (\ref{eq:problem}) and 
let $h=\xi - x_0$. 
Let $A_\Delta = P_{\Delta} U^*\left(q_1^{-1} P_{\Omega_1} \oplus \ldots \oplus 
q_r^{-1} P_{\Omega_r}\right)U P_{\Delta}|_{P_{\Delta}(\cH)}$. Then, it follows 
from (ii) and observations (\ref{eq:inv_exists}), (\ref{first_est}), 
(\ref{eq:side_est}) that
\begin{equation}
\begin{split}\label{eq:err_on_supp}
&\|P_\Delta h\| = \|A_\Delta^{-1} A_\Delta P_\Delta h\|\\
&\leq \|A_\Delta^{-1}\| \|P_\Delta U^*\left(q_1^{-1} P_{\Omega_1} \oplus 
\ldots \oplus q_r^{-1} P_{\Omega_r}\right)U (I-P_\Delta^\perp)h\|\\
&\leq \frac{4}{3} \|P_\Delta U^*\left(q_1^{-1} P_{\Omega_1} \oplus \ldots 
\oplus q_r^{-1} P_{\Omega_r}\right)\| \| P_\Omega U h\| \\
&+ \frac{4}{3}  \max_{i\in\Delta^c}\| P_\Delta U^*\left(q_1^{-1} P_{\Omega_1} 
\oplus \ldots \oplus q_r^{-1} P_{\Omega_r}\right) U e_i\|  \|P_\Delta^\perp 
h\|_{l^1} \\
&\leq \frac{4}{3} \|P_\Delta U^*\left(q_1^{-1} P_{\Omega_1} \oplus \ldots 
\oplus q_r^{-1} P_{\Omega_r}\right)\| \| P_\Omega U h\|\\
&+ \frac{4}{3} \nm{P_\Delta U^*\left(q_1^{-1/2} P_{\Omega_1} \oplus \ldots 
\oplus q_r^{-1/2}\right)}  \max_{i\in\Delta^c}\nm{ \left(q_1^{-1/2} 
P_{\Omega_1} \oplus \ldots \oplus q_r^{-1/2} P_{\Omega_r}\right) U e_i}  
\|P_\Delta^\perp h\|_{l^1} \\
&\leq  \frac{4\sqrt{5}}{3\sqrt{q}} \delta +  \frac{5}{3}\|P_\Delta^\perp 
h\|_{l^1},
\end{split}
\end{equation}
where in the final step we use $\| P_{\Omega} U h \| \leq \| P_{\Omega} U 
\zeta - y \| + \| z \| \leq 2 \delta$.
We will now obtain a bound for $\|P_\Delta^\perp h\|_{l^1}$. First note that
\begin{equation}
\begin{split}
\|h+x_0\|_{l^1} &= \|P_\Delta h + P_\Delta x_0\|_{l^1} + \|P_\Delta^\perp 
(h+x_0)\|_{l^1}\\
&\geq \Re \ip{P_\Delta h}{\mathrm{sgn}(P_\Delta x_0)} + \|P_\Delta 
x_0\|_{l^1}+ \|P_\Delta^\perp h\|_{l^1} - \|P_\Delta^\perp x_0\|_{l^1}\\
&\geq \Re \ip{P_\Delta h}{\mathrm{sgn}(P_\Delta x_0)} + \|x_0\|_{l^1}+ 
\|P_\Delta^\perp h\|_{l^1} - 2\|P_\Delta^\perp x_0\|_{l^1}.
\end{split}
\end{equation}
Since $\|x_0\|_{l^1} \geq \|h+x_0\|_{l^1}$, we have that 
\begin{equation}\label{eq:outside_supp_interm}
\|P_\Delta^\perp h\|_{l^1}\leq \abs{\ip{P_\Delta h}{\mathrm{sgn}(P_\Delta 
x_0)}} +  2\|P_{\Delta}^\perp x_0\|_{l^1}.
\end{equation}
We will use this equation later on in the proof, but before we do that observe 
that some basic adding and subtracting yields
\begin{equation}
\begin{split}
\abs{\ip{P_\Delta h}{\mathrm{sgn}(x_0)}}
&\leq \abs{\ip{P_\Delta h}{\mathrm{sgn}(P_\Delta x_0) - P_\Delta \rho}} + 
\abs{\ip{h}{\rho}}+\abs{\ip{P_\Delta^\perp h}{P_\Delta^\perp \rho}} \\
&\leq \| P_\Delta h\| \|\mathrm{sgn}(P_\Delta x_0) - P_\Delta \rho \| + 
\abs{\ip{P_\Omega U h}{w}} + \|P_\Delta^\perp h\|_{l^1}\|P_\Delta^\perp 
\rho\|_{l^\infty}\\
&\leq \frac{q}{8}\| P_\Delta h\| +  2L 
\delta\sqrt{s}+\frac{1}{2}\|P_\Delta^\perp h\|_{l^1}\\
&\leq \frac{\sqrt{5q}}{6} \delta + \frac{5q}{24}\|P_\Delta^\perp h\|_{l^1}  +  
2 L \delta\sqrt{s} + \frac{1}{2}\|P_\Delta^\perp h\|_{l^1}
\end{split}
\end{equation}
where the last inequality utilises (\ref{eq:err_on_supp}) and the penultimate 
inequality follows from properties (iii), (iv) and (v) of the dual vector 
$\rho$. Combining this with (\ref{eq:outside_supp_interm}) and the fact that 
$q \leq 1$ gives that
\begin{equation}\label{eq:outside_supp}
\|P_\Delta^\perp h\|_{l^1}\leq 
\delta\left(\frac{4\sqrt{5q}}{3}+8L\sqrt{s}\right) +  8\|P_{\Delta}^\perp 
x_0\|_{l^1}.
\end{equation}
Thus, (\ref{eq:err_on_supp}) and (\ref{eq:outside_supp}) yields:
\begin{equation}
\begin{split}
\|h\| &\leq \nm{P_\Delta h} + \nm{P_\Delta^\perp h} \leq   \frac{8}{3} \| 
P_{\Delta}^\perp h\|_{l^1} + \frac{4\sqrt{5}}{3\sqrt{q}} \delta \leq 
\left(8\sqrt{q} + 22 L \sqrt{s} + \frac{3}{\sqrt{q}}\right)\cdot \delta + 22 
\nm{P_\Delta^\perp x_0}_{l^1}.
\end{split}
\end{equation}
The proof of the second part of this proposition follows the proof as outlined 
above and we omit the details.
\end{proof}


The next two propositions give sufficient conditions for Proposition 
\ref{sufficient} to be true. But before we state them we need to define the 
following.

\begin{definition}\label{def:kappa}
Let $U$ be an isometry of either $\mathbb{C}^{N \times N}$ or 
$\cB(l^2(\bbN))$.  For $\mathbf{N} = (N_1,\ldots,N_r) \in \bbN^r$, $\mathbf{M} 
= (M_1,\ldots,M_r) \in \bbN^r$ with $1 \leq N_1 < \ldots < N_r$ and $1 \leq 
M_1 < \ldots < M_r $, $\mathbf{s} = (s_1,\ldots,s_r) \in \bbN^r$ and $1 \leq k 
\leq r$, let
$$
\kappa_{\mathbf{N},\mathbf{M}}(k,l) =  \max_{\eta 
\in\Theta}\|P^{N_{k-1}}_{N_{k}}UP^{M_{l-1}}_{M_{l}} \eta\|_{l^{\infty}} \cdot 
\sqrt{\mu(P^{N_{k-1}}_{N_{k}}U)}.
$$
where
\bes{
\Theta = \{\eta : \|\eta\|_{l^{\infty}} \leq 1, 
|\mathrm{supp}(P_{M_l}^{M_{l-1}}\eta)| = s_l, \, l=1,\hdots, r-1, \, 
|\mathrm{supp}(P_{M_{r-1}}^\perp \eta)| = s_r,\},
}
and $N_0 = M_0 = 0$.  
We also define
$$
\kappa_{\mathbf{N},\mathbf{M}}(k,\infty) =  \max_{\eta \in\Theta 
}\|P^{N_{k-1}}_{N_{k}}UP_{M_{r-1}}^\perp  \eta\|_{l^{\infty}} \cdot 
\sqrt{\mu(P^{N_{k-1}}_{N_{k}}U)}.
$$
\end{definition}

\begin{proposition}\label{main_prop}
Let $U \in \mathcal{B}(l^2(\mathbb{N}))$ be an isometry and $x \in 
l^1(\bbN)$.  Suppose that $\Omega = \Omega_{\mathbf{N},\mathbf{m}}$ is a 
multilevel sampling scheme, where $\mathbf{N} = (N_1,\ldots,N_r) \in \bbN^r$ 
and $\mathbf{m} = (m_1,\ldots,m_r) \in \bbN^r$.  Let 
$(\mathbf{s},\mathbf{M})$, where $\mathbf{M} = (M_1,\ldots,M_r) \in \bbN^r$, 
$M_1 < \ldots < M_r$, and $\mathbf{s} = (s_1,\ldots,s_r) \in \bbN^r$, be any 
pair such that the following holds:
\begin{itemize}
\item[(i)] The parameters
$N:=N_r,$ and $K := \max_{k=1,\ldots,r} (N_{k}-N_{k-1})/m_k,$
satisfy the weak balancing property with respect to $U$, $M:= M_r$ and $s : = 
s_1+\ldots + s_r$;
\item[(ii)] for $\epsilon > 0$ and $1 \leq k \leq r$,
\begin{equation}\label{conditions31}
 1 \gtrsim (\log(s\epsilon^{-1}) + 1) \cdot \frac{N_k-N_{k-1}}{m_k} \cdot 
 \left(
\sum_{l=1}^r \kappa_{\mathbf{N},\mathbf{M}}(k,l)\right) \cdot \log\left( K M 
\sqrt{s}\right),
\end{equation} 
\item[(iii)]
\begin{equation}\label{conditions32}
m_k \gtrsim(\log(s\epsilon^{-1}) + 1)  \cdot \hat m_k \cdot \log\left(K M 
\sqrt{s}\right),
\end{equation} 
where $\hat m_k$ satisfies
\bes{
 1 \gtrsim \sum_{k=1}^r \left(\frac{N_k-N_{k-1}}{\hat m_k} - 1\right) \cdot 
 \mu_{\mathbf{N},\mathbf{M}}(k,l)\cdot \tilde s_k, \qquad \forall \, l = 1, 
 \hdots, r,
 }
where
$
\tilde s_{1}+ \hdots + \tilde s_{r}  \leq s_1+ \hdots + s_r,
$ $\tilde s_k \leq S_k(s_1,\hdots, s_r)$
and $S_k$ is defined in (\ref{S}).
\end{itemize}
Then (i)-(v) in Proposition \ref{sufficient} follow with probability exceeding 
$1-\epsilon$, with 
 (ii) replaced by
 \begin{equation}\label{replaced}
 \max_{i\in\{1,\ldots,M\}\cap \Delta^c}\|\left(q_1^{-1/2} P_{\Omega_1} \oplus 
 \ldots \oplus q_r^{-1/2} P_{\Omega_r}\right) U e_i \| \leq \sqrt{\frac{5}{4}},
\end{equation}
(iv)  replaced by  $\|P_MP_\Delta^\perp \rho\|_{l^\infty} \leq \frac{1}{2}$ 
and $L$ in (v) is given by 
\begin{equation}\label{L}
L= C \cdot \sqrt{K}\cdot \left(1+ 
\frac{\sqrt{\log_2\left(6\epsilon^{-1}\right)}}{\log_2(4KM\sqrt{s})}\right).
\end{equation}
If 
$m_k = N_{k}-N_{k-1}$ for all $1 \leq k \leq r$ then (i)-(v) follow with 
probability one (with the alterations suggested above).
\end{proposition}

\begin{proposition}\label{main_prop2}
Let $U \in \mathcal{B}(l^2(\mathbb{N}))$ be an isometry and $x \in 
l^1(\bbN)$.  Suppose that $\Omega = \Omega_{\mathbf{N},\mathbf{m}}$ is a 
multilevel sampling scheme, where $\mathbf{N} = (N_1,\ldots,N_r) \in \bbN^r$ 
and $\mathbf{m} = (m_1,\ldots,m_r) \in \bbN^r$.  Let 
$(\mathbf{s},\mathbf{M})$, where $\mathbf{M} = (M_1,\ldots,M_r) \in \bbN^r$, 
$M_1 < \ldots < M_r$, and $\mathbf{s} = (s_1,\ldots,s_r) \in \bbN^r$, be any 
pair such that the following holds:
\begin{itemize}
\item[(i)] The parameters $N$ and $K$ (as in Proposition \ref{main_prop})
satisfy the strong balancing property with respect to $U$, $M = M_r$ and $s : 
= s_1+\ldots + s_r$;
\item[(ii)]  for $\epsilon > 0$ and $1 \leq k \leq r$,
\begin{equation}\label{conditions41}
 1 \gtrsim (\log(s\epsilon^{-1}) + 1) \cdot \frac{N_k-N_{k-1}}{m_k} \cdot 
 \left(  \kappa_{\mathbf{N},\mathbf{M}}(k,\infty) +
\sum_{l=1}^{r-1} \kappa_{\mathbf{N},\mathbf{M}}(k,l)\right) \cdot \log\left( K 
\tilde M \sqrt{s}\right),
\end{equation} 
\item[(iii)]
\begin{equation}\label{conditions42}
m_k \gtrsim(\log(s\epsilon^{-1}) + 1)  \cdot \hat m_k \cdot \log\left(K \tilde 
M \sqrt{s}\right),
\end{equation} 
where 
$\tilde{M} = \min\{i\in\bbN: \| \max_{j\geq i} P_N U P_{\{j\}} \| \leq 
1/(K32\sqrt{s})\}$,
and
$\hat m_k$ is as in Proposition \ref{main_prop}.
\end{itemize}
Then (i)-(v) in Proposition \ref{sufficient} follow with probability exceeding 
$1-\epsilon$ with $L$ as in (\ref{L}). If 
$m_k = N_{k}-N_{k-1}$ for all $1 \leq k \leq r$ then (i)-(v) follow with 
probability one.
\end{proposition}


\begin{lemma}[Bounds for $\kappa_{\mathbf{N},\mathbf{M}}(k,l)$]
For $k,l= 1, \ldots, r$
\label{bounds}
\be{\label{eq:kappa_bound}
\kappa_{\mathbf{N},\mathbf{M}}(k,l) \leq \min\left\{ 
\mu_{\mathbf{N},\mathbf{M}}(k,l) \cdot s_l, \, \sqrt{s_l \cdot 
\mu(P^{N_{k-1}}_{N_{k}}U)}\cdot \nm{P^{N_{k-1}}_{N_{k}}UP^{M_{l-1}}_{M_{l}}} 
\right\}.
}
Also, for $k= 1, \ldots, r$
\be{\label{eq:kappa_bound_infty}
\kappa_{\mathbf{N},\mathbf{M}}(k,\infty) \leq \min \left\{ 
\mu_{\mathbf{N},\mathbf{M}}(k,\infty) \cdot s_r, \, \sqrt{s_r \cdot 
\mu(P^{N_{k-1}}_{N_{k}}U)}\cdot \nm{P^{N_{k-1}}_{N_{k}}UP^\perp_{M_{r-1}}} 
\right\}.
}
\end{lemma}

\begin{proof}
For $k,l=1,\ldots,r$
\eas{
\kappa_{\mathbf{N},\mathbf{M}}(k,l) &=  \max_{\eta 
\in\Theta}\|P^{N_{k-1}}_{N_{k}}UP^{M_{l-1}}_{M_{l}} \eta\|_{l^{\infty}} \cdot 
\sqrt{\mu(P^{N_{k-1}}_{N_k}U)}\\
&= \max_{\eta \in\Theta} \max_{N_{k-1}< i\leq  N_k} \abs{\sum_{M_{l-1} < j 
\leq M_l} \eta_j u_{ij}} \cdot \sqrt{\mu(P^{N_{k-1}}_{N_k}U)} 
\\
& \leq s_l \cdot \sqrt{\mu(P^{N_{k-1}}_{N_k} U P^{M_{l-1}}_{M_l})} 
\cdot\sqrt{\mu(P^{N_{k-1}}_{N_k}U)} 
\leq s_l \cdot  \mu_{\mathbf{N},\mathbf{M}}(k,l)
}
since $|u_{ij} | \leq 1$, and similarly,
\eas{
&\kappa_{\mathbf{N},\mathbf{M}}(k,\infty) =  \max_{\eta 
\in\Theta}\|P^{N_{k-1}}_{N_{k}}UP_{M_{r-1}}^\perp \eta\|_{l^{\infty}} \cdot 
\sqrt{\mu(P^{N_{k-1}}_{N_k}U)}\\
&= \max_{\eta \in\Theta} \max_{N_{k-1}< i\leq  N_k} \abs{\sum_{M_{r-1} < j} 
\eta_j u_{ij}} \cdot \sqrt{\mu(P^{N_{k-1}}_{N_k}U)} \leq s_r \cdot  
\mu_{\mathbf{N},\mathbf{M}}(k,\infty).
}
Finally, it is straightforward to show that for $k,l=1,\ldots,r$,
$$
\kappa_{\mathbf{N},\mathbf{M}}(k,l)\leq \sqrt{s_l} \cdot 
\nm{P^{N_{k-1}}_{N_{k}}UP^{M_{l-1}}_{M_{l}}} \sqrt{\mu(P^{N_{k-1}}_{N_k}U)}
$$
and
$$
\kappa_{\mathbf{N},\mathbf{M}}(k,\infty)\leq \sqrt{s_r} \cdot 
\nm{P^{N_{k-1}}_{N_{k}}UP^\perp_{M_{r-1}}} \sqrt{\mu(P^{N_{k-1}}_{N_k}U)}.
$$
\end{proof}

We are now ready to prove the main theorems.

\begin{proof}[Proof of Theorems \ref{main_two_level_fin_dim2} and 
\ref{main_two_level_weak}]
It is clear that Theorem \ref{main_two_level_fin_dim2} follows from Theorem 
\ref{main_two_level_weak}, thus it remains to prove the latter.
We will apply Proposition \ref{main_prop} to a two-level sampling scheme 
$\Omega = \Omega_{\mathbf{N},\mathbf{m}}$, where $\mathbf{N} = (N_1,N_2)$ and 
$\mathbf{m} = (m_1,m_2)$ with $m_1 = N_1$ and $m_2 = m$. Also, consider 
$(\mathbf{s},\mathbf{M})$, where $\mathbf{s}=(M_1,s_2)$, $\mathbf{M} = 
(M_1,M_2)$. Thus, if $N_1,N_2, m_1,m_2 \in \bbN$ are such that
\bes{
N=N_2,\quad K = \max \left \{ \frac{N_2-N_1}{m_2} , \frac{N_1}{m_1} \right \}
}
satisfy the weak balancing property with respect to $U$, $M  = M_2$ and $s = 
M_1 + s_2$, we have that 
(i) - (v) in Proposition \ref{sufficient} follow with probability exceeding 
$1-s\epsilon$, with (ii) replaced by
 $$
 \max_{i\in\{1,\ldots,M\}\cap \Delta^c}\|\left( P_{N_1} \oplus  
 \frac{N_2-N_1}{m_2} P_{\Omega_2}\right) U e_i \| \leq \sqrt{\frac{5}{4}},
 $$ 
(iv)  replaced by  $\|P_MP_\Delta^\perp \rho\|_{l^\infty} \leq \frac{1}{2}$ 
and $L$ in (v) is given by (\ref{L}), if 
\begin{equation}\label{conditions3211}
 1 \gtrsim (\log(s\epsilon^{-1}) + 1) \cdot \frac{N-N_1}{m_2} \cdot 
 (\kappa_{\mathbf{N}, \mathbf{M}}(2,1)+\kappa_{\mathbf{N}, \mathbf{M}}(2,2))  
 \cdot \log\left( K M \sqrt{s}\right),
\end{equation}
\begin{equation}\label{conditions3222}
m_2 \gtrsim(\log(s\epsilon^{-1}) + 1)  \cdot \hat m_2 \cdot \log\left(K M 
\sqrt{s}\right),
\end{equation} 
where $\hat m_2$ satisfies
$
 1  \gtrsim ((N_2-N_1)/\hat m_2 - 1)\cdot \mu_{N_1} \cdot \tilde s_2,
 $
and $\tilde s_{2}  \leq S_2$ (recall $S_2$ from Definition \ref{S}). Recall 
from (\ref{eq:kappa_bound}) that
$$
\kappa_{\mathbf{N}, \mathbf{M}}(2,1) \leq \sqrt{s_1 \cdot  \mu_{N_1}} \cdot 
\nm{P_{N_1}^\perp U P_{M_1}}, \quad
\kappa_{\mathbf{N}, \mathbf{M}}(2,2) \leq s_2 \cdot \mu_{N_1}.
$$
 Also, it follows directly from Definition \ref{S}  that
$$
S_2 \leq \left( \nm{P_{N_1}^\perp U P_{M_1}}\cdot\sqrt{M_1} + \sqrt{s_2} 
\right)^2.
$$
Thus, provided that $\nm{P_{N_1}^\perp U P_{M_1}} \leq \gamma/\sqrt{M_1}$ 
where $\gamma$ is as in (i) of Theorem \ref{main_two_level_weak}, we observe 
that (iii) of Theorem \ref{main_two_level_weak} implies (\ref{conditions3211}) 
and (\ref{conditions3222}). Thus, the theorem now follows from Proposition 
\ref{sufficient}.
\end{proof}

\begin{proof}[Proof of Theorem \ref{main_full_fin_noise2} and Theorem 
\ref{main_full_inf_noise2}]
It is straightforward  that Theorem \ref{main_full_fin_noise2} follows from 
Theorem \ref{main_full_inf_noise2}. Now, recall from Lemma 
\ref{eq:kappa_bound}  that
\eas{
&\kappa_{\mathbf{N},\mathbf{M}}(k,l) \leq s_l \cdot  
\mu_{\mathbf{N},\mathbf{M}}(k,l), \quad
\kappa_{\mathbf{N},\mathbf{M}}(k,\infty) \leq s_r \cdot  
\mu_{\mathbf{N},\mathbf{M}}(k,\infty), \qquad k, l=1,\ldots,r.
}
Thus, a direct application of Proposition \ref{main_prop2} and Proposition 
\ref{sufficient} completes the proof.

\end{proof}

It remains now to prove Propositions \ref{main_prop} and \ref{main_prop2}.  
This is the content of the next sections.

\subsection{Preliminaries}


Before we commence on the rather length proof of these propositions, let us 
recall one of the monumental results in probability theory that will be of 
greater use later on.

\begin{theorem}\label{Talagrand}(Talagrand \cite{Talagrand, Ledoux})
There exists a number $K$ with the following property. Consider $n$
independent random variables $X_i$ valued in a measurable space
$\Omega $ and let $\mathcal{F}$ be a (countable) class  of measurable
functions on $\Omega.$ Let $Z$ be the random variable $Z = \sup_{f \in
  \mathcal{F}}\sum_{i \leq n} f(X_i)$ and define
$$
S = \sup_{f \in \mathcal{F}}\|f\|_{\infty}, \qquad V = \sup_{f \in
  \mathcal{F}} \mathbb{E}\left( \sum_{i \leq n}f(X_i)^2  
 \right).
$$
If $ \mathbb{E}(f(X_i)) = 0$ for all $f \in \mathcal{F}$ and $i\leq n$, then, 
for each $t > 0$, 
we have
$$
\mathbb{P}(|Z - \mathbb{E}(Z)| \geq t) \leq 3 \exp \left(
-\frac{1}{K}\frac{t}{S} \log\left( 1 + \frac{tS}{V+S\mathbb{E}(\overline{Z})}  
\right)\right),
$$
where $\overline{Z} = \sup_{f\in\mathcal{F}}|\sum_{i \leq n} f(X_i)|$.
\end{theorem}

Note that this version of Talagrand's theorem is found in \cite[Cor.\ 
7.8]{Ledoux}.  We next present a theorem and several technical propositions 
that will serve as the main tools in our proofs of Propositions 
\ref{main_prop} and \ref{main_prop2}.  A crucial tool herein is the Bernoulli 
sampling model. We will use the notation $\{a,\hdots,b\} \supset \Omega \sim 
\mathrm{Ber}(q)$, where $a<b$ $a,b \in\mathbb{N}$, 
when $\Omega$ is given by $\Omega = \{k: \delta_k = 1\}$ and 
$\{\delta_k\}_{k=1}^N$ is a sequence of Bernoulli variables with 
$\mathbb{P}(\delta_k = 1) = q$.

\defn{
\label{multi_level_Bernoulli_dfn}
Let $r \in \bbN$, $\mathbf{N} = (N_1,\ldots,N_r) \in \bbN^r$ with $1 \leq N_1 
< \ldots < N_r$, $\mathbf{m} = (m_1,\ldots,m_r) \in \bbN^r$, with $m_k \leq 
N_k-N_{k-1}$, $k=1,\ldots,r$, and suppose that
\bes{
\Omega_k \subseteq \{ N_{k-1}+1,\ldots,N_{k} \},\quad  \Omega_k  \sim 
\mathrm{Ber}\left(\frac{m_k}{N_{k}-N_{k-1}}\right),    \quad k=1,\ldots,r,
}
 where $N_0 = 0$.  We refer to the set
$
\Omega = \Omega_{\mathbf{N},\mathbf{m}} := \Omega_1 \cup \ldots \cup \Omega_r.
$
as an $(\mathbf{N},\mathbf{m})$-multilevel Bernoulli sampling scheme.
}

\begin{theorem}\label{inverse_bound}
Let $U \in \mathcal{B}(l^2(\mathbb{N}))$ be an isometry.  Suppose that $\Omega 
= \Omega_{\mathbf{N},\mathbf{m}}$ is a multilevel Bernoulli sampling scheme, 
where $\mathbf{N} = (N_1,\ldots,N_r) \in \bbN^r$ and $\mathbf{m} = 
(m_1,\ldots,m_r) \in \bbN^r$.  Consider $(\mathbf{s},\mathbf{M})$, where 
$\mathbf{M} = (M_1,\ldots,M_r) \in \bbN^r$, $M_1 < \ldots < M_r$, and 
$\mathbf{s} = (s_1,\ldots,s_r) \in \bbN^r$, and let
$$
\Delta = \Delta_1 \cup \hdots \cup \Delta_r, \qquad \Delta_k \subset 
\{M_{k-1}+1,\hdots, M_k\}, \qquad |\Delta_k| = s_k
$$
where $M_0 = 0$.
If 
$
\|P_{M_r}U^*P_{N_r}UP_{M_r} - P_{M_r}\| \leq 1/8
$
 then, for $\gamma \in (0,1),$
\begin{equation}\label{prob_bound_isometry}
\mathbb{P}(\|P_{\Delta}U^* (q_1^{-1}P_{\Omega_1} \oplus \hdots \oplus 
q_r^{-1}P_{\Omega_r})UP_{\Delta} - P_{\Delta}\|  \geq 1/4) \leq \gamma, 
\end{equation} 
where $q_k = m_k/(N_k-N_{k-1}),$
provided that 
\be{ \label{eq:norm_condition}
1 \gtrsim \frac{N_k-N_{k-1}}{m_k} \cdot \left(\sum_{l=1}^r 
\kappa_{\mathbf{N},\mathbf{M}}(k,l)\right) \cdot \left(\log\left(\gamma^{-1} 
\, s\right) +1\right).}
In addition, if $q = \min \{ q_k \}^{r}_{k=1} = 1$ then 
$$\mathbb{P}(\|P_{\Delta}U^* (q_1^{-1}P_{\Omega_1} \oplus \hdots \oplus 
q_r^{-1}P_{\Omega_r})UP_{\Delta} - P_{\Delta}\|  \geq 1/4)=0.$$ 
\end{theorem}

In proving this theorem we deliberately avoid the use of the Matrix Bernstein 
inequality \cite{Gross}, as Talagrand's theorem is more convenient for our infinite-dimensional
setting. Before we can prove this theorem, we need the following technical 
lemma.

\begin{lemma}\label{mean_value}
Let  $U \in \mathcal{B}(l^2(\mathbb{N}))$  with $\|U\| \leq 1$, and consider 
the setup in Theorem \ref{inverse_bound}. Let $N = N_r$ and
let $\{\delta_j\}_{j=1}^N$ be independent  random Bernoulli variables with 
$\mathbb{P}(\delta_j = 1) = \tilde q_j,$ $\tilde q_j = m_k/(N_k-N_{k-1})$ and 
$j \in \{N_{k-1}+1, \hdots, N_k\},$
and define
$
Z = \sum_{j=1}^{N}Z_j,$ $Z_j =\left(\tilde 
q_j^{-1}\delta_j-1\right)\eta_j\otimes \bar\eta_j$ and $\eta_j = 
P_{\Delta}U^*e_j.$ 
Then 
\begin{equation*}
\mathbb{E}\left(\|Z\|\right)^2 \leq 48\max\{\log(|\Delta|),1\}     \,
\max_{1\leq j \leq
  N} \left\{\tilde q_j^{-1}\|\eta_j\|^2\right\},
\end{equation*}
when
$
\left(\max\{\log(|\Delta|), 1\}\right)^{-1} \geq 18\max_{1\leq j \leq
  N} \left\{\tilde q_j^{-1}\|\eta_j\|^2\right\}.
$
\end{lemma}
The proof of this lemma involves essentially reworking an argument due to 
Rudelson \cite{Rudelson}, and is similar to arguments given previously in 
\cite{BAACHGSCS} (see also \cite{Candes_Romberg}).  We include it here for 
completeness as the setup deviates slightly.  We shall also require the 
following result:

\begin{lemma}\label{Rudde_Lem}(Rudelson)
Let $\eta_1, \hdots, \eta_M \in \mathbb{C}^n$ and let $\varepsilon_1,
\hdots \varepsilon_M$ be independent Bernoulli
variables taking values $1,-1$ with probability $1/2$. Then 
$$
\mathbb{E}\left(\left \|\sum_{i=1}^M \varepsilon_i \bar \eta_i \otimes \eta_i
\right\|\right) \leq \frac{3}{2}\sqrt{p} \max_{i \leq M}
\|\eta_i\|\sqrt{\left\|\sum_{i=1}^M \bar \eta_i \otimes \eta_i  \right\|},
$$
where $p= \max\{ 2, 2\log(n) \}$.
\end{lemma}

Lemma \ref{Rudde_Lem} is often referred to as Rudelson's Lemma 
\cite{Rudelson}.  However, we use the above complex version that was proven by 
Tropp \cite[Lem.\ 22]{Tropp2}.

\begin{proof}[Proof of Lemma \ref{mean_value}]
We commence by letting $\tilde \delta = \{\tilde \delta_j\}_{j=1}^N$ be
independent copies of $\delta = \{\delta_j\}_{j=1}^N.$ Then, since $\bbE(Z) = 
0$,
\begin{equation}\label{Jensen}
\begin{split}
\mathbb{E}_{\delta}\left(\|Z\|\right) &=
\mathbb{E}_{\delta}\left(\left\|Z
- \mathbb{E}_{\tilde 
  \delta}\left(\sum_{j=1}^N
\left( \tilde q_j^{-1}\tilde \delta_j-1\right)\eta_j\otimes \bar\eta_j
\right) \right\|\right)  \\
&\leq 
\mathbb{E}_{\delta}\left(\mathbb{E}_{\tilde
  \delta}\left(\left\| Z - \sum_{j=1}^N
\left(\tilde q_j^{-1} \tilde \delta_j-1\right) \eta_j\otimes \bar\eta_j
\right \|\right)\right),
\end{split}
\end{equation}
by Jensen's inequality.  Let $\varepsilon = \{\varepsilon_j\}_{j=1}^N$ be a 
sequence of
Bernoulli variables taking values $\pm 1$ with probability $1/2$. Then, 
by (\ref{Jensen}), symmetry, Fubini's Theorem and the triangle inequality, it 
follows that  
\begin{equation}\label{iterated} 
\begin{split}
&\mathbb{E}_{\delta}\left(\| Z\|\right)  \leq
\mathbb{E}_{\varepsilon}\left(\mathbb{E}_{\delta}
\left(\mathbb{E}_{\tilde \delta}\left(\left\|
\sum_{j=1}^N
\varepsilon_j
\left(\tilde q_j^{-1}\delta_j -
\tilde q_j^{-1}\tilde \delta_j\right)
\eta_j\otimes \bar\eta_j
\right\|
\right)\right)\right)\\
& \quad \leq
2 \mathbb{E}_{\delta}
\left(\mathbb{E}_{\varepsilon}\left(\left\|
\sum_{j=1}^N
\varepsilon_j
\tilde q_j^{-1}\delta_j
\eta_j\otimes \bar\eta_j
\right\|
\right)\right).
\end{split}
\end{equation}
We are now able to apply Rudelson's Lemma (Lemma \ref{Rudde_Lem}).  However, 
as specified before, it is the complex
version that is crucial here. By Lemma \ref{Rudde_Lem} we get that 
\begin{equation}\label{rudde2}
\mathbb{E}_{\varepsilon}\left(\left\|
\sum_{j=1}^N
\varepsilon_j \tilde q_j^{-1} \delta_j
\eta_j\otimes \bar\eta_j
\right\|
\right) 
 \leq 
\frac{3}{2}\sqrt{\max\{2\log(s),2\}}\max_{1\leq j \leq
  N}\tilde q_j^{-1/2}\|\eta_j\|\sqrt{
\left\|\sum_{j=1}^Nq_j^{-1} \tilde q_j^{-1}\delta_j
\eta_j\otimes \bar\eta_j\right\|},
\end{equation}
where $s = |\Delta|$.
And hence, by using (\ref{iterated})  and (\ref{rudde2}), it follows that
$$
\mathbb{E}_{\delta}\left(\| Z\|\right) \leq
3\sqrt{\max\{2\log(s),2\}}\max_{1\leq j \leq
  N}\tilde q_j^{-1/2}\|\eta_j\|   
  \sqrt{\mathbb{E}_{\delta}\left(\left \|Z + \sum_{j=1}^N
\eta_j\otimes \bar\eta_j\right \|\right)}.
$$
Note that $\|\sum_{j=1}^N
\eta_j\otimes \bar\eta_j\| \leq 1$, since $U$ is an isometry.  The result now 
follows from the straightforward calculus fact that if $r >0$, $c\leq 1$  and 
$r \leq c\sqrt{r+1}$ then we have that $r \leq c(1+\sqrt{5})/2$. 
\end{proof}

\begin{proof}[Proof of Theorem \ref{inverse_bound}]
Let $N = N_r$ just to be clear here.
Let $\{\delta_j\}_{j=1}^N$ be random Bernoulli variables as defined in Lemma 
\ref{mean_value} and 
define $Z = \sum_{j=1}^{N}Z_j,$  $Z_j =\left(\tilde 
q_j^{-1}\delta_j-1\right)\eta_j\otimes \bar\eta_j$ with 
$\eta_j = P_{\Delta}U^*e_j.$
Now observe that 
\begin{equation}\label{decomposition}
P_{\Delta}U^* (q_1^{-1}P_{\Omega_1} \oplus \hdots \oplus 
q_r^{-1}P_{\Omega_r})UP_{\Delta} = \sum_{j=1}^N \tilde q_j^{-1}\delta_j 
\eta_j\otimes \bar\eta_j, \quad P_{\Delta}U^* P_{N}UP_{\Delta} = \sum_{j=1}^N 
\eta_j\otimes \bar\eta_j.
\end{equation}
Thus, it follows that 
\begin{equation}\label{lou22}
\begin{split}
\|P_{\Delta}U^* (q_1^{-1}P_{\Omega_1} \oplus \hdots \oplus 
q_r^{-1}P_{\Omega_r})UP_{\Delta} - P_{\Delta}\| &\leq \|Z \| + 
\left\|(P_{\Delta}U^* P_{N}UP_{\Delta}
-P_{\Delta})\right\| \leq \|Z\| + \frac{1}{8},
\end{split}
\end{equation}
by the assumption that $
\|P_{M_r}U^*P_{N_r}UP_{M_r} - P_{M_r}\| \leq 1/8
$. Thus, to prove the assertion we need to estimate $\|Z\|$, and 
Talagrand's Theorem (Theorem \ref{Talagrand}) will be our main tool.
Note that clearly, since $Z$ is self-adjoint, 
we have that 
$
\|Z\| = \sup_{\zeta \in \mathcal{G}}|\langle Z\zeta,\zeta\rangle|,
$ where
  $\mathcal{G}$ is a countable set of vectors in the unit ball of 
  $P_{\Delta}(\cH)$ .  For $\zeta \in 
 \mathcal{G}$ define the mappings
$$
\hat \zeta_1(T) =  \langle T\zeta,\zeta\rangle, \quad \hat \zeta_2(T) =  
-\langle T\zeta,\zeta\rangle,  \qquad T \in \mathcal{B}(\mathcal{H}). 
$$
In order to use Talagrand's Theorem \ref{Talagrand} we restrict the domain 
$\mathcal{D}$ of the mappings $\zeta_i$ to 
$$
\mathcal{D} = \{T \in \mathcal{B}(\mathcal{H}): \|T\| \leq \max_{1\leq j \leq 
N} \{\tilde q^{-1}_j\|\eta_j\|^2\}\}.
$$
Let $\mathcal{F}$ denote the family of mappings $\hat \zeta_1, \hat \zeta_2$ 
for $\zeta \in \mathcal{G}$. Then $\|Z\| = \sup_{\hat \zeta \in \mathcal{F}} 
\hat \zeta(Z)$,  and for $i=1,2$ we have  
$$
 |\hat \zeta_i(Z_j)| = \left|\left(\tilde 
 q^{-1}_j\delta_j-1\right)\right||\langle
\left(\eta_j\otimes
\bar \eta_j\right)\zeta,\zeta\rangle| \leq \max_{1\leq j \leq N} \{\tilde 
q^{-1}_j\|\eta_j\|^2\}. 
$$
Thus, $Z_j \in \mathcal{D}$ for $1\leq j \leq N$ and 
$
S :=  \sup_{\zeta \in \mathcal{F}}\|\hat \zeta\|_{\infty} 
= \max_{1\leq j \leq N} \{\tilde q^{-1}_j\|\eta_j\|^2\}.
$
Note that 
\begin{equation*}
\|\eta_j\|^2 =  \langle P_{\Delta}U^*e_j, P_{\Delta}U^*e_j \rangle = 
\sum_{k=1}^r 
\langle P_{\Delta_k}U^*e_j, P_{\Delta_k}U^*e_j \rangle.
\end{equation*}
Also, note that an easy application of Holder's inequality gives the following 
(note that the $l^1$ and $l^{\infty}$ bounds are finite because all the 
projections have finite rank),
\begin{equation*}
\begin{split}
&|\langle P_{\Delta_k}U^*e_j, P_{\Delta_k}U^*e_j \rangle|  
\leq \|P_{\Delta_k}U^*e_j\|_{l^1} \|P_{\Delta_k}U^*e_j\|_{l^{\infty}}\\
&\leq \|P_{\Delta_k}U^*P_{N_l}^{N_{l-1}}  \|_{l^1 \rightarrow l^1} 
\|P_{\Delta_k}U^*e_j\|_{l^{\infty}} \leq 
 \| P_{N_l}^{N_{l-1}} U  P_{\Delta_k}\|_{l^{\infty} \rightarrow l^{\infty}} 
 \cdot \sqrt{\mu(P_{N_l}^{N_{l-1}} U)} \leq 
 \kappa_{\mathbf{N},\mathbf{M}}(l,k),
\end{split}
\end{equation*}
for $j \in \{N_{l-1}+1,\hdots, N_l\}$ and $l \in \{1,\hdots,r\}$.
Hence, it follows that 
\begin{equation}\label{bach_fugue_dmin}
\|\eta_j\|^2 \leq \max_{1\leq k \leq r}(\kappa_{\mathbf{N},\mathbf{M}}(k,1) + 
\hdots + \kappa_{\mathbf{N},\mathbf{M}}(k,r)),
\end{equation}
and therefore
$
S \leq  \max_{1\leq k \leq r}\left(q_k^{-1} \sum_{j=1}^r 
\kappa_{\mathbf{N},\mathbf{M}}(k,j) \right).
$ 
Finally, note that by (\ref{bach_fugue_dmin}) and the reasoning above, it 
follows that 
\begin{equation}\label{V_ref}
\begin{split}
V &:= \sup_{\hat \zeta_i\in \mathcal{F}} \mathbb{E} \left(\sum_{j=1}^{N}\hat
\zeta_i(Z_j)^2\right) = \sup_{\zeta \in \mathcal{G}} \mathbb{E}\left(
\sum_{j=1}^{N}
\left(\tilde q^{-1}_j\delta_j-1\right)^2|\langle 
P_{\Delta}U^*e_j,\zeta\rangle|^4 
\right)\\
&\leq  \max_{1\leq k \leq r}\|\eta_k\|^2 
\left(\frac{N_k-N_{k-1}}{m_k}-1\right) \sup_{
  \zeta \in \mathcal{G}} \sum_{j = 1}^{N}|\langle 
  e_j,UP_{\Delta}\zeta\rangle|^2,\\  
 & \leq \max_{1\leq k \leq r}\frac{N_k-N_{k-1}}{m_k}\left(\sum_{l=1}^r  
 \kappa_{\mathbf{N},\mathbf{M}}(k,l) \right)\sup_{
  \zeta \in \mathcal{G}} \|U\zeta\|^2 = \max_{1\leq k \leq 
  r}\frac{N_k-N_{k-1}}{m_k}\left(\sum_{l=1}^r  
  \kappa_{\mathbf{N},\mathbf{M}}(k,l) \right),
\end{split}
\end{equation}
where we used the fact that $U$ is an isometry to deduce that $\|U\| = 1$. 
Also, by Lemma \ref{mean_value} and (\ref{bach_fugue_dmin}) , it follows that 
\begin{equation}\label{m_mean_value}
\mathbb{E}\left(\|Z\|\right)^2 \leq 48\, \max_{1\leq k \leq 
r}\frac{N_k-N_{k-1}}{m_k}\left(\sum_{l=1}^r  
\kappa_{\mathbf{N},\mathbf{M}}(k,l) \right)\cdot \log(s)
\end{equation}
when
\begin{equation}\label{eq;m}
1 \geq 18 \max_{1\leq k \leq r}\frac{N_k-N_{k-1}}{m_k}\left(\sum_{l=1}^r  
\kappa_{\mathbf{N},\mathbf{M}}(k,l) \right) \cdot \log(s),
\end{equation}
(recall that we have assumed $s \geq 3$).
Thus, by (\ref{lou22}) and Talagrand's Theorem \ref{Talagrand}, it follows 
that 
\begin{align}\label{exp_bound2}
&\mathbb{P}\left(\|P_{\Delta}U^* (q_1^{-1}P_{\Omega_1} \oplus \hdots \oplus 
q_r^{-1}P_{\Omega_r})UP_{\Delta} - P_{\Delta}\|  \geq 1/4\right) \notag \\
&\leq  \mathbb{P}\left(\|Z\| \geq \frac{1}{16} + \sqrt{24\, \max_{1\leq k \leq 
r}\frac{N_k-N_{k-1}}{m_k}\left(\sum_{l=1}^r  
\kappa_{\mathbf{N},\mathbf{M}}(k,l) \right) \cdot \log(s)} \right)\notag\\
&\leq 3\exp\left(-\frac{1}{16K}\left(\max_{1\leq k \leq 
r}\frac{N_k-N_{k-1}}{m_k}\left(\sum_{l=1}^r  
\kappa_{\mathbf{N},\mathbf{M}}(k,l) \right)\right)^{-1}
\log\left(1 + 1/32\right) \right),
\end{align}
when $m_k$'s are chosen such that the right hand side of (\ref{m_mean_value}) 
is less than or equal to $1$.
Thus, by (\ref{lou22}) and Talagrand's Theorem \ref{Talagrand}, it follows 
that 
\begin{align}\label{exp_bound3}
&\mathbb{P}\left(\|P_{\Delta}U^* (q_1^{-1}P_{\Omega_1} \oplus \hdots \oplus 
q_r^{-1}P_{\Omega_r})UP_{\Delta} - P_{\Delta}\|  \geq 1/4\right) \notag \\
&\leq \mathbb{P}\left(\|Z\|   \geq 1/8\right) 
\leq  \mathbb{P}\left(\|Z\| \geq \frac{1}{16} + \bbE \|Z\| \right)
\leq \mathbb{P}\left(\abs{\|Z\| - \bbE \|Z\| } \geq \frac{1}{16} 
\right)\notag\\
&\leq 3\exp\left(-\frac{1}{16K}\left(\max_{1\leq k \leq 
r}\frac{N_k-N_{k-1}}{m_k}\left(\sum_{l=1}^r  
\kappa_{\mathbf{N},\mathbf{M}}(k,l) \right)\right)^{-1}
\log\left(1 + 1/32\right) \right),
\end{align}
when $m_k$'s are chosen such that the right hand side of (\ref{m_mean_value}) 
is less than or equal to $1/16^2$.
Note that this condition is implied by the assumptions of the theorem as is 
(\ref{eq;m}). This yields the first part of the theorem. The second claim of 
this theorem follows from the assumption that $
\|P_{M_r}U^*P_{N_r}UP_{M_r} - P_{M_r}\| \leq 1/8.
$
\end{proof}

\begin{proposition}\label{jup}
Let $U \in \mathcal{B}(l^2(\mathbb{N}))$ be an isometry.  Suppose that $\Omega 
= \Omega_{\mathbf{N},\mathbf{m}}$ is a multilevel Bernoulli sampling scheme, 
where $\mathbf{N} = (N_1,\ldots,N_r) \in \bbN^r$ and $\mathbf{m} = 
(m_1,\ldots,m_r) \in \bbN^r$.  Consider $(\mathbf{s},\mathbf{M})$, where 
$\mathbf{M} = (M_1,\ldots,M_r) \in \bbN^r$, $M_1 < \ldots < M_r$, and 
$\mathbf{s} = (s_1,\ldots,s_r) \in \bbN^r$, and let
$
\Delta = \Delta_1 \cup \hdots \cup \Delta_r, 
$
$
\Delta_k \subset 
\{M_{k-1},\hdots, M_k\},
$
$
|\Delta_k| = s_k,
$
where $M_0 = 0$.
 Let $\beta \geq 1/4$.
\begin{enumerate}
\item[(i)] 
If
\bes{
N:=N_r,\quad K := \max_{k=1,\ldots,r} \left \{ \frac{N_{k}-N_{k-1}}{m_k} 
\right \},
}
satisfy the weak balancing property with respect to $U$, $M:= M_r$ and $s : = 
s_1+\ldots + s_r$,
then,
for $\xi \in \mathcal{H}$ and $\beta, \gamma >0$, 
we have that 
\begin{equation}\label{ima}
\mathbb{P}\left(\|P_MP_{\Delta}^{\perp}U^* (q_1^{-1}P_{\Omega_1} \oplus \hdots 
\oplus q_r^{-1}P_{\Omega_r})UP_{\Delta}\xi \|_{l^{\infty}} >
   \beta\|\xi\|_{l^{\infty}} \right) \leq \gamma,
\end{equation}
provided that
\begin{equation}\label{condition_perp1}
 \frac{ \beta}{\log\left( \frac{4}{ \gamma}(M-s)\right)} \geq C \ \Lambda, 
 \qquad 
  \frac{ \beta^2}{\log\left( \frac{4}{ \gamma}(M-s)\right)} \geq 
 C \ \Upsilon,
\end{equation} 
for some constant $C > 0$,
where $q_k = m_k/(N_k - N_{k-1})$ for $k=1,\ldots, r$,
\begin{equation}\label{Lambda_def}
\Lambda = \max_{1\leq k \leq r} \left\{ \frac{N_k-N_{k-1}}{m_k} \cdot 
\left(\sum_{l=1}^r  \kappa_{\mathbf{N},\mathbf{M}}(k,l) \right)\right\},
\end{equation}
\begin{equation}\label{Upsilon_def}
\Upsilon = \max_{1 \leq l \leq r}
\sum_{k=1}^r \left(\frac{N_k-N_{k-1}}{m_k} - 1\right) \cdot 
\mu_{\mathbf{N},\mathbf{M}}(k,l)\cdot \tilde s_k,
\end{equation}
for all $\{\tilde{s}_k\}_{k=1}^r$ such that
$\tilde s_{1}+ \hdots + \tilde s_{r}  \leq s_1+ \hdots + s_r$ and $\tilde s_k 
\leq S_k(s_1,\hdots, s_r).$
Moreover, if $q_k = 1$ for all $k=1,\ldots,r$, then (\ref{condition_perp1}) is 
trivially satisfied for any $\gamma>0$ and the left-hand side of (\ref{ima}) 
is equal to zero.
\item[(ii)] If $N$ satisfies the strong Balancing Property with respect to 
$U,$ $M$ and $s$, then,
for $\xi \in \mathcal{H}$ and $\beta, \gamma >0$, 
we have that 
\begin{equation}\label{ima2}
\mathbb{P}\left(\|P_{\Delta}^{\perp}U^* (q_1^{-1}P_{\Omega_1} \oplus \hdots 
\oplus q_r^{-1}P_{\Omega_r})UP_{\Delta}\xi \|_{l^{\infty}} >
   \beta\|\xi\|_{l^{\infty}} \right) \leq \gamma,
\end{equation}  provided that
\begin{equation}\label{condition_perp2}
 \frac{ \beta}{\log\left( \frac{4}{ \gamma}(\tilde \theta-s)\right)} \geq C \ 
 \Lambda, \qquad 
  \frac{ \beta^2}{\log\left( \frac{4}{ \gamma}(\tilde \theta-s)\right)} \geq 
 C \ \Upsilon,
\end{equation} 
for some constant $C > 0$, $\tilde \theta = \tilde \theta(\{q_k\}_{k=1}^r, 
1/8, \{N_k\}_{k=1}^r,s, M )$ and $\Upsilon$, $\Lambda$ as defined in (i) and 
\begin{align*}
&\tilde \theta(\{q_k\}_{k=1}^r, t, \{N_k\}_{k=1}^r,s, M ) \\
&= \abs{\left\lbrace i\in\bbN : \max_{\substack{\Gamma_1 \subset 
\{1,\ldots,M\},\quad \abs{\Gamma_1} = s \\ \Gamma_{2,j}\subset 
\{N_{j-1}+1,\ldots,N_j\}, \quad j=1,\ldots,r}} \| P_{\Gamma_1} U^* 
(q_1^{-1}P_{\Gamma_{2,1}} \oplus \hdots \oplus q_r^{-1}P_{\Gamma_{2,r}}) U 
e_i\| > \frac{t}{\sqrt{s}} \right\rbrace}.
\end{align*}
Moreover, if $q_k = 1$ for all $k=1,\ldots,r$, then (\ref{condition_perp2}) is 
trivially satisfied for any $\gamma>0$ and the left-hand side of (\ref{ima2}) 
is equal to zero.
\end{enumerate}
\end{proposition}

\begin{proof}
To prove (i) we note that, 
without loss of generality, we can assume that $\|\xi\|_{l^\infty}=1$. Let 
$\{\delta_j\}_{j=1}^N$ be random Bernoulli variables with 
$
\mathbb{P}(\delta_j = 1) = \tilde q_j = q_k,
$
for $j \in \{N_{k-1}+1, \hdots, N_k\}$ and $1 \leq k \leq r.$ 
A key observation that will be crucial below is that 
\begin{equation}\label{paco} 
\begin{split}
&P_{\Delta}^{\perp}U^* (q_1^{-1}P_{\Omega_1} \oplus \hdots \oplus 
q_r^{-1}P_{\Omega_r})UP_{\Delta}\xi  = 
 \sum_{j=1}^{N}P_{\Delta}^{\perp}U^* \tilde q_j^{-1} \delta_j (e_j \otimes e_j)
UP_{\Delta}\xi\\ 
& \qquad = \sum_{j=1}^{N}P_{\Delta}^{\perp}U^*  (\tilde q_j^{-1}\delta_j 
-1)(e_j
\otimes e_j)   UP_{\Delta}\xi +  
P_{\Delta}^{\perp}U^* P_{N}  UP_{\Delta}\xi.
\end{split}
\end{equation}
We will use this equation at the end of the argument, but first we will 
estimate the size of the individual components of 
$\sum_{j=1}^{N}P_{\Delta}^{\perp}U^*  (\tilde q_j^{-1}\delta_j -1)(e_j
\otimes e_j)   UP_{\Delta}\xi$. To do that define, for $1 \leq j \leq N$,  
the random variables 
$$
X^i_j = \langle U^*  (\tilde q_j^{-1}\delta_j -1)(e_j \otimes e_j)
UP_{\Delta}\xi,e_i  \rangle, \qquad i \in \Delta^c.
$$
We will show using Bernstein's inequality that, for each $i\in\Delta^c$ and $t 
> 0$, 
\begin{equation}\label{claim_Bernstein}
\mathbb{P}\left(\left|\sum_{j=1}^{N} X_j^i\right|  > t \right) \leq
4\exp\left(-\frac{t^2/4}{\Upsilon
+  \Lambda t/3}\right).
\end{equation}
To prove the claim, we need to estimate $\mathbb{E}\left(|X_j^i|^2\right)$  
and $|X_j^i|$.   First note that,
$$
\mathbb{E}\left(|X_j^i|^2\right) = (\tilde q_j^{-1}-1)|\langle e_j, 
UP_{\Delta} \xi\rangle|^2  |\langle e_j,Ue_i \rangle|^2,
$$
and note that $| \ip{e_j}{U e_i} |^2 \leq \mu_{\mathbf{N},\mathbf{M}} (k,l)$ 
for $j \in \{ N_{k-1}+1,\ldots,N_k \}$ and $i \in \{ M_{l-1}+1,\ldots,M_l 
\}$.  Hence
\eas{
\sum_{j=1}^{N} \mathbb{E}\left(|X_j^i|^2\right) &\leq \sum^{r}_{k=1} 
(q^{-1}_k-1) \mu_{\mathbf{N},\mathbf{M}}(k,l) \| P^{N_{k-1}}_{N_k} U 
P_{\Delta} \xi \|^2 
\\
& \leq \sup_{\zeta \in \Theta} \left \{ \sum^{r}_{k=1} (q^{-1}_{k}-1) 
\mu_{\mathbf{N},\mathbf{M}}(k,l) \| P^{N_{k-1}}_{N_k} U \zeta \|^2 \right \}, 
}
where
\bes{
\Theta = \{\eta : \|\eta\|_{l^{\infty}} \leq 1, 
|\mathrm{supp}(P_{M_l}^{M_{l-1}}\eta)| = s_l, \, l=1,\hdots, r\}.
}
The supremum in the above bound is attained for some $\tilde \zeta \in 
\Theta$.  If $\tilde s_k = \|P_{N_k}^{N_{k-1}} U \tilde \zeta\|^2$, then we 
have
\begin{equation}\label{bound_sum_E}
\sum_{j=1}^{N} \mathbb{E}\left(|X_j^i|^2\right) \leq  \sum_{k=1}^r (q_k^{-1} 
-1)\mu_{\mathbf{N},\mathbf{M}}(k,l)  \tilde s_k.
\end{equation}
Note that it is clear from the definition that $s_k \leq S_k(s_1,\hdots, s_r)$ 
for $1 \leq k \leq r$. Also, using the fact that $\|U\| \leq 1$ and the 
definition of $\Theta$, we note that
\begin{equation*}
\tilde s_1+ \hdots + \tilde s_r = \sum_{k=1}^r 
\|P^{N_{k-1}}_{N_k}UP_{\Delta}\zeta\|^2 \leq \|UP_{\Delta}\zeta\|^2 = 
\|\zeta\|^2 \leq s_1+ \hdots + s_r.
\end{equation*}
To estimate $|X_j^i|$ we start by observing that, by the triangle inequality, 
the fact that $\|\xi\|_{l^{\infty}} = 1$ and Holder's inequality, it follows 
that
$
|\langle \xi,   P_{\Delta}U^*e_j\rangle| \leq\sum_{k=1}^r |\langle 
P_{M_k}^{M_{k-1}}\xi,   P_{\Delta}U^*e_j\rangle|,
$
and
$$
|\langle P_{M_k}^{M_{k-1}}\xi,   P_{\Delta}U^*e_j\rangle| \leq \| 
P_{N_l}^{N_{l-1}} U  P_{\Delta_k}\|_{l^{\infty} \rightarrow l^{\infty}}, \quad 
j \in \{N_{l-1}+1,\hdots, N_l\}, \quad l \in \{1,\hdots,r\}.
$$
Hence, it follows that for $1\leq j \leq N$ and $i \in \Delta^c$,
\begin{equation}\label{useful_bound}
\begin{split}
&|X_j^i| = \tilde q_j^{-1}|(\delta_j -\tilde q_j)||\langle \xi,   
P_{\Delta}U^*e_j\rangle| |\langle e_j,Ue_i \rangle |,  \\
&\leq  \max_{1\leq k \leq r}\left\{\frac{N_k-N_{k-1}}{m_k} \cdot 
\left(\kappa_{\mathbf{N},\mathbf{M}}(k,1)+ \hdots +  
\kappa_{\mathbf{N},\mathbf{M}}(k,r) \right)\right\}.
\end{split}
\end{equation}
Now, clearly $\mathbb{E}(X^i_j) = 0$ for $1 \leq j \leq N$ and $i \in 
\Delta^c$. Thus, by applying Bernstein's inequality to $\mathrm{Re}(X_j^i)$ and
$\mathrm{Im}(X_j^i )$ for $j=1,\ldots,N$, via (\ref{bound_sum_E}) and 
(\ref{useful_bound}), the claim (\ref{claim_Bernstein}) follows.

Now, by (\ref{claim_Bernstein}), (\ref{paco}) and the assumed weak Balancing 
property (wBP), 
it follows that 
\begin{equation*}
\begin{split}
&\mathbb{P}\left(\|P_MP_{\Delta}^{\perp}U^* (q_1^{-1}P_{\Omega_1} \oplus 
\hdots \oplus q_r^{-1}P_{\Omega_r})UP_{\Delta}\xi \|_{l^{\infty}} >  \beta 
\right) \\
& \leq \sum_{i \in \Delta^c \cap \{1,\hdots,M\}}
\mathbb{P}\left(\left|\sum_{j=1}^{N} X_j^i + \langle
P_MP_{\Delta}^{\perp}U^* P_{N}^{\perp}  UP_{\Delta}\xi, e_i \rangle
\right|  > \beta \right)\\ 
&\leq \sum_{i \in \Delta^c \cap \{1,\hdots,M\}}
\mathbb{P}\left(\left|\sum_{j=1}^{N} X_j^i \right|  >  \beta -  
\|P_MP_{\Delta}^{\perp}U^* P_{N} UP_{\Delta}\|_{l^\infty} \right) \\
&\leq 4(M-s)\exp\left(-\frac{t^2/4}{\Upsilon
+  \Lambda t/3}\right), \quad t = \frac{1}{2}  \beta, \qquad \text{by} \, 
(\ref{claim_Bernstein}), (\text{wBP}),
\end{split}
\end{equation*}
 Also, 
\begin{equation*}
4(M-s)\exp\left(-\frac{t^2/4}{\Upsilon
+  \Lambda t/3}\right) \leq \gamma
\end{equation*}
when 
$$
\log\left( \frac{4}{ \gamma}(M-s)\right)^{-1} \geq \left(\frac{4 
\Upsilon}{t^2} + \frac{4\Lambda}{3t}\right).
$$
And this concludes the proof of (i). 
To prove (ii), for $t > 0$, suppose that there is a set $\Lambda_t  \subset 
\mathbb{N}$ such that  
$$
\mathbb{P}\left(\sup_{i \in \Lambda_t}|\langle 
P_{\Delta}^{\perp}U^*(q_1^{-1}P_{\Omega_1} \oplus \hdots \oplus 
q_r^{-1}P_{\Omega_r})UP_{\Delta}\eta,e_i\rangle | > t \right) = 0, \qquad 
|\Lambda_t^c| < \infty.
$$
Then, as before, by (\ref{claim_Bernstein}), (\ref{paco}) and the assumed 
strong Balancing property (sBP), 
it follows that 
\begin{equation*}
\begin{split}
&\mathbb{P}\left(\|P_{\Delta}^{\perp}U^* (q_1^{-1}P_{\Omega_1} \oplus \hdots 
\oplus q_r^{-1}P_{\Omega_r})UP_{\Delta}\xi \|_{l^{\infty}} >  \beta \right) \\
&\leq \sum_{i \in \Delta^c \cap \Lambda_t^c}
\mathbb{P}\left(\left|\sum_{j=1}^{N} X_j^i + \langle P_{\Delta}^{\perp}U^* 
P_{N}^{\perp}  UP_{\Delta}\xi, e_i \rangle
\right|  > \beta \right),
\end{split}
\end{equation*}
yielding
\begin{equation*}
\begin{split}
&\mathbb{P}\left(\|P_{\Delta}^{\perp}U^* (q_1^{-1}P_{\Omega_1} \oplus \hdots 
\oplus q_r^{-1}P_{\Omega_r})UP_{\Delta}\xi \|_{l^{\infty}} >  \beta \right) \\
&\leq \sum_{i \in \Delta^c \cap \Lambda_t^c}
\mathbb{P}\left(\left|\sum_{j=1}^{N} X_j^i \right|  >  \beta -  
\|P_{\Delta}^{\perp}U^* P_{N} UP_{\Delta}\|_{l^\infty} \right) \\
&\leq 4(\abs{\Lambda_t^c}-s)\exp\left(-\frac{t^2/4}{\Upsilon
+  \Lambda t/3}\right) <\gamma, \quad t = \frac{1}{2}  \beta, \qquad \text{by} 
\, (\ref{claim_Bernstein}), (\text{sBP}),\\
\end{split}
\end{equation*}
whenever
$$
\log\left( \frac{4}{ \gamma}(\abs{\Lambda_t^c}-s)\right)^{-1} \geq 
\left(\frac{4 \Upsilon}{t^2} + \frac{4\Lambda}{3t}\right).
$$
Hence, it remains to obtain a bound on $\abs{\Lambda_t^c}$. 
Let
$$
 \theta(q_1,\ldots,q_r, t,s) = \left\lbrace i\in\bbN : 
 \max_{\substack{\Gamma_1 \subset \{1,\ldots,M\},\quad \abs{\Gamma_1} = s \\ 
 \Gamma_{2,j}\subset \{N_{j-1}+1,\ldots,N_j\}, \quad j=1,\ldots,r}} \| 
 P_{\Gamma_1} U^* (q_1^{-1}P_{\Gamma_{2,1}} \oplus \hdots \oplus 
 q_r^{-1}P_{\Gamma_{2,r}}) U e_i\| > \frac{t}{\sqrt{s}} \right\rbrace.
$$
Clearly, $\Delta_t^c \subset  \theta(q_1,\ldots,q_r, t,s)$ and
\begin{align*}
\| P_{\Gamma_1} U^* (q_1^{-1}P_{\Gamma_{2,1}} \oplus \hdots \oplus 
q_r^{-1}P_{\Gamma_{2,r}}) U e_i\|
&\leq \max_{1\leq j\leq r}q_j^{-1} \| P_N U P_{i-1}^\perp \| \to 0
\end{align*}
as $i\to \infty$. So, $\abs{ \theta(q_1,\ldots,q_r, t,s)}<\infty.$ 
Furthermore, since $\tilde \theta(\{q_k\}_{k=1}^r, t, \{N_k\}_{k=1}^r,s, M )$ 
is a decreasing function in $t$, for all $t\geq \frac{1}{8}$,
$$\abs{ \theta(q_1,\ldots,q_r, t,s)}<\tilde \theta(\{q_k\}_{k=1}^r, 1/8, 
\{N_k\}_{k=1}^r,s, M )$$
thus, we have proved (ii). The statements at the end of (i) and (ii) are clear 
from the reasoning above.
\end{proof}

\begin{proposition}\label{gould_l_inf}
Consider the same setup as in Proposition \ref{jup}.
If $N$ and $K$ satisfy the weak Balancing Property with respect to 
$U,$ $M$ and $s$, then,
for $\xi \in \mathcal{H}$ and $\gamma >0$, 
we have
\begin{equation}\label{exp_bound32}
\begin{split}
\mathbb{P}(\|P_{\Delta}U^* (q_1^{-1}P_{\Omega_1} \oplus \hdots \oplus 
q_r^{-1}P_{\Omega_r})UP_{\Delta} - P_{\Delta})\xi\|_{l^{\infty}}  > \tilde  
\alpha \|\xi\|_{l^{\infty}}) \leq \gamma,
\end{split}
\end{equation}
with
$
\tilde \alpha = (2\log_2^{1/2}\left(4 \sqrt{s}KM\right))^{-1}, 
$
provided that 
\begin{equation*}
\begin{split}
1 &\gtrsim \Lambda \cdot \left(\log\left(s\gamma^{-1}\right)+1\right) 
\cdot\log\left(\sqrt{s}KM\right),\\
1 &\gtrsim \Upsilon \cdot \left(\log\left(s\gamma^{-1}\right)+1\right) 
\cdot\log\left(\sqrt{s}KM\right), 
\end{split}
\end{equation*}
where $\Lambda$ and $\Upsilon$ are defined in (\ref{Lambda_def}) and 
(\ref{Upsilon_def}).
Also,
\begin{equation}\label{alpha_half}
\begin{split}
\mathbb{P}(\|P_{\Delta}U^* (q_1^{-1}P_{\Omega_1} \oplus \hdots \oplus 
q_r^{-1}P_{\Omega_r})UP_{\Delta} - P_{\Delta})\xi\|_{l^{\infty}}  >\frac{1}{2} 
\|\xi\|_{l^{\infty}}) \leq \gamma
\end{split}
\end{equation}
provided that 
\begin{equation*}
1 \gtrsim  \Lambda \cdot \left(\log\left(s\gamma^{-1}\right)+1\right), \quad 1 
\gtrsim  \Upsilon \cdot \left(\log\left(s\gamma^{-1}\right)+1\right).
\end{equation*} 
Moreover, if $q_k = 1$ for all $k=1,\ldots,r$, then  the left-hand sides of 
(\ref{exp_bound32}) and (\ref{alpha_half}) are equal to zero.
\end{proposition}

\begin{proof}
Without loss of generality we may assume that $\|\xi\|_{l^{\infty}} = 1.$
Let $\{\delta_j\}_{j=1}^N$ be random Bernoulli variables with 
$
\mathbb{P}(\delta_j = 1) = \tilde q_j := q_k,  
$
with $j \in \{N_{k-1}+1, \hdots, N_k\}$ and $1 \leq k \leq r$.
Let also, for $j \in \mathbb{N},$ $\eta_j =
(UP_{\Delta})^*e_j.$ 
Then, after observing that 
$$
P_{\Delta}U^* (q_1^{-1}P_{\Omega_1} \oplus \hdots \oplus 
q_r^{-1}P_{\Omega_r})UP_{\Delta} = \sum_{j=1}^N q_j^{-1}\delta_j \eta_j\otimes 
\bar\eta_j, \quad 
P_{\Delta}U^* P_{N}UP_{\Delta} = \sum_{j=1}^N \eta_j\otimes \bar\eta_j,
$$
it follows immediately that
\begin{equation}\label{tale}
P_{\Delta}U^* (q_1^{-1}P_{\Omega_1} \oplus \hdots \oplus 
q_r^{-1}P_{\Omega_r})UP_{\Delta} - P_{\Delta} = \sum_{j=1}^{N}  (\tilde 
q_j^{-1}\delta_j - 1) \eta_j\otimes  \bar\eta_j - 
(P_{\Delta}U^*P_{N}UP_{\Delta} - P_{\Delta}).
\end{equation}
As in the proof of Proposition \ref{jup} 
our goal is to eventually use Bernstein's inequality and the following
is therefore a setup for that.  
Define, for $1 \leq j \leq N$,  
the random variables 
$
Z^i_j = \langle (\tilde q_j^{-1}\delta_j - 1) (\eta_j\otimes  \bar\eta_j) 
\xi,e_i  \rangle, 
$ for $i \in \Delta.$ 
We claim that, for $t > 0$, 
\begin{equation}\label{claim_Bernstein21}
\mathbb{P}\left(\left|\sum_{j=1}^{N} Z_j^i\right|  > t \right) \leq
4\exp\left(-\frac{t^2/4}{\Upsilon
+  \Lambda t/3}\right), \qquad i \in \Delta. 
\end{equation}
Now, clearly $\mathbb{E}(Z^i_j) = 0$, so we may use Bernstein's inequality. 
Thus, we need to estimate 
$\mathbb{E}\left(|Z_j^i|^2\right)$ and $|Z_j^i|$. 
 We will start with $\mathbb{E}\left(|Z_j^i|^2\right)$. Note that 
\begin{equation}\label{E1}
\mathbb{E}\left(|Z_j^i|^2\right) = (\tilde q_j^{-1}-1)|\langle e_j, 
UP_{\Delta} \xi\rangle|^2  |\langle e_j,Ue_i \rangle|^2.
\end{equation}
Thus, we can argue exactly as in the proof of Proposition \ref{jup} and deduce 
that 
\begin{equation}\label{bound_sum_E2}
\sum_{j=1}^{N} \mathbb{E}\left(|Z_j^i|^2\right) \leq  \sum_{k=1}^r (q_k^{-1} 
-1) \mu_{N_{k-1}} \tilde s_k,
\end{equation}
where 
 $s_k \leq S_k(s_1,\hdots, s_r)$ for $1 \leq k \leq r$ and 
$\tilde s_1+ \hdots + \tilde s_r  \leq s_1+ \hdots + s_r.$
To estimate $|Z_j^i|$ we argue as in the proof of Proposition \ref{jup} and 
obtain 
\be{\label{useful_bound122}
\begin{split}
|Z_j^i| \leq  \max_{1\leq k \leq r}\left\{\frac{N_k-N_{k-1}}{m_k} \cdot 
(\kappa_{\mathbf{N},\mathbf{M}}(k,1) + \hdots + 
\kappa_{\mathbf{N},\mathbf{M}}(k,r))\right\}.
\end{split}
}
Thus, by applying Bernstein's inequality to $\mathrm{Re}(Z_1^i), \hdots,
\mathrm{Re}(Z_N^i)$ and  
$\mathrm{Im}(Z_1^i ), \hdots, \mathrm{Im}(Z_N^i)$ we obtain, via 
(\ref{bound_sum_E2}) and (\ref{useful_bound122}) the estimate 
(\ref{claim_Bernstein21}), 
and we have proved the claim.

Now armed with (\ref{claim_Bernstein21}) we can deduce that , by (\ref{paco}) 
and the assumed weak Balancing property (wBP), 
it follows that 
\begin{equation}\label{prob_ineq}
\begin{split}
&\mathbb{P}\left(\|P_{\Delta}U^* (q_1^{-1}P_{\Omega_1} \oplus \hdots \oplus 
q_r^{-1}P_{\Omega_r})UP_{\Delta} - P_{\Delta})\xi\|_{l^{\infty}} > \tilde 
\alpha \right) \\
  & \leq \sum_{i \in \Delta}
\mathbb{P}\left(\left|\sum_{j=1}^{N} Z_j^i + \langle 
(P_{\Delta}U^*P_{N}UP_{\Delta} - P_{\Delta})\xi, e_i \rangle
\right|  > \tilde \alpha \right)\\ 
&\leq \sum_{i \in \Delta}
\mathbb{P}\left(\left|\sum_{j=1}^{N} Z_j^i \right|  > \tilde \alpha -  
\| P_MU^*P_{N}UP_M - P_M   \|_{l^1} \right),\\
&\leq 4\,s\exp\left(-\frac{t^2/4}{\Upsilon
+  \Lambda t/3}\right), \quad t = \tilde \alpha, \qquad \text{by} \, 
(\ref{claim_Bernstein21}), (\text{wBP}).
\end{split}
\end{equation}
Also, 
\begin{equation}\label{exp_bound4321}
4s\exp\left(-\frac{t^2/4}{\Upsilon
+  \Lambda t/3}\right) \leq \gamma,
\end{equation}
when 
$$
1 \geq \left(\frac{4 \Upsilon}{t^2} + \frac{4}{3t}\Lambda\right) \cdot 
\log\left( \frac{4s}{ \gamma}  \right).
$$
And this gives the first part of the proposition. Also, the fact that the left
hand side of (\ref{exp_bound32}) is zero when $q_k=1$ for $1 \leq k \leq r$ is 
clear from (\ref{exp_bound4321}).
Note that  (ii) follows by arguing exactly as above and replacing $\tilde 
\alpha$ by 
$\frac{1}{4}$.

\end{proof}

\begin{proposition}\label{thm:max_column_bound}
Let $U \in \mathcal{B}(l^2(\mathbb{N}))$ such that $\|U\| \leq 1$.  
Suppose that $\Omega = \Omega_{\mathbf{N},\mathbf{m}}$ is a multilevel 
Bernoulli sampling scheme, where $\mathbf{N} = (N_1,\ldots,N_r) \in \bbN^r$ 
and $\mathbf{m} = (m_1,\ldots,m_r) \in \bbN^r$.  Consider 
$(\mathbf{s},\mathbf{M})$, where $\mathbf{M} = (M_1,\ldots,M_r) \in \bbN^r$, 
$M_1 < \ldots < M_r$, and $\mathbf{s} = (s_1,\ldots,s_r) \in \bbN^r$, and let
$
\Delta = \Delta_1 \cup \hdots \cup \Delta_r, 
$ where $\Delta_k \subset \{M_{k-1}+1,\hdots, M_k\},$ $|\Delta_k| = s_k,$
and $M_0 = 0$. Then, for any $t\in(0,1)$ and $\gamma \in (0,1),$
\eas{
&\bbP\left(\max_{i\in\{1,\ldots,M\} \cap \Delta^c } \|P_{\{i\}}U^* 
(q_1^{-1}P_{\Omega_1} \oplus \hdots \oplus q_r^{-1}P_{\Omega_r})UP_{\{i\}}\|  
\geq 1+t\right) \leq \gamma
}
provided that
\be{\label{eq:req}
\frac{t^2}{4} \geq 
\log\left(\frac{2M}{\gamma}\right)\cdot \max_{1 \leq k \leq r} 
\left\{\left(\frac{N_k - N_{k-1}}{m_k}-1\right) \cdot 
\mu_{\mathbf{N},\mathbf{M}}(k,l) \right\} 
}
for all $l = 1,\hdots, r$ when $M = M_r$ and for all $l = 1,\hdots, r-1, 
\infty$ when $M > M_r.$
In addition, if $m_k = N_k-N_{k-1}$ for each $k=1,\ldots r$, then 
\be{\label{eq:q=1}
\mathbb{P}(\|P_{\{i\}}U^* (q_1^{-1}P_{\Omega_1} \oplus \hdots \oplus 
q_r^{-1}P_{\Omega_r})UP_{\{i\}} \|  \geq 1+t)=0, \quad \forall i \in 
\mathbb{N}.
}
\end{proposition}

\begin{proof}
Fix $i\in\{1,\ldots,M\}$. Let  $\{\delta_j\}_{j=1}^N$ be random independent 
Bernoulli variables with 
$\mathbb{P}(\delta_j = 1) = \tilde q_j := q_k $  for $j \in \{N_{k-1}+1, 
\hdots, N_k\}.$
Define  $Z = \sum_{j=1}^{N}Z_j$ and  $Z_j =\left(\tilde 
q_j^{-1}\delta_j-1\right) \abs{u_{ji}}^2.$ Now observe that 
\eas{
P_{\{i\}}U^* (q_1^{-1}P_{\Omega_1} \oplus \hdots \oplus 
q_r^{-1}P_{\Omega_r})UP_{\{i\}} &= \sum_{j=1}^N \tilde q_j^{-1}\delta_j 
\abs{u_{ji}}^2
=  \sum_{j=1}^N Z_j + \sum_{j=1}^N \abs{u_{ji}}^2,
}
where we interpret $U$ as the infinite matrix $U = \{u_{ij}\}_{i,j \in 
\mathbb{N}}$.
Thus, since $\|U\| \leq 1$,
\be{\label{eq:decomp_i}
\|P_{\{i\}}U^* (q_1^{-1}P_{\Omega_1} \oplus \hdots \oplus 
q_r^{-1}P_{\Omega_r})UP_{\{i\}}\|  \leq \abs{\sum_{j=1}^N Z_j } + 1
}
and it is clear that (\ref{eq:q=1}) is true.
For the case where $q_k <1$ for some $k\in\{ 1,\ldots,r\}$, observe that for 
$i \in \{M_{l-1}+1,\hdots, M_l\}$ (recall that $Z_j$ depend on $i$),  we have 
that $\bbE(Z_j) =0$. Also, 
$$
\abs{Z_j} \leq 
\begin{cases} 
\max_{1 \leq k \leq r}\{ \max\{  q_k^{-1} -1, 1\} \cdot 
\mu_{\mathbf{N},\mathbf{M}}(k,l)\} := B_i & i \in \{M_{l-1}+1,\hdots, M_l\}\\
\max_{1 \leq k \leq r}\{ \max\{  q_k^{-1} -1, 1\} \cdot 
\mu_{\mathbf{N},\mathbf{M}}(k,\infty)\} := B_{\infty} & i > M_r,
\end{cases}
$$
and,  by again using the assumption that $\|U\| \leq 1$,
\begin{equation*}
\begin{split}
&\sum_{j=1}^N\bbE(\abs{Z_j}^2) = \sum_{j=1}^N ( \tilde q_j^{-1} -1) 
\abs{u_{ji}}^4 \\
& \qquad \quad \leq 
 \begin{cases}
  \max_{1 \leq k \leq r} \{( q_k^{-1} -1) \, \mu_{\mathbf{N},\mathbf{M}}(k,l) 
  \} =: \sigma_i^2 & i \in \{M_{l-1}+1,\hdots, M_l\}\\
  \max_{1 \leq k \leq r} \{( q_k^{-1} -1) \, 
  \mu_{\mathbf{N},\mathbf{M}}(k,\infty) \} =: \sigma_{\infty}^2 & i > M_r.
\end{cases}
\end{split}
\end{equation*}
Thus, by Bernstein's inequality and (\ref{eq:decomp_i}),
\eas{
&\bbP(\|P_{\{i\}}U^* (q_1^{-1}P_{\Omega_1} \oplus \hdots \oplus 
q_r^{-1}P_{\Omega_r})UP_{\{i\}}\|  \geq 1+t)\\
& \qquad \qquad \qquad \leq
\bbP\left(\abs{\sum_{j=1}^N Z_j} \geq t\right) \leq 2 
\exp\left(-\frac{t^2/2}{\sigma^2 + B t/3}\right),
\\
&B = 
\begin{cases}
\max_{1\leq i \leq r} B_i & M = M_r,\\
\max_{i\in \{1,\hdots,r-1,\infty\}}B_i & M > M_r
\end{cases},
\quad 
\sigma^2 = 
\begin{cases}
\max_{1\leq i \leq r} \sigma^2_i & M = M_r,\\
\max_{i\in \{1,\hdots,r-1,\infty\}}\sigma^2_1 & M > M_r.
\end{cases}
}
Applying the union bound yields
\eas{
&\bbP\left(\max_{i\in\{1,\ldots,M\} } \|P_{\{i\}}U^* (q_1^{-1}P_{\Omega_1} 
\oplus \hdots \oplus q_r^{-1}P_{\Omega_r})UP_{\{i\}}\|  \geq 1+t\right) \leq 
\gamma
}
whenever (\ref{eq:req}) holds.
\end{proof}

\subsection{Proofs of Propositions \ref{main_prop} and \ref{main_prop2}}

The proof of the propositions relies on an idea that originated in a paper by 
D. Gross \cite{Gross}, namely, the golfing scheme. The variant we are using 
here is based on an idea from \cite{BAACHGSCS} as well as uneven section techniques from \cite{Hansen_JAMS, Hansen20082092}, see also \cite{strohmer}. However, the informed reader 
will recognise that the setup here differs substantially from both 
\cite{Gross} and \cite{BAACHGSCS}. See also \cite{Candes_Plan} for other 
examples of the use of the golfing scheme.
Before we embark on the proof, we will state and prove a useful lemma. 
\begin{lemma}\label{binom_lemma}
Let $\tilde X_k$ be independent binary variables taking values $0$ and $1$, 
such that $\tilde X_k = 1$ with probability $P$. Then, 
\be{\label{eq:binom_bound}
\bbP\left(\sum_{i=1}^{N} \tilde X_i \geq k\right) \geq \left(\frac{N\cdot 
e}{k}\right)^{-k} \binom{N}{k} P^k.
}
\end{lemma}
\begin{proof}
First observe that
\eas{
\bbP\left(\sum_{i=1}^{N} \tilde X_i \geq k\right)
&= \sum_{i=k}^N \binom{N}{i} P^i (1-P)^{N-i}
= \sum_{i=0}^{N-k} \binom{N}{i+k} P^{i+k} (1-P)^{N-k-i}\\
&= \binom{N}{k} P^k \sum_{i=0}^{N-k} \frac{(N-k)!k!}{(N-i-k)! (i+k)!} P^i 
(1-P)^{N-k-i}\\
&= \binom{N}{k} P^k \sum_{i=0}^{N-k} \binom{N-k}{i} P^i (1-P)^{N-k-i} 
\left[\binom{i+k}{k}\right]^{-1}.
} 
The result now follows because 
$
\sum_{i=0}^{N-k} \binom{N-k}{i} P^i (1-P)^{N-k-i} = 1
$ and for $i=0,\ldots, N-k$, we have that 
$$
\binom{i+k}{k}\leq \left(\frac{(i+k)\cdot e}{k}\right)^k\leq 
\left(\frac{N\cdot e}{k}\right)^k,
$$
where the first inequality follows from Stirling's approximation (see 
\cite{Intro_to_Algorithms}, p. 1186).
\end{proof}

\begin{proof}[Proof of Proposition \ref{main_prop}]
We start by mentioning that converting from the Bernoulli sampling model and 
uniform sampling model has become standard in the literature. In particular, 
one can do this by showing that the Bernoulli model implies (up to a constant) 
the uniform sampling model in each of the conditions in Proposition 
\ref{sufficient}. This is straightforward and the reader may consult 
\cite{CandesRombergTao,Candes_Romberg,FoucartRauhutBook} for details. We will 
therefore consider (without loss of generality) only the multilevel Bernoulli 
sampling scheme.

Recall that we are using the following Bernoulli sampling model:
Given $N_0 = 0$, $N_1,\hdots, N_r \in \mathbb{N}$ we let
$$
\{N_{k-1}+1,\hdots, N_k\} \supset \Omega_k \sim \mathrm{Ber}\left(q_k\right), 
\quad q_k = \frac{m_k}{N_k-N_{k-1}}.
$$
Note that we may replace this Bernoulli sampling model 
with the following equivalent  sampling model (see \cite{BAACHGSCS}):
$$
\Omega_k = \Omega^1_k \cup \Omega^2_k \cup \cdots \cup \Omega^u_k, 
\qquad \Omega^j_k \sim \mathrm{Ber}(q^j_k), \qquad 1 \leq k \leq r,
$$
for some $u \in \mathbb{N}$ with
\begin{equation}\label{the_qs}
(1-q^1_k)(1-q^2_k)\cdots (1-q^u_k) = (1 - q_k).
\end{equation}
The latter model is the one we will use throughout the proof and the specific 
value of $u$ will 
be chosen later.  
Note also that because of overlaps we will have
\begin{equation}\label{overlaps}
q^1_k + q^2_k + \hdots + q^u_k \geq q_k, \qquad 1 \leq k \leq r.
\end{equation}
The strategy of the proof is to show the validity of (i) and (ii), and the 
existence of a $\rho \in \mathrm{ran}(U^*(P_{\Omega_1} \oplus \hdots \oplus 
P_{\Omega_r}))$ that satisfies (iii)-(v) in Proposition \ref{sufficient} with 
probability exceeding $1-\epsilon$, where  (iii) is replaced by 
(\ref{replaced}), (iv) is  replaced by  $\|P_MP_\Delta^\perp \rho\|_{l^\infty} 
\leq \frac{1}{2}$ and $L$ in (v) is given by (\ref{L}).

{\bf Step I: The construction of $\rho$:} 
 We start by defining $\gamma = \epsilon/6$ (the reason for this particular 
 choice will become clear later). We also define a number of quantities (and 
 the reason for these choices will become clear later in the proof):
\begin{equation}\label{u_and_v}
u = 8\lceil 3v + \log(\gamma^{-1})\rceil, \qquad  v = 
\lceil\log_2(8KM\sqrt{s}) \rceil, 
\end{equation}
as well as 
$$
\{q^i_k: 1\leq k \leq r, 1\leq i \leq u\}, \quad  \{\alpha_i\}_{i=1}^{u}, 
\quad \{\beta_i\}_{i=1}^{u}
$$ by 
\begin{equation}\label{the_q}
q_k^1 = q_k^2 = \frac{1}{4}q_k, \qquad  \tilde q_k = q_k^3 = \hdots = q_k^u, 
\qquad q_k = (N_k-N_{k-1})m_k^{-1}, \quad 1\leq k\leq r, 
\end{equation}
with
$$
(1-q_k^1)(1-q_k^2)\cdots (1-q_k^u) = (1 - q_k)
$$
and
\begin{equation}\label{alpha_def}
\alpha_1 = \alpha_2 = (2\log_2^{1/2}(4KM \sqrt{s}))^{-1}, \qquad \alpha_i =  
1/2, \quad 3 \leq i \leq u,
\end{equation}
as well as
\begin{equation}\label{beta_def}
\beta_1 = \beta_2 = \frac{1}{4}, \qquad \beta_i =  
\frac{1}{4}\log_2(4 KM \sqrt{s}), \quad 3 \leq i \leq u.
\end{equation}
Consider now the following construction 
of $\rho$. We will 
define recursively the sequences 
$\{Z_i\}_{i=0}^{u} \subset \mathcal{H}$, $\{Y_i\}_{i=1}^{u} \subset 
\mathcal{H}$ and 
$\{\omega_i\}_{i=0}^{u} \subset \mathbb{N}$
as follows: first let $\omega_0 = \{0\}$, $\omega_1 = \{0,1\}$ and $\omega_2 = 
\{0,1,2\}$. Then define
recursively, for $i \geq 3$, the following:
\begin{equation}\label{thetas}
\omega_i = 
\begin{cases}
\omega_{i-1} \cup \{i\} &   \text{if} \,  \|(P_{\Delta} -  
P_{\Delta}U^*(\frac{1}{q^i_1}P_{\Omega^i_1} \oplus \hdots \oplus 
\frac{1}{q^i_r} P_{\Omega^i_r})UP_{\Delta})Z_{i-1}\|_{l^{\infty}} \leq 
\alpha_i\|P_{\Delta_k} Z_{i-1}\|_{l^{\infty}},\\
&\text{and} \, \|P_M P_{\Delta}^{\perp}U^*(\frac{1}{q^i_1}P_{\Omega^i_1} 
\oplus \hdots \oplus \frac{1}{q^i_r} 
P_{\Omega^i_r})UP_{\Delta}Z_{i-1}\|_{l^{\infty}} \leq
\beta_i\|Z_{i-1}\|_{l^{\infty}},             \\
\omega_{i-1} &  \text{otherwise},
\end{cases}
\end{equation}
$$
Y_i = 
\begin{cases}
\sum_{j \in \omega_i} U^*(\frac{1}{q^j_1}P_{\Omega^j_1} \oplus \hdots \oplus 
\frac{1}{q^j_r} P_{\Omega^j_r})UZ_{j-1} &   \text{if} \, i \in  \omega_i,\\
Y_{i-1} &  \text{otherwise},
\end{cases}
\qquad i \geq 1,
$$
$$
Z_i = 
\begin{cases}
\mathrm{sgn}(x_0) - P_{\Delta}Y_i &   \text{if} \, i \in  \omega_i,\\
Z_{i-1} &   \text{otherwise},
\end{cases}
\qquad i \geq 1, \qquad Z_0 = \mathrm{sgn}(x_0).
$$
Now, let $\{A_i\}_{i=1}^2$ and $\{B_i\}_{i=1}^5$ denote the following events
\begin{equation}\label{events}
\begin{split}
&A_i: \qquad \|(P_{\Delta} - U^*(\frac{1}{q^i_1}P_{\Omega^i_1} \oplus \hdots 
\oplus \frac{1}{q^i_r} P_{\Omega^i_r})UP_{\Delta})Z_{i-1}\|_{l^{\infty}} 
\leq \alpha_i\left\|Z_{i-1}\right\|_{l^{\infty}} , \qquad i = 1,2,\\
&B_i: \qquad \|P_MP_{\Delta}^{\perp}U^*(\frac{1}{q^i_1}P_{\Omega^i_1} \oplus 
\hdots \oplus \frac{1}{q^i_r} P_{\Omega^i_r})UP_{\Delta}Z_{i-1}\|_{l^{\infty}} 
\leq
\beta_i\|Z_{i-1}\|_{l^{\infty}}, \qquad i = 1,2,\\
&B_3: \qquad \|P_{\Delta}U^*(\frac{1}{q_1}P_{\Omega_1} \oplus \hdots \oplus 
\frac{1}{q_r} P_{\Omega_r})UP_{\Delta}
-P_{\Delta}\| \leq 1/4, \\
&\qquad\qquad  \max_{i\in\Delta^c \cap \{1,\ldots,M\}}\|\left(q_1^{-1/2} 
P_{\Omega_1} \oplus \ldots \oplus q_r^{-1/2} P_{\Omega_r}\right) U e_i \| \leq 
\sqrt{5/4} \\
&B_4: \qquad |\omega_{u}| \geq v,\\
&B_5:  \qquad (\cap_{i=1}^2 A_i) \cap (\cap_{i=1}^4 B_i).
\end{split}
\end{equation}
Also, let $\tau(j)$ denote
 the $j^{\rth}$ element in $\omega_{u}$ (e.g. $\tau(0) =0, \tau(1) = 1, 
 \tau(2) = 2$ etc.)
and finally define $\rho$ by 
$$
\rho = 
\begin{cases}
Y_{\tau(v)} & \text{if $B_5$ occurs,}\\
0 & \text{otherwise}.
\end{cases}
$$
Note that, clearly, $\rho \in \mathrm{ran}(U^*P_{\Omega})$, and we just need 
to show that when the event $B_5$ occurs, then (i)-(v) in Proposition 
\ref{sufficient} will follow.

{\bf Step II: $B_5 \Rightarrow (\text{i}), (\text{ii})$.} To see that the 
assertion is true, note that if $B_5$ occurs then $B_3$ occurs, which 
immediately (i) and (ii).

{\bf Step III: $B_5 \Rightarrow (\text{iii}), (\text{iv})$}. 
To show the assertion, we start by making the following observations:  By the 
construction of $Z_{\tau(i)}$ and the fact that $Z_0 =  \mathrm{sgn}(x_0)$, it 
follows that 
\begin{equation*}
\begin{split}
Z_{\tau(i)} &= Z_0 - 
(P_{\Delta}U^*(\frac{1}{q^{\tau(1)}_1}P_{\Omega^{\tau(1)}_1} \oplus \hdots 
\oplus \frac{1}{q^{\tau(1)}_r} P_{\Omega^{\tau(i)}_r})UP_{\Delta})Z_0\\
& \quad  + \hdots + 
P_{\Delta}U^*(\frac{1}{q^{\tau(i)}_1}P_{\Omega^{\tau(i)}_1} \oplus \hdots 
\oplus \frac{1}{q^{\tau(i)}_r} 
P_{\Omega^{\tau(i)}_r})UP_{\Delta})Z_{\tau(i-1)})\\
& = Z_{\tau(i-1)} - 
P_{\Delta}U^*(\frac{1}{q^{\tau(i)}_1}P_{\Omega^{\tau(i)}_1} \oplus \hdots 
\oplus \frac{1}{q^{\tau(i)}_r} 
P_{\Omega^{\tau(i)}_r})UP_{\Delta})Z_{\tau(i-1)}\qquad i \leq |\omega_{u}|, 
\end{split}
\end{equation*}
so we immediately get that 
$$
Z_{\tau(i)} = (P_{\Delta} - 
P_{\Delta}U^*(\frac{1}{q^{\tau(i)}_1}P_{\Omega^{\tau(i)}_1} \oplus \hdots 
\oplus \frac{1}{q^{\tau(i)}_r} 
P_{\Omega^{\tau(i)}_r})UP_{\Delta})Z_{\tau(i-1)}, \qquad i \leq |\omega_{u}|. 
$$
Hence, if the event $B_5$ occurs, we have, by the choices in (\ref{alpha_def}) 
and (\ref{beta_def})
\begin{equation}\label{gust}
\|\rho - \mathrm{sgn}(x_0)\| = \|Z_{\tau(v)}\| \leq \sqrt{s} 
\|Z_{\tau(v)}\|_{l^{\infty}} \leq \sqrt{s}\prod_{i=1}^{v} \alpha_{\tau(i)} 
\leq \frac{\sqrt{s}}{2^v} \leq \frac{1}{8K},
\end{equation}
since we have chosen $v = \lceil\log_2(8KM\sqrt{s}) \rceil$. Also, 
\begin{equation}\label{gust2}
\begin{split}
\|P_MP_{\Delta}^{\perp}\rho\|_{l^{\infty}} &\leq \sum_{i=1}^{v} 
\|P_MP_{\Delta}^{\perp}U^*(\frac{1}{q^{\tau(i)}_1}P_{\Omega^{\tau(i)}_1} 
\oplus \hdots \oplus \frac{1}{q^{\tau(i)}_r} 
P_{\Omega^{\tau(i)}_r})UP_{\Delta}Z_{\tau(i-1)}\|_{l^{\infty}}\\
&\leq \sum_{i=1}^{v} \beta_{\tau(i)}\|Z_{\tau(i-1)}\|_{l^{\infty}}
\leq \sum_{i=1}^{v} \beta_{\tau(i)}\prod_{j=1}^{i-1}\alpha_{\tau(j)} \\
&\leq \frac{1}{4}(1+ 
\frac{1}{2\log_2^{1/2}(a)} + \frac{\log_2(a)}{2^3 \log_2(a)} +    \hdots  + 
\frac{1}{2^{v-1}}) \leq \frac{1}{2}, \qquad a = 4KM\sqrt{s}.
\end{split}
\end{equation}
In particular, (\ref{gust}) and (\ref{gust2}) imply (iii) and (iv) in 
Proposition \ref{sufficient}.

{\bf Step IV: $B_5 \Rightarrow (\text{v})$}. To show that, note that we may 
write the already constructed $\rho$ as $\rho = U^* P_\Omega w$ where 
$$
w= \sum_{i=1}^v w_i,\quad w_i = 
\left(\frac{1}{q^{\tau(i)}_1}P_{\Omega_1}\oplus \ldots \oplus 
\frac{1}{q^{\tau(i)}_r}P_{\Omega_r}\right)UP_\Delta Z_{\tau(i-1)}.
$$
To estimate $\|w\|$ we simply compute
\begin{equation*}
\begin{split}
\|w_i\|^2 &= 
\left\langle\left(\frac{1}{q^{\tau(i)}_1}P_{\Omega_1^{\tau(i)}}\oplus \ldots 
\oplus \frac{1}{q^{\tau(i)}_r}P_{\Omega_r^{\tau(i)}}\right)UP_\Delta 
Z_{\tau(i-1)},\left(\frac{1}{q^{\tau(i)}_1}P_{\Omega_1^{\tau(i)}}\oplus \ldots 
\oplus \frac{1}{q^{\tau(i)}_r}P_{\Omega_r^{\tau(i)}}\right)UP_\Delta 
Z_{\tau(i-1)}\right\rangle\\
&= \sum_{k=1}^r \left(\frac{1}{q^{\tau(i)}_k}\right)^2 
\|P_{\Omega_k^{\tau(i)}} U Z_{\tau(i-1)} \|^2,
\end{split}
\end{equation*}
and then use the assumption that  the event $B_5$ holds to deduce that 
\begin{equation*}
\begin{split}
&\sum_{k=1}^r \left(\frac{1}{q^{\tau(i)}_k}\right)^2 \|P_{\Omega_k^{\tau(i)}} 
U Z_{\tau(i-1)} \|^2
\leq \max_{1\leq k \leq r}\left\{ \frac{1}{q^{\tau(i)}_k}\right\} 
\ip{\sum_{k=1}^r \frac{1}{q^{\tau(i)}_k} P_\Delta U^* P_{\Omega_k^{\tau(i)}} U 
Z_{\tau(i-1)} }{Z_{\tau(i-1)}}\\
&= \max_{1\leq k \leq r}\left\{ \frac{1}{q^{\tau(i)}_k}\right\} \ip{ 
\left(\sum_{k=1}^r \frac{1}{q^{\tau(i)}_k}  P_\Delta U^* 
P_{\Omega_k^{\tau(i)}} U - P_\Delta\right) Z_{\tau(i-1)}}{Z_{\tau(i-1)}}+ \| 
Z_{\tau(i-1)}\|^2\\
&\leq  \max_{1\leq k \leq r}\left\{ \frac{1}{q^{\tau(i)}_k}\right\}  \left(\| 
Z_{\tau(i-1)}\|\|Z_{\tau(i)}\| + \| Z_{\tau(i-1)}\|^2\right) \\
&\leq  \max_{1\leq k \leq r}\left\{ \frac{1}{q^{\tau(i)}_k}\right\} s  
\left(\| Z_{\tau(i-1)}\|_{l^{\infty}}\|Z_{\tau(i)}\|_{l^{\infty}} + \| 
Z_{\tau(i-1)}\|_{l^{\infty}}^2\right) \leq \max_{1\leq k \leq r}\left\{ 
\frac{1}{q^{\tau(i)}_k}\right\} s (\alpha_i 
+1)\left(\prod_{j=1}^{i-1}\alpha_j\right)^2,
\end{split}
\end{equation*}
where the last inequality follows from the assumption that the event $B_5$ 
holds.
Hence
\begin{equation}\label{eq:w_bound}
\begin{split}
\|w\| \leq \sqrt{s}  \sum_{i=1}^v \left( \max_{1\leq k \leq r}\left\{ 
\frac{1}{\sqrt{q^{\tau(i)}_k}}\right\} \sqrt{\alpha_i 
+1}\prod_{j=1}^{i-1}\alpha_j\right)
\end{split}
\end{equation}
Note that, due to the fact that $q_k^1 + \ldots+ q_k^u \geq q_k$, we have that
$$
\tilde{q}_k \geq \frac{m_k}{2(N_k-N_{k-1})}\frac{1}{8\left\lceil 
\log(\gamma^{-1})+ 3\lceil\log_2(8KM\sqrt{s}) \rceil\right\rceil -2 }.
$$
This gives, in combination with the chosen values of $\{\alpha_j\}$ and 
(\ref{eq:w_bound}) that
\begin{equation}
\begin{split}
\|w\| &\leq 2\sqrt{s} \max_{1\leq k \leq r} \sqrt{\frac{N_k-N_{k-1}}{m_k}} 
\left(1+ \frac{1}{2\log_2^{1/2}\left(4KM\sqrt{s}\right)}\right)^{3/2}\\
& + \sqrt{s} \max_{1\leq k \leq r} \sqrt{\frac{N_k-N_{k-1}}{m_k}}\cdot 
\frac{\sqrt{3}}{2}\cdot \frac{\sqrt{8\left\lceil \log(\gamma^{-1})+ 
3\lceil\log_2(8KM\sqrt{s}) \rceil\right\rceil -2 
}}{\log_2\left(4KM\sqrt{s}\right)}\cdot\sum_{i=3}^v \frac{1}{2^{i-3}}\\
&\leq 2\sqrt{s} \max_{1\leq k \leq r} 
\sqrt{\frac{N_k-N_{k-1}}{m_k}}\left(\left(\frac{3}{2}\right)^{3/2} + 
\sqrt{\frac{6}{{\log_2(4KM\sqrt{s})}}}\sqrt{1+\frac{\log_2\left(\gamma^{-1}\right)
 +6}{\log_2(4KM\sqrt{s})}}\right)\\
&\leq \sqrt{s} \max_{1\leq k \leq r} 
\sqrt{\frac{N_k-N_{k-1}}{m_k}}\left(\frac{3\sqrt{3}}{\sqrt{2}} + 
\frac{2\sqrt{6}}{{\sqrt{\log_2(4KM\sqrt{s})}}}\sqrt{1+\frac{\log_2\left(\gamma^{-1}\right)
 +6}{\log_2(4KM\sqrt{s})}}\right).
\end{split}
\end{equation}

{\bf Step V: The weak balancing property, (\ref{conditions31}) and 
(\ref{conditions32}) $\Rightarrow \mathbb{P}(A_1^c \cup A_2^c \cup B_1^c \cup 
B_2^c \cup B_3^c) \leq 5\gamma$}.

To see this, note that by Proposition \ref{gould_l_inf} we immediately get 
(recall that $q^1_k = q^2_k = 1/4q_k$) that $\mathbb{P}(A_1^c) \leq \gamma$ 
and $\mathbb{P}(A_2^c) \leq \gamma$ as long as the weak balancing property and 
\begin{equation}\label{bound_from_prop}
\begin{split}
1 &\gtrsim \Lambda \cdot \left(\log\left(s\gamma^{-1}\right)+1\right) 
\cdot\log\left(\sqrt{s}KM\right), \quad 1 \gtrsim  \Upsilon \cdot 
\left(\log\left(s\gamma^{-1}\right)+1\right) \cdot\log\left(\sqrt{s}KM\right),
\end{split}
\end{equation}
are satisfied, where $K = \max_{1\leq k \leq r} (N_k-N_{k-1})/m_k$,

\begin{equation}\label{Lambda_def2}
\Lambda = \max_{1\leq k \leq r} \left\{ \frac{N_k-N_{k-1}}{m_k} \cdot 
\left(\sum_{l=1}^r  \kappa_{\mathbf{N},\mathbf{M}}(k,l) \right)\right\},
\end{equation}
\begin{equation}\label{Upsilon_def2}
\Upsilon = \max_{1 \leq l \leq r}
\sum_{k=1}^r \left(\frac{N_k-N_{k-1}}{m_k} - 1\right) \cdot 
\mu_{\mathbf{N},\mathbf{M}}(k,l)\cdot \tilde s_k,
\end{equation}
and where
$\tilde s_{1}+ \hdots + \tilde s_{r}  \leq s_1+ \hdots + s_r$ and $\tilde s_k 
\leq S_k(s_1,\hdots, s_r)$.
 However, clearly, (\ref{conditions31}) and (\ref{conditions32}) imply 
 (\ref{bound_from_prop}).  Also,  Proposition \ref{jup} yields that 
 $\mathbb{P}(B_1^c) \leq \gamma$ and $\mathbb{P}(B_2^c) \leq \gamma$ as long 
 as the weak balancing property and 
\begin{equation}\label{bound_from_prop2}
 1 \gtrsim \Lambda \cdot \log\left( \frac{4}{ \gamma}(M-s)\right), \qquad 
 1 \gtrsim \Upsilon \cdot \log\left( \frac{4}{ \gamma}(M-s)\right),  
\end{equation}
are satisfied. However, again, (\ref{conditions31}) and (\ref{conditions32}) 
imply (\ref{bound_from_prop2}).  Finally, it remains to bound 
$\mathbb{P}(B_3^c)$. First note that by Theorem \ref{inverse_bound}, we may 
deduce that 
$$
\mathbb{P}\left(\|P_{\Delta}U^*(\frac{1}{q_1}P_{\Omega_1} \oplus \hdots \oplus 
\frac{1}{q_r} P_{\Omega_r})UP_{\Delta}
-P_{\Delta}\| > 1/4,\right) \leq \gamma/2,
$$ when 
the weak balancing property and 
\begin{equation}\label{bound_from_prop3}
1 \gtrsim \Lambda \cdot \left(\log\left(\gamma^{-1} \, s\right) +1\right)
\end{equation}
holds and (\ref{conditions31})  implies (\ref{bound_from_prop3}).

For the second part of $B_3$, we may deduce from Proposition 
\ref{thm:max_column_bound} that
$$ \bbP\left(\max_{i\in\Delta^c\cap\{1,\ldots,M\}}\|\left(q_1^{-1/2} 
P_{\Omega_1} \oplus \ldots \oplus q_r^{-1/2} P_{\Omega_r}\right) U e_i \| > 
\sqrt{5/4}\right) \leq \frac{\gamma}{2},$$
whenever
\be{\label{eq:cond_mu_1}
1 \gtrsim \log\left(\frac{2M}{\gamma}\right)\cdot \max_{1 \leq k \leq r} 
\left\{\left(\frac{N_k - N_{k-1}}{m_k}-1\right) \cdot  
\mu_{\mathbf{N},\mathbf{M}}(k,l) \right\}, \qquad l=1,\ldots,r.
}
which is true whenever (\ref{conditions31}) holds. Indeed,  recalling the 
definition of $\kappa_{\mathbf{N},\mathbf{M}}(k,j)$ and $\Theta$ in Definition 
\ref{def:kappa}, observe that
\be{  \label{kappa_to_mu}
\max_{\eta \in\Theta, \| \eta\|_\infty=1} \sum_{l=1}^r \nm{P_{N_k}^{N_{k-1}} U 
P_{M_l}^{M_{l-1}}\eta}_\infty \geq \max_{\eta \in\Theta, \| \eta\|_\infty=1} 
\nm{P_{N_k}^{N_{k-1}} U \eta}_\infty \geq \sqrt{\mu(P_{N_k}^{N_{k-1}} U 
P_{M_l}^{M_{l-1}})}
}
for each $l=1,\ldots, r$ which implies that 
$
\sum_{j=1}^r \kappa_{\mathbf{N},\mathbf{M}}(k,j) \geq  
\mu_{\mathbf{N},\mathbf{M}}(k,l),
$
for $l=1,\ldots, r.$ 
Consequently,  (\ref{eq:cond_mu_1}) follows from (\ref{conditions31}).
Thus, $\bbP(B_3^c)\leq \gamma$.

{\bf Step VI: The weak balancing property, (\ref{conditions31}) and 
(\ref{conditions32}) $\Rightarrow  \mathbb{P}(B_4^c) \leq \gamma$}. 
To see this, define
the random variables $X_1, \hdots X_{u-2}$ by 
\begin{equation}\label{theX}
X_j = 
\begin{cases}
0 & \omega_{j+2} \neq \omega_{j+1},\\
1 & \omega_{j+2} = \omega_{j+1}.
\end{cases}
\end{equation}
We immediately observe that 
\begin{equation}\label{bound_B4}
\mathbb{P}(B_4^c) = \mathbb{P}(|\omega_{u}| < v) = \mathbb{P}(X_1+\hdots 
+X_{u-2} > u-v).
\end{equation}
However, the random variables $X_1, \hdots X_{u-2}$ are not independent, and 
we therefore cannot directly apply the standard Chernoff bound. In particular, 
we must adapt the setup slightly. Note that 
\begin{equation}\label{conditioning}
\begin{split}
&\mathbb{P}(X_1+\hdots +X_{u-2} > u-v) \\&\leq \sum_{l=1}^{\binom{u-2}{u-v}} 
\mathbb{P}(X_{\pi(l)_1} =1, X_{\pi(l)_2} =1, \hdots, X_{\pi(l)_{u-v}} = 1)\\
&= \sum_{l=1}^{\binom{u-2}{u-v}} \mathbb{P}(X_{\pi(l)_{u-v}} = 1 \, \vert \, 
X_{\pi(l)_1} =1, \hdots,  X_{\pi(l)_{u-v-1}} =1)\mathbb{P}(X_{\pi(l)_1} =1, 
\hdots,  X_{\pi(l)_{u-v-1}} = 1)\\
&= \sum_{l=1}^{\binom{u-2}{u-v}} \mathbb{P}(X_{\pi(l)_{u-v}} = 1 \, \vert \, 
X_{\pi(l)_1} =1, \hdots,  X_{\pi(l)_{u-v-1}} =1)\\
&\qquad \times  
\mathbb{P}(X_{\pi(l)_{u-v-1}} = 1 \, \vert \, X_{\pi(l)_1} =1, \hdots,  
X_{\pi(l)_{u-v-2}} =1) \cdots \mathbb{P}(X_{\pi(l)_1} =1)
\end{split}
\end{equation}
where $\pi: \{1,\hdots, \binom{u-2}{u-v}\} \rightarrow \mathbb{N}^{u-v}$ 
ranges over all $\binom{u-2}{u-v}$ ordered subsets of  $\{1,\hdots, u-2\}$
of size $u-v$.
Thus, if we can provide a bound $P$ such that 
\begin{equation}\label{prob_bound}
\begin{split}
P &\geq \mathbb{P}(X_{\pi(l)_{u-v-j}} = 1 \, \vert \, X_{\pi(l)_1} =1, 
\hdots,  X_{\pi(l)_{u-v-(j+1)}} =1), \\
P &\geq \mathbb{P}(X_{\pi(l)_1} = 1)
\end{split}
\end{equation} 
$$
l = 1, \hdots, \binom{u-2}{u-v}, \quad j = 0,\hdots, u-v-2,
$$
then, by (\ref{conditioning}),  
\begin{equation}\label{final_bound}
\mathbb{P}(X_1+\hdots +X_{u-2} > u-v) \leq \binom{u-2}{u-v} P^{u-v}.
\end{equation}
We will continue assuming that (\ref{prob_bound}) is true, and then return to 
this inequality below.

Let $\{\tilde X_k\}_{k=1}^{u-2}$ be independent binary variables taking values 
$0$ and $1$, such that $\tilde X_k = 1$ with probability $P$. Then, by Lemma 
\ref{binom_lemma}, (\ref{final_bound}) and (\ref{bound_B4}) it follows that 
\begin{equation}\label{bound_B4_II}
\mathbb{P}(B_4^c) \leq \mathbb{P}\left(\tilde X_1+\hdots + \tilde X_{u-2} \geq 
u-v \right) \left(\frac{(u-2)\cdot e}{u-v} \right)^{u-v}.
\end{equation}
Then, by the standard Chernoff bound (\cite[Theorem 2.1, equation 
2]{McDiarmid}), it follows that, for $t > 0$,
\begin{equation}\label{tony}
\mathbb{P}\left(\tilde X_1+\hdots + \tilde X_{u-2} \geq  (u -2)(t + P)\right) 
\leq e^{-2(u-2) t^2}.
\end{equation}
Hence, if we let $t = (u-v)/(u-2) - P$, it follows from (\ref{bound_B4_II}) 
and (\ref{tony}) that 
$$
\mathbb{P}(B_4^c) \leq e^{-2(u-2) t^2 + (u-v)(\log(\frac{u-2}{u-v})+1)} \leq 
e^{-2(u-2) t^2 + u-2}.
$$
Thus, by choosing $P = 1/4$ we get that 
$
\mathbb{P}(B_4^c) \leq \gamma
$
whenever $u \geq x$ and $x$ is the largest root satisfying 
$$
(x-u)\left(\frac{x-v}{u-2} - \frac{1}{4} \right) - \log(\gamma^{-1/2}) - 
\frac{x-2}{2} = 0,
$$
and this yields $u \geq 8\lceil 3v + \log(\gamma^{-1/2})\rceil$ which is 
satisfied by the choice of $u$ in (\ref{u_and_v}).
Thus, we would have been done with Step VI if we could verify 
(\ref{prob_bound}) with $P = 1/4$, and this is the theme in the following 
claim.

{\bf Claim: The weak balancing property, (\ref{conditions31}) and 
(\ref{conditions32}) $\Rightarrow$ (\ref{prob_bound}) with $P = 1/4$.}
To prove the claim we first observe that 
$X_j = 0$ when
$$
\|(P_{\Delta} -  P_{\Delta}U^*(\frac{1}{q^i_1}P_{\Omega^i_1} \oplus \hdots 
\oplus \frac{1}{q^i_r} P_{\Omega^i_r})UP_{\Delta})Z_{i-1}\|_{l^{\infty}} \leq 
\frac{1}{2}\|Z_{i-1}\|_{l^{\infty}}
$$
$$
\|P_M P_{\Delta}^{\perp}U^*(\frac{1}{q^i_1}P_{\Omega^i_1} \oplus \hdots \oplus 
\frac{1}{q^i_r} P_{\Omega^i_r})UP_{\Delta}Z_{i-1}\|_{l^{\infty}} \leq
\frac{1}{4}\log_2(4 K M \sqrt{s})\|Z_{i-1}\|_{l^{\infty}}, \qquad i = j+2,
$$
where we recall from (\ref{the_q}) that 
$$
q_k^3 = q_k^4 = \hdots = q^u_k = \tilde q_k, \qquad 1\leq k \leq r.
$$
Thus, by choosing $\gamma = 1/8$ in (\ref{alpha_half}) in Proposition 
\ref{gould_l_inf} and $\gamma = 1/8$ in (i) in Proposition \ref{jup}, it 
follows that  
$
\frac{1}{4}\geq \mathbb{P}(X_j = 1), 
$ for 
$j = 1,\hdots, u-2,$
when the weak balancing property is satisfied and 
  \begin{align}\label{conditions_on_q1}
 \left(\log\left(8s\right)+1\right)^{-1} &\gtrsim \tilde q_k^{-1} \cdot 
 \sum_{l=1}^r \kappa_{\mathbf{N},\mathbf{M}}(k,l), \quad 1 \leq k \leq r \\
 \label{conditions_on_q12}
 \left(\log\left(8s\right)+1\right)^{-1}  & \gtrsim \left( \sum_{k=1}^r 
 \left(\tilde q_k^{-1} - 1\right) \cdot \mu_{\mathbf{N},\mathbf{M}}(k,l)\cdot 
 \tilde s_k \right), \quad 1 \leq l \leq r,
  \end{align}
as well as
\begin{align}\label{conditions_on_q2}
 \frac{\log_2(4KM\sqrt{s})}{\log\left( 32(M-s)\right)} & \gtrsim \tilde 
 q_k^{-1} \cdot \sum_{l=1}^r\kappa_{\mathbf{N},\mathbf{M}}(k,l), \quad 1 \leq 
 k \leq r\\
\label{conditions_on_q22}
 \frac{\log_2(4KM\sqrt{s})}{\log\left( 32(M-s)\right)} &
 \gtrsim \left( \sum_{k=1}^r\left(\tilde q_k^{-1} - 1\right) \cdot 
 \mu_{\mathbf{N},\mathbf{M}}(k,l) \cdot \tilde s_k \right), \quad 1 \leq l 
 \leq r,
\end{align}
with
$
K = \max_{1\leq k \leq r} (N_k-N_{k-1})/m_k.
$
Thus, to prove the claim we must demonstrate that (\ref{conditions31}) and 
(\ref{conditions32}) $\Rightarrow$ (\ref{conditions_on_q1}), 
(\ref{conditions_on_q12}),   (\ref{conditions_on_q2}) and 
(\ref{conditions_on_q22}). We split this into two stages:

{\bf Stage 1:} (\ref{conditions32}) $\Rightarrow$ (\ref{conditions_on_q22}) 
and (\ref{conditions_on_q12}). 
To show the assertion we must demonstrate that if, for $1 \leq k \leq r$,
\begin{equation}\label{m_k}
m_k \gtrsim (\log(s\epsilon^{-1}) + 1)  \cdot \hat m_k \cdot \log\left(K M 
\sqrt{s}\right),
\end{equation} 
where $\hat m_k$ satisfies
\begin{equation}\label{hatm_k}
 1 \gtrsim \sum_{k=1}^r  \left(\frac{N_k-N_{k-1}}{\hat m_k} - 1\right) \cdot 
 \mu_{\mathbf{N}, \mathbf{M}}(k,l) \cdot \tilde s_k, \qquad l=1,\ldots,r,
\end{equation}
we get (\ref{conditions_on_q22}) and (\ref{conditions_on_q12}). To see this, 
note that by (\ref{overlaps}) we have that 
\begin{equation}\label{q}
q^1_k + q^2_k + (u-2)\tilde q_k \geq q_k, \qquad 1 \leq k \leq r,
\end{equation}
so since $q^1_k = q^2_k = \frac{1}{4}q_k$, and by (\ref{q}), (\ref{m_k}) and 
the choice of $u$ in (\ref{u_and_v}), it follows that  

\begin{align*}
2(8(\lceil\log(\gamma^{-1}) + &3\lceil\log_2(8KM\sqrt{s}) 
\rceil\rceil)-2)\tilde q_k \geq q_k =  \frac{m_k}{N_k-N_{k-1}} \\
& \geq C \frac{\hat m_k}{N_k-N_{k-1}}(\log(s\epsilon^{-1}) +1) \log\left(K M 
\sqrt{s}\right)\\
&\geq C \frac{\hat m_k}{N_k-N_{k-1}} (\log(s)+1)(\log\left(K M 
\sqrt{s}\right)+ \log(\epsilon^{-1})),
\end{align*}
for some constant $C$ (recall that we have assumed that $\log(s)\geq 1$). And 
this gives (by recalling that $\gamma= \epsilon/6$) that 
$
\tilde q_k \geq \hat C \frac{\hat m_k}{N_k-N_{k-1}} (\log(s)+1),
$
for some constant $\hat C$. 
Thus, (\ref{conditions32}) implies that for $1 \leq l \leq r$,
\eas{
 1 &\gtrsim \left(\log\left(s\right)+1\right) \left( \sum_{k=1}^r 
 \left(\frac{N_k-N_{k-1}}{m_k(\log(s)+1)} - \frac{1}{\log(s)+1}\right) \cdot 
 \mu_{\mathbf{N},\mathbf{M}}(k,l)\cdot \tilde s_k \right)\\
 &\gtrsim  \left(\log\left(s\right)+1\right) \left( \sum_{k=1}^r \left(\tilde 
 q_k^{-1} - 1\right) \cdot \mu_{\mathbf{N},\mathbf{M}}(k,l)\cdot \tilde s_k 
 \right),
}
and this implies (\ref{conditions_on_q22}) and (\ref{conditions_on_q12}), 
given an appropriate choice of the constant $C$.

{\bf Stage 2:} (\ref{conditions31}) $\Rightarrow$ (\ref{conditions_on_q2}) and 
(\ref{conditions_on_q1}).  
To show the assertion we must demonstrate that if, for $1 \leq k \leq r$,
\begin{equation}\label{m_k2}
 1 \gtrsim (\log(s\epsilon^{-1}) + 1) \cdot \frac{N_k-N_{k-1}}{m_k} \cdot ( 
 \sum_{l=1}^r \kappa_{\mathbf{N}, \mathbf{M}}(k,l) )  \cdot \log\left(K M 
 \sqrt{s}\right), 
\end{equation} 
we obtain (\ref{conditions_on_q2}) and (\ref{conditions_on_q1}). To see this, 
note that by arguing as above via the fact that  $q^1_k = q^2_k = 
\frac{1}{4}q_k$, and by (\ref{q}), (\ref{m_k2}) and the choice of $u$ in 
(\ref{u_and_v}) we have that  
\begin{align*}
2(8(\lceil\log(\gamma^{-1}) +& 3\lceil\log_2(8KM\sqrt{s}) 
\rceil\rceil)-2)\tilde q_k \geq q_k =  \frac{m_k}{N_k-N_{k-1}} \\&\geq C  
\cdot ( \sum_{l=1}^r \kappa_{\mathbf{N}, \mathbf{M}}(k,l) ) \cdot 
(\log(s\epsilon^{-1}) + 1) \cdot \log\left(K M \sqrt{s}\right)\\
&\geq C  \cdot ( \sum_{l=1}^r \kappa_{\mathbf{N}, \mathbf{M}}(k,l) ) \cdot 
(\log(s)+1) \left( \log(\epsilon^{-1})  + \log\left(K M \sqrt{s}\right)\right),
\end{align*}
for some constant $C$.  Thus, we have that for some appropriately chosen 
constant $\hat C$,
$
\tilde q_k \geq \hat C \cdot (\log(s)+1)\cdot\sum_{l=1}^r \kappa_{\mathbf{N}, 
\mathbf{M}}(k,l).
$
So, (\ref{conditions_on_q2}) and (\ref{conditions_on_q1}) holds given an 
appropriately chosen $C$.
This yields the last puzzle of the proof, and we are done.
\end{proof}

\begin{proof}[Proof of Proposition \ref{main_prop2}]
The proof is very close to the proof of Proposition \ref{main_prop} and we 
will simply point out the differences.
The strategy of the proof is to show the  validity of (i) and (ii), and the 
existence of a $\rho \in \mathrm{ran}(U^*(P_{\Omega_1} \oplus \hdots \oplus 
P_{\Omega_r}))$ that satisfies (iii)-(v) in Proposition \ref{sufficient} with 
probability exceeding $1-\epsilon$.

{\bf Step I: The construction of $\rho$:} The construction is almost identical 
to the construction in the proof of Proposition \ref{main_prop}, except that 
\begin{equation}\label{u_and_v2}
u = 8\lceil\log(\gamma^{-1}) + 3v\rceil, \qquad  v = \lceil\log_2(8K\tilde 
M\sqrt{s}) \rceil, 
\end{equation}
\begin{equation*}
\alpha_1 = \alpha_2 = (2\log_2^{1/2}(4K\tilde M \sqrt{s}))^{-1}, \qquad 
\alpha_i =  1/2, \quad 3 \leq i \leq u,
\end{equation*}
as well as
\begin{equation*}
\beta_1 = \beta_2 = \frac{1}{4}, \qquad \beta_i =  
\frac{1}{4}\log_2(4 K\tilde M \sqrt{s}), \quad 3 \leq i \leq u,
\end{equation*}
and (\ref{thetas}) gets changed to 
\begin{equation*}
\omega_i = 
\begin{cases}
\omega_{i-1} \cup \{i\} &   \text{if} \,  \|(P_{\Delta} -  
P_{\Delta}U^*(\frac{1}{q^i_1}P_{\Omega^i_1} \oplus \hdots \oplus 
\frac{1}{q^i_r} P_{\Omega^i_r})UP_{\Delta})Z_{i-1}\|_{l^{\infty}} \leq 
\alpha_i\|P_{\Delta_k} Z_{i-1}\|_{l^{\infty}},\\
&\text{and} \, \|P_{\Delta}^{\perp}U^*(\frac{1}{q^i_1}P_{\Omega^i_1} \oplus 
\hdots \oplus \frac{1}{q^i_r} P_{\Omega^i_r})UP_{\Delta}Z_{i-1}\|_{l^{\infty}} 
\leq
\beta_i\|Z_{i-1}\|_{l^{\infty}},             \\
\omega_{i-1} &  \text{otherwise},
\end{cases}
\end{equation*}
the events $B_i$, $i = 1,2$ in (\ref{events}) get replaced by 
\begin{equation*}
\begin{split}
&\widetilde B_i: \qquad \|P_{\Delta}^{\perp}U^*(\frac{1}{q^i_1}P_{\Omega^i_1} 
\oplus \hdots \oplus \frac{1}{q^i_r} 
P_{\Omega^i_r})UP_{\Delta}Z_{i-1}\|_{l^{\infty}} \leq
\beta_i\|Z_{i-1}\|_{l^{\infty}}, \qquad i = 1,2.\\
\end{split}
\end{equation*}
and the second part of $B_3$ becomes
$$ \max_{i\in\Delta^c}\|\left(q_1^{-1/2} P_{\Omega_1} \oplus \ldots \oplus 
q_r^{-1/2} P_{\Omega_r}\right) U e_i \| \leq \sqrt{5/4} .
$$

{\bf Step II: $B_5 \Rightarrow (\text{i}), (\text{ii})$}. This step is 
identical to Step II in the proof of Proposition \ref{main_prop}. 

{\bf Step III: $B_5 \Rightarrow (\text{iii}), (\text{iv})$}. 
Equation (\ref{gust2}) gets changed to
\begin{equation*}
\begin{split}
\|P_{\Delta}^{\perp}\rho\|_{l^{\infty}} &\leq \sum_{i=1}^{v} 
\|P_{\Delta}^{\perp}U^*(\frac{1}{q^{\tau(i)}_1}P_{\Omega^{\tau(i)}_1} \oplus 
\hdots \oplus \frac{1}{q^{\tau(i)}_r} 
P_{\Omega^{\tau(i)}_r})UP_{\Delta}Z_{\tau(i-1)}\|_{l^{\infty}}\\
&\leq \sum_{i=1}^{v} \beta_{\tau(i)}\|Z_{\tau(i-1)}\|_{l^{\infty}}
\leq \sum_{i=1}^{v} \beta_{\tau(i)}\prod_{j=1}^{i-1}\alpha_{\tau(j)} \\
&\leq \frac{1}{4}(1+ 
\frac{1}{2\log_2^{1/2}(a)} + \frac{\log_2(a)}{2^3 \log_2(a)} +    \hdots  + 
\frac{1}{2^{v-1}}) \leq \frac{1}{2}, \qquad a = 4\tilde MK\sqrt{s}.
\end{split}
\end{equation*}

{\bf Step IV: $B_5 \Rightarrow (\text{v})$}. This step is identical to Step IV 
in the proof of Proposition \ref{main_prop}.

{\bf Step V: The strong balancing property, (\ref{conditions41}) and 
(\ref{conditions42}) $\Rightarrow \mathbb{P}(A_1^c \cup A_2^c \cup \widetilde 
B_1^c \cup \widetilde B_2^c \cup B_3^c) \leq 5\gamma$}. 
We will start by bounding $\mathbb{P}(\widetilde B_1^c )$ and  
$\mathbb{P}(\widetilde B_2^c ).$ Note that by 
 Proposition \ref{jup} (ii) it follows that $\mathbb{P}(\widetilde B_1^c) \leq 
 \gamma$ and $\mathbb{P}(\widetilde B_2^c) \leq \gamma$ as long as the strong 
 balancing property is satisfied and 
\begin{equation}\label{bound_from_prop2strong}
 1 \gtrsim \Lambda \cdot \log\left( \frac{4}{ \gamma}(\tilde \theta-s)\right), 
 \qquad 
 1 \gtrsim \Upsilon \cdot \log\left( \frac{4}{ \gamma}(\tilde \theta-s)\right)
\end{equation} 
where $\tilde \theta = \tilde \theta(\{q^i_k\}_{k=1}^r, 1/8, 
\{N_k\}_{k=1}^r,s, M )$ for $i = 1,2$ and where $\tilde \theta$ is defined in 
Proposition \ref{jup} (ii) and $\Lambda$ and $\Upsilon$ are defined in 
(\ref{Lambda_def2}) and (\ref{Upsilon_def2}).
Note that it is easy to see that we have 
$$
\abs{\left\lbrace j \in\bbN : \max_{\substack{\Gamma_1 \subset 
\{1,\ldots,M\},\quad \abs{\Gamma_1} = s \\ \Gamma_{2,j}\subset 
\{N_{j-1}+1,\ldots,N_j\}, \quad j=1,\ldots,r}} \| P_{\Gamma_1} U^* 
((q^i_1)^{-1}P_{\Gamma_{2,1}} \oplus \hdots \oplus 
(q^i_r)^{-1}P_{\Gamma_{2,r}}) U e_j\| > \frac{1}{8\sqrt{s}} \right\rbrace} 
\leq \tilde M,
$$
where 
$$
\tilde{M} = \min\{i\in\bbN: \max_{j\geq i}\| P_N U P_{\{j\}} \| \leq 
1/(K32\sqrt{s})\}, 
$$
and this follows from the choice in (\ref{the_q}) where $q_k^1 = q_k^2 = 
\frac{1}{4}q_k$ for $1 \leq k \leq r$. Thus, it immediately follows that 
(\ref{conditions41}) and (\ref{conditions42}) imply 
(\ref{bound_from_prop2strong}).
To bound $\mathbb{P}(B_3^c)$, we first deduce as in Step V of the proof of  
Proposition \ref{main_prop} that 
$$
\mathbb{P}\left(\|P_{\Delta}U^*(\frac{1}{q_1}P_{\Omega_1} \oplus \hdots \oplus 
\frac{1}{q_r} P_{\Omega_r})UP_{\Delta}
-P_{\Delta}\| > 1/4,\right) \leq \gamma/2
$$ when 
the strong balancing property and (\ref{conditions41}) holds.
For the second part of $B_3$, we know from the choice of $\tilde M$ that
$$\max_{i\geq \tilde M}\| \left(q_1^{-1/2} P_{\Omega_1} \oplus \ldots \oplus 
q_r^{-1/2} P_{\Omega_r}\right) U e_i \| \leq \sqrt{\frac{5}{4}}
$$
and we
may deduce from Proposition \ref{thm:max_column_bound}  that
$$ \bbP\left(\max_{i\in\Delta^c\cap\{1,\ldots,\tilde M\}}\|\left(q_1^{-1/2} 
P_{\Omega_1} \oplus \ldots \oplus q_r^{-1/2} P_{\Omega_r}\right) U e_i \| > 
\sqrt{5/4}\right) \leq \frac{\gamma}{2},$$
whenever
\bes{
1 \gtrsim \log\left(\frac{2 \tilde M}{\gamma}\right)\cdot \max_{1 \leq k \leq 
r} \left\{\left(\frac{N_k - N_{k-1}}{m_k}-1\right) \, \mu_{\mathbf{N}, 
\mathbf{M}}(k,l) \right\}, \quad l=1,\ldots,r-1,\infty,
}
which is true whenever (\ref{conditions41}) holds, since by a similar argument 
to (\ref{kappa_to_mu}),
\bes{
 \kappa_{\mathbf{N},\mathbf{M}}(k,\infty) + \sum_{j=1}^{r-1} 
 \kappa_{\mathbf{N},\mathbf{M}}(k,j) \geq  \mu_{\mathbf{N},\mathbf{M}}(k,l), 
 \qquad l=1,\ldots, r-1,\infty. 
} Thus, $\bbP(B_3^c) \leq \gamma$. As for bounding $\mathbb{P}(A_1^c)$ and  
$\mathbb{P}(A_2^c)$, observe that by the strong balancing property $\tilde M 
\geq M$, thus this is done exactly as in Step V of the proof of Proposition 
\ref{main_prop}. 

{\bf Step VI: The strong balancing property, (\ref{conditions41}) and 
(\ref{conditions42}) $\Rightarrow  \mathbb{P}(B_4^c) \leq \gamma$}. 
To see this, define
the random variables $X_1, \hdots X_{u-2}$ as in (\ref{theX}). Let $\pi$ be 
defined as in Step VI of the proof of Proposition \ref{main_prop}.  Then it 
suffices to show that (\ref{conditions41}) and (\ref{conditions42}) imply  
that for $l=1,\ldots \binom{u-2}{u-v}$ and $ j = 0,\hdots, u-v-2$, we have
\begin{equation}\label{prob_bound2}
\begin{split}
\frac{1}{4}  &\geq \mathbb{P}(X_{\pi(l)_{u-v-j}} = 1 \, \vert \, X_{\pi(l)_1} 
=1, \hdots,  X_{\pi(l)_{u-v-(j+1)}} =1), \\
\frac{1}{4}  &\geq \mathbb{P}(X_{\pi(l)_1} = 1).
\end{split}
\end{equation}

{\bf Claim: The strong balancing property, (\ref{conditions41}) and 
(\ref{conditions42}) $\Rightarrow$ (\ref{prob_bound2}).}
To prove the claim we first observe that 
$X_j = 0$ when
$$
\|(P_{\Delta} -  P_{\Delta}U^*(\frac{1}{q^i_1}P_{\Omega^i_1} \oplus \hdots 
\oplus \frac{1}{q^i_r} P_{\Omega^i_r})UP_{\Delta})Z_{i-1}\|_{l^{\infty}} \leq 
\frac{1}{2}\|Z_{i-1}\|_{l^{\infty}}
$$
$$
\|P_{\Delta}^{\perp}U^*(\frac{1}{q^i_1}P_{\Omega^i_1} \oplus \hdots \oplus 
\frac{1}{q^i_r} P_{\Omega^i_r})UP_{\Delta}Z_{i-1}\|_{l^{\infty}} \leq
\frac{1}{4}\log_2(4 K \tilde M \sqrt{s})\|Z_{i-1}\|_{l^{\infty}}, \qquad i = 
j+2.
$$
Thus, by again recalling from (\ref{the_q}) that 
$
q_k^3 = q_k^4 = \hdots = q^u_k = \tilde q_k,
$
$1\leq k \leq r$,
and by choosing $\tilde \gamma = 1/4$ in (\ref{alpha_half}) in Proposition 
\ref{gould_l_inf} and $\tilde \gamma = 1/4$ in (ii) in Proposition \ref{jup}, 
we conclude that  (\ref{prob_bound2}) follows 
when the strong balancing property is satisfied as well as 
(\ref{conditions_on_q1}) and (\ref{conditions_on_q12}).
 and
\begin{align}\label{conditions_on_q42}
 \frac{\log_2(4K \tilde M\sqrt{s})}{\log\left( 16(\tilde M-s)\right)} &\geq 
 C_2\cdot \tilde q_k^{-1} \cdot \left(\sum_{l=1}^{r-1} \kappa_{\mathbf{N}, 
 \mathbf{M}}(k,l)+  \kappa_{\mathbf{N}, \mathbf{M}}(k,\infty)\right), \quad k 
 = 1, \ldots, r\\
\label{conditions_on_q422}
 \frac{\log_2(4K\tilde M\sqrt{s})}{\log\left( 16(\tilde M-s)\right)} &\geq 
 C_2 \cdot \left( \sum_{l=1}^{r}  \left(\tilde q_k^{-1} - 1\right) \cdot 
 \mu_{\mathbf{N}, \mathbf{M}}(k,l) \cdot \tilde s_k \right), \quad 
 l=1,\ldots,r-1,\infty
\end{align}
for
$
K = \max_{1\leq k \leq r} (N_k-N_{k-1})/m_k.
$
for some constants $C_1$ and $C_2$.
Thus, to prove the claim we must demonstrate that (\ref{conditions41}) and 
(\ref{conditions42}) $\Rightarrow$ (\ref{conditions_on_q1}), 
(\ref{conditions_on_q12}),   (\ref{conditions_on_q42}) and 
(\ref{conditions_on_q422}). 
This is done by repeating Stage 1 and Stage 2 in Step VI of the proof of 
Proposition \ref{main_prop} almost verbatim, except replacing $M$ by $\tilde 
M$.
\end{proof}

\subsection{Proof of Theorem \ref{Four_to_wave}}\label{ss:Four_wave}
Throughout this section, we use the notation
\begin{equation}\label{Fourier_transform}
\hat f(\xi ) = \int_\bbR f(x) e^{-ix\xi} \mathrm{d}x,
\end{equation}
to denote the Fourier transform of a function $f\in L^1(\bbR)$.

\subsubsection{Setup}\label{Setup}
We first introduce the wavelet sparsity and  Fourier sampling bases that we 
consider, and in particular, their orderings.
Consider an orthonormal basis of compactly supported wavelets with an MRA 
\cite{DaubechiesCPAM,daubechies1992ten}.  For simplicity, suppose that 
$\mathrm{supp}(\Psi) = \mathrm{supp}(\Phi) = [0,a]$ for some $a \geq 1$, where 
$\Psi$ and $\Phi$ are the mother wavelet and scaling function respectively.  
For later use, we recall the following three properties of any such wavelet 
basis:
\begin{enumerate}
\item There exist $ \alpha\geq 1$, $C_\Psi$ and $ C_\Phi>0$, such that
\begin{align} \label{wavelet_decay}
\abs{\hat \Phi(\xi )} \leq \frac{C_\Phi}{(1+\abs{\xi})^\alpha}, \quad 
\abs{\hat \Psi(\xi )} \leq \frac{C_\Psi}{(1+\abs{\xi})^\alpha}. 
\end{align}
See \cite[Eqn.\  (7.1.4)]{daubechies1992ten}.  We will denote 
$\max\{C_\Psi,C_\Phi\}$ by $C_{\Phi,\Psi}$.
\item $\Psi$ has $v\geq 1$ vanishing moments and $\hat \Psi(z) = (-iz)^v 
\theta_\Psi(z)$ for some bounded function $\theta_\Psi$ (see \cite[p.208 \& 
p.284]{mallat09wavelet}). 
\item $\|\hat \Phi\|_{L^\infty}, \|\hat \Psi \|_{L^\infty} \leq 1$.
\end{enumerate}
\begin{remark}
The three properties above are based on the standard setup for an MRA, however, we also consider a stronger assumption on the decay of the Fourier transform of derivatives of the scaling function and the mother wavelet. In particular, in addition, we sometimes assume that for  $C > 0$ and $\alpha \geq 1.5$,
\be{\label{eq:cond2_fdecay_mthm}
\abs{\hat{\Phi}^{(k)}(\xi)}\leq \frac{C}{(1+\abs{\xi})^\alpha}, \quad 
\abs{\hat{\Psi}^{(k)}(\xi)}\leq \frac{C}{(1+\abs{\xi})^\alpha}, \quad 
\xi\in\mathbb{R}, \quad k=0,1,2,
}
where $\hat{\Phi}^{(k)}$ and $\hat{\Psi}^{(k)}$ denotes the $k^{\rth}$ 
derivative of the Fourier transform of $\Phi$ and $\Psi$ respectively.  As is evident from Theorem \ref{Four_to_wave}, the faster decay, the closer the relationship between $N$ and $M$ in the balancing property gets to linear. Also, faster decay and more vanishing moments yield a closer to block-diagonal structure of the matrix $U$. 
\end{remark}

We now wish to construct a wavelet basis for the compact interval $[0,a]$.  
The most standard approach is to consider the following collection of functions
\eas{
\Lambda_a = \{&\Phi_k, \Psi_{j,k}: \mathrm{supp}(\Phi_k)^o \cap [0,a] \neq 
\emptyset,  \, \mathrm{supp}(\Psi_{j,k})^o \cap [0,a] \neq \emptyset, j \in 
\mathbb{Z}_+, k \in \mathbb{Z}, \},
}
where 
$
\Phi_k = \Phi(\cdot - k),
$
and 
$
\Psi_{j,k} = 2^{\frac{j}{2}}\Psi(2^j \cdot - k).
$
(the notation $K^o$ denotes the interior of a set $K \subseteq \mathbb{R}$).  
This  gives
$$
\left \{ f \in \rL^2(\bbR) : \mathrm{supp}(f) \subseteq [0,a] \right \} 
\subseteq \overline{\mathrm{span}\{\varphi: \varphi \in \Lambda_a\}} \subseteq 
\left \{ f \in \rL^2(\bbR) : \mathrm{supp}(f) \subseteq [-T_1,T_2] \right \},
$$
where $T_1, T_2 > 0$ are such that $[-T_1,T_2]$ contains the support of all 
functions in $\Lambda_a$.
Note that the inclusions may be proper (but not always, as is the case with 
the Haar wavelet).  It is easy to see that 
\eas{
 \Psi_{j,k} \notin \Lambda_a   \Longleftrightarrow \frac{a+k}{2^j} \leq 0, 
 \quad a \leq \frac{k}{2^j},
\\
\Phi_k \notin \Lambda_a   \Longleftrightarrow a+k \leq 0, \quad a \leq k,
}
and therefore
\eas{
\Lambda_a =& \{\Phi_k: |k| = 0,\hdots, \lceil a\rceil-1\}  \cup \{\Psi_{j,k}: j 
\in \mathbb{Z}_+, k \in \mathbb{Z},   -\lceil a\rceil < k < 2^j\lceil 
a\rceil\}.
}
We order $\Lambda_a$ in increasing order of wavelet resolution as follows:
\begin{equation}\label{ordering}
\begin{split}
&\{ \Phi_{-\lceil a\rceil+1}, \hdots, \Phi_{-1}, \Phi_0 , \Phi_1, \hdots, 
\Phi_{\lceil a\rceil-1}, \\
&\Psi_{0,-\lceil a\rceil+1}, \hdots, \Psi_{0,-1}, \Psi_{0,0},  \Psi_{0,1},   
\hdots, \Psi_{0,\lceil a\rceil-1},  \Psi_{1,-\lceil a\rceil+1}, \hdots \}, 
\end{split}
\end{equation}
and then we finally denote the functions according to this ordering by $\{\varphi_j\}_{j\in\mathbb{N}}.$
By the definition of $\Lambda_a$, we let $T_1=\lceil a\rceil -1$ and 
$T_2=2\lceil a\rceil -1$.  Finally, for $R\in\bbN$, let $\Lambda_{R,a}$ contain 
all wavelets in $\Lambda_a$ with resolution less than $R$, so that
\be{\label{eq:omega_r}
\Lambda_{R,a}  =\{ \varphi \in\Lambda_a: \varphi = \Psi_{j,k}, \ 0 \leq j < R, \text{ or 
} \varphi = \Phi_{k}\}.
}
We also denote the size of $\Lambda_{R,a}$ by $W_R$.  It is easy to verify that 
\be{ \label{eq:W_R}
W_R = 2^R\lceil a\rceil +(R+1)(\lceil a\rceil -1).
}

Having constructed an orthonormal wavelet system for $[0,a]$, we now introduce 
the appropriate Fourier sampling basis.  We must sample at a rate that is at least that of the 
Nyquist rate.  Hence we let $\omega \leq 1/(T_1+T_2)$ be the \textit{sampling 
density} (note that $1/(T_1+T_2)$ is the Nyquist criterion for functions 
supported on $[-T_1,T_2]$). For simplicity, we assume throughout that
\be{
\label{eq:sampling_den_assump}
\omega\in (0,1/(T_1+T_2)), \quad \omega^{-1}\in\bbN,
}
and remark that this assumption is an artefact of our proofs and is not 
necessary in practice.  
The Fourier sampling vectors are now defined as follows. 
\be{\label{Fourier_samples2}
\psi_j(x) = \sqrt{\omega}e^{-2\pi i j \omega x}\chi_{[-T_1/(\omega(T_1 
+T_2)),T_2/(\omega(T_1 +T_2))]}(x), \qquad j \in \mathbb{Z}.
}
This gives an orthonormal sampling basis for the space $\{ f \in \rL^2(\bbR) : 
\mathrm{supp}(f) \subseteq [-T_1,T_2] \}$.  Since $\Lambda_a$ is an orthonormal 
system in for this space, it follows that the infinite matrix 
\begin{equation}\label{inf_U}
U = \left(
\begin{matrix}
 u_{11}         & u_{12}    & u_{13}  & \hdots \\
 u_{21}         & u_{22}    & u_{23}  & \hdots \\
 u_{31}         & u_{32}    & u_{33}  & \hdots \\
\vdots          & \vdots    & \vdots  & \ddots\\
\end{matrix}
\right), \qquad u_{ij} = \langle \varphi_j, \tilde \psi_i \rangle,
\end{equation}
 is an isometry, where 
$\{\varphi_j\}_{j \in \mathbb{N}}$ represents the wavelets ordered according 
to (\ref{ordering}) and $\{\tilde \psi_j\}_{j\in\mathbb{N}}$ is the standard 
ordering of the Fourier basis (\ref{Fourier_samples2}) over $\mathbb{N}$ 
($\tilde \psi_1 = \psi_0$, $\tilde \psi_{2n} = \psi_n$ and $\tilde \psi_{2n+1} 
= \psi_{-n}$). With slight abuse of notation it is this ordering that we are using in Theorem \ref{Four_to_wave}.

\subsubsection{Some preliminary estimates}
Throughout this section, we assume the setup and notation introduced above.

\begin{theorem}
\label{t:Wavelet_Asy_Inc2}
Let $U$ be the matrix of the Fourier/wavelets pair introduced in (\ref{inf_U}) with sampling density $\omega$ as in (\ref{eq:sampling_den_assump}) .  Suppose that $\Phi$ and $\Psi$ satisfy the decay estimate (\ref{wavelet_decay}) with $\alpha \geq 1$ and that $\Psi$ has $v \geq 1$ vanishing moments. Then the following holds.

\begin{itemize}
\item[(i)]  We have $\mu(U) \geq \omega $.  
\item[(ii)] We have that 
\begin{equation*}
\begin{split}
\mu(P_{N}^\perp U ) &\leq   \frac{C_{\Phi, \Psi}^2}{\pi N(2\alpha 
-1)(1+1/(2\alpha -1))^{2\alpha}}, \qquad N \in \mathbb{N},\\
\mu(U P_N^\perp ) &\leq \|\Psi\|_{L^\infty}^2  \frac{4\omega 
\lceil a\rceil}{N}, \qquad N \geq 2\lceil a\rceil +2(\lceil a\rceil -1),
\end{split}
\end{equation*}
and consequently  $\mu(P^{\perp}_N U) , \mu(U P^{\perp}_N ) = \ord{N^{-1}}$. 
\item[(iii)] If the wavelet and scaling function satisfy the decay estimate 
(\ref{wavelet_decay}) with $\alpha> 1/2$, then,  for $R$ and $N$ such that
$\omega^{-1} 2^R \leq N$ and $M=\abs{\Lambda_{R,a}}$ (recall the definition 
of $\Lambda_{R,a}$ from (\ref{eq:omega_r})), $$\mu(P_N^\perp U P_{M})\leq 
\frac{C_{\Phi,\Psi}^2}{\pi^{2\alpha} \omega^{2\alpha -1}} ( 2^{R-1} 
N^{-1})^{2\alpha-1} N^{-1}.$$
\item[(iv)] If the wavelet has $v\geq 1$ vanishing moments, $\omega^{-1}  2^R 
\geq N$ and $M=\abs{\Lambda_{R,a}}$ with $R\geq 1$, then $$\mu(P_N U P_M^\perp) 
\leq \frac{\omega}{2^R} \cdot\left(\frac{\pi \omega N}{ 2^R}\right)^{2v}\cdot 
\nm{\theta_\Psi }_{L^\infty}^2,
$$
where $\theta_\Psi$ is the function such that $\hat \Psi(z) = (-iz)^v 
\theta_\Psi(z)$ (see above).
\end{itemize}
\end{theorem}

\begin{proof}
Note that $\mu(U) \geq \abs{\ip{\Phi}{\psi_0}}^2 = \omega \abs{\hat 
\Phi(0)}^2$, moreover, it is known that $\hat{\Phi}(0) =1$ \cite[Ch. 2, Thm.\ 
1.7]{eugenio1996first}. Thus, (i) follows.

To show (ii), let $R\in\bbN$,  
$-\lceil a\rceil < j < 2^R\lceil a\rceil$ and $k\in\bbZ$. Then, by the choice of $j$, we have that $\Psi_{R,j}$ is supported on $[-T_1,T_2]$.  Also, $\psi_k(x) = \sqrt{\omega}e^{-2\pi i k \omega x}\chi_{[-T_1/(\omega(T_1 
+T_2)),T_2/(\omega(T_1 +T_2))]}(x)$. Thus, since by (\ref{eq:sampling_den_assump}) we have $\omega\in (0,1/(T_1+T_2)),$ it follows that 
\begin{equation}\label{Psi_estimate}
\begin{split}
\ip{\Psi_{R,j}}{\psi_k} &= \sqrt{\omega}\int_{-\frac{T_1}{\omega(T_1+T_2)}}^{\frac{T_2}{\omega(T_1+T_2)}}\Psi_{R,j}(x) e^{2\pi i\omega k x} \mathrm{d}x\\
& = \sqrt{\omega} \hat \Psi_{R,j}(-2\pi \omega k) = 
\sqrt{\frac{\omega}{2^R}}\hat \Psi\left(\frac{-2\pi 
k\omega}{ 2^R}\right)e^{2\pi i \omega kj/2^R}.
\end{split}
\end{equation}
Also, similarly, it follows that 
\begin{equation}\label{Phi_estimate}
\begin{split}
\ip{\Phi_{j}}{\psi_k} =  \sqrt{\omega}\int_{-\frac{T_1}{\omega(T_1+T_2)}}^{\frac{T_2}{\omega(T_1+T_2)}}\Phi_{j}(x) e^{2\pi i\omega k x} \mathrm{d}x = \sqrt{\omega}\hat \Phi_j\left(-2\pi k\omega\right) = \sqrt{\omega}\hat \Phi\left(-2\pi k\omega\right)e^{2\pi i \omega k j}.
\end{split}
\end{equation}
Thus, the decay estimate in (\ref{wavelet_decay}) yields 
\begin{equation*}
\begin{split}
 \mu(P_{N}^\perp U ) &\leq \sup_{\abs{k} \geq \frac{N}{2} 
 }\max_{\varphi\in\Lambda_a} \abs{\ip{ \varphi}{\psi_k}}^2\\ &=   
 \max\left\{\sup_{\abs{k} \geq \frac{N}{2} 
 }\max_{R\in\bbZ_+}\frac{\omega}{2^R} \abs{\hat{\Psi}\left(\frac{-2\pi
 \omega k}{2^R}\right)}^2, \omega\sup_{\abs{k} \geq \frac{N}{2} 
 }\abs{\hat{\Phi}\left(-2\pi
 \omega k\right)}^2 \right\}\\
 &\leq  \max_{\abs{k} \geq \frac{N}{2}}\max_{R\in\bbZ_+} \frac{\omega}{2^R} 
 \frac{C_{\Phi, \Psi}^2}{\left(1+ \abs{2\pi \omega k 2^{-R}}\right)^{2\alpha}} 
 \leq \max_{R\in\bbZ_+} \frac{\omega}{2^R}\frac{C_{\Phi, \Psi}^2}{\left(1+ 
 \abs{\pi \omega N 2^{-R}}\right)^{2\alpha}}.
\end{split}
\end{equation*}
The function $f(x) = x^{-1}(1+\pi\omega N/x)^{-2\alpha}$ on $[1,\infty)$ 
satisfies $f'(\pi\omega N(2\alpha -1)) =0$. Hence
\bes{
\begin{split}
\mu(P_{N}^\perp U ) \leq   \frac{C_{\Phi, \Psi}^2}{\pi N(2\alpha 
-1)(1+1/(2\alpha -1))^{2\alpha}},
\end{split}
}
which gives the first part of (ii). For the second part, we first recall the 
definition of $W_R$ for $R\in\bbN$ from (\ref{eq:W_R}). Then, given any 
$N\in\bbN$ such that $N\geq W_1 =  2\lceil a\rceil +2(\lceil a\rceil -1)$, let $R$ be such that
$
W_R \leq N < W_{R+1}.
$
Then, for each $n\geq N$, there exists some $j \geq R$ and $l\in\bbZ$ such 
that the $n^{th}$ element via the ordering (\ref{ordering}) is $\varphi_n = 
\Psi_{j,l}$ (note that we only need $\Psi_{j,l}$ here and not $\Phi_j$ as we have chosen $N \geq W_1$).  Hence, by using (\ref{Psi_estimate}),
\begin{equation*}
\begin{split}
 \mu(U P_N^\perp ) &= \max_{n\geq N} \max_{k \in\bbZ 
 }\abs{\ip{\varphi_n}{\psi_k}}^2 =  \max_{j\geq R} \max_{k \in\bbZ 
 }\frac{\omega}{2^j}\abs{ \hat{\Psi}\left(\frac{-2\pi
 \omega k}{2^j}\right)}^2\\
 &\leq \|\hat\Psi\|_{L^\infty}^2 \frac{\omega}{2^R} \leq 4 
 \|\hat\Psi\|_{L^\infty}^2 \frac{\omega \lceil a\rceil}{N},
\end{split}
\end{equation*}
where the last line follows because $N < W_{R+1} = 2^{R+1}\lceil a\rceil 
+(R+2)(\lceil a\rceil-1)$ implies that
$$
2^{-R} < \frac{1}{N}\left( 2\lceil a\rceil +(R+2)(\lceil a\rceil 
-1)2^{-R}\right) \leq \frac{4\lceil a\rceil}{N}.
$$
This concludes the proof of (ii).

To show (iii), let $R$ and $N$ be such that
$\omega^{-1} 2^R \leq N$ and $M=\abs{\Lambda_{R,a}}$. Observe that (\ref{Psi_estimate}) and (\ref{Phi_estimate}) together with the decay estimate in (\ref{wavelet_decay}) yield
\eas{
 \mu(P_{N}^\perp U P_{W_R} ) &\leq \max_{\abs{k} \geq \frac{N}{2} 
 }\max_{\varphi\in\Lambda_{R,a}} \abs{\ip{ \varphi}{\psi_k}}^2\\ &=   
 \max\left\{\max_{\abs{k} \geq \frac{N}{2} }\max_{j< R}\frac{\omega}{2^j} 
 \abs{\hat{\Psi}\left(\frac{-2\pi
 \omega k}{2^j}\right)}^2, \max_{\abs{k} \geq \frac{N}{2} 
 }\abs{\hat{\Phi}\left(-2\pi
 \omega k\right)}^2 \right\}\\
 &\leq  \max_{\abs{k} \geq \frac{N}{2}}\max_{j< R} \frac{\omega}{2^j} 
 \frac{C_{\Phi,\Psi}^2}{\left(1+ \abs{2\pi \omega k 2^{-j}}\right)^{2\alpha}}
 \leq \max_{k\geq \frac{N}{2}}\max_{j< R} 
 \frac{C_{\Phi,\Psi}^2}{\pi^{2\alpha}\omega^{2\alpha -1}} 
 \frac{2^{j(2\alpha-1)}}{(2k)^{2\alpha}} \\
&=\frac{C_{\Phi,\Psi}^2}{\pi^{2\alpha} \omega^{2\alpha -1}} ( 2^{R-1} 
N^{-1})^{2\alpha-1} N^{-1},
}
and this colludes the proof of (iii).

To show (iv), first note that because  $R\geq 1$, for all $n>  W_R$ , 
$\varphi_n = \Psi_{j,k}$ for some $j\geq 0$ and $k\in\bbZ$. Then, recalling 
the properties of Daubechies wavelets with $v$ vanishing moments, and by using (\ref{Psi_estimate}) we get that 
\begin{equation*}
\begin{split}
 \mu(P_N U P_{W_R}^\perp ) &= \max_{n>  W_R} \max_{\abs{k} \leq \frac{N}{2} 
 }\abs{\ip{\varphi_n}{\psi_k}}^2
 =  \max_{j\geq R} \max_{\abs{k} \leq \frac{N}{2}  }\frac{\omega}{2^j}\abs{ 
 \hat{\Psi}\left(\frac{-2\pi
 \omega k}{2^j}\right)}^2\\
 &\leq \frac{\omega}{2^R} \cdot \left(\frac{\pi \omega N}{ 
 2^R}\right)^{2v}\cdot \nm{\theta_\Psi }_{L^\infty}^2,
\end{split}
\end{equation*}
as required.
\end{proof}

\begin{corollary}\label{lem:inc_bd}
Let $\mathbf{N}$ and $\mathbf{M}$ be as in Theorem \ref{Four_to_wave} and
recall the definition of $\mu_{\mathbf{N},\mathbf{M}}(k,j)$ in 
(\ref{loc_coherence}). Suppose that $\Phi$ and $\Psi$ satisfy the decay estimate (\ref{wavelet_decay}) with $\alpha \geq 1$ and that $\Psi$ has $v \geq 1$ vanishing moments. Then,
\be{ \label{eq:estimate_mu_wavelet}
\text{for} \,\, k \geq 2,  \quad \mu_{\mathbf{N},\mathbf{M}}(k,j) \leq B_{\Phi,\Psi}\cdot \begin{cases}
\frac{ \sqrt{\omega}}{\sqrt{N_{k-1}2^{R_{j-1}}}}\cdot \left(\frac{\omega 
N_{k}}{2^{R_{j-1}}}\right)^{v}  &j\geq k+1\\
 \frac{1}{N_{k-1}}\left(\frac{2^{R_{j -1}}}{\omega N_{k-1}}\right)^{\alpha 
 -1/2} &j\leq k-1\\
 \frac{1 }{N_{k-1}} &j=k,
\end{cases}
}
\be{\label{eq:estimate_mu_inf}
\text{for} \,\, k \geq 2,  \quad \mu_{\mathbf{N},\mathbf{M}}(k,\infty) \leq B_{\Phi,\Psi}\cdot \begin{cases}
\frac{ \sqrt{\omega}}{\sqrt{N_{k-1}2^{R_{r-1}}}} \cdot  \left(\frac{\omega 
N_{k}}{2^{R_{r-1}}}\right)^{v}& k\leq r-1\\
 \frac{1 }{N_{r-1}} &k=r,
\end{cases}
}
\begin{equation}\label{eq:estimate_mu_wavelet2}
\mu_{\mathbf{N},\mathbf{M}}(1,j) \leq B_{\Phi,\Psi}\cdot \begin{cases}
\frac{ \sqrt{\omega}}{\sqrt{2^{R_{j-1}}}}\cdot \left(\frac{\omega N_1}{2^{R_{j-1}}}\right)^{v}  &j\geq 2\\
  1& j=1,
\end{cases}
\end{equation}
\begin{equation}\label{eq:estimate_mu_inf2}
\mu_{\mathbf{N},\mathbf{M}}(1,\infty) \leq B_{\Phi,\Psi}\cdot 
\frac{ \sqrt{\omega}}{\sqrt{2^{R_{r-1}}}} \cdot  \left(\frac{\omega 
N_{1}}{2^{R_{r-1}}}\right)^{v},
 \end{equation}
where $B_{\Phi,\Psi}$ is a constant which depends only on $\Phi$ and $\Psi$ and $R_0 = 0$.
\end{corollary}

\begin{proof}
Throughout this proof, $B_{\Phi,\Psi}$ is a constant which depends only on 
$\Phi$ and $\Psi$, although its value may change from instance to instance.
Note that
\begin{equation}
\label{eq:mu_1}
\begin{split}
\mu_{\mathbf{N},\mathbf{M}}(k,j) &= \sqrt{\mu(P_{N_k}^{N_{k-1}} U 
P_{M_j}^{M_{j-1}})\cdot \mu(P_{N_k}^{N_{k-1}} U)} \\
&\leq B_{\Phi,\Psi} 
N_{k-1}^{-1/2}\sqrt{\mu(P_{N_k}^{N_{k-1}} U P_{M_j}^{M_{j-1}})}, \qquad k \geq 2, \, j \in \{1,\hdots, r\},
\end{split}
\end{equation}
since we have $\mu(P_{N_{k-1}}^\perp 
U)\leq B_{\Phi,\Psi} N_{k-1}^{-1}$ by (ii) of Theorem \ref{t:Wavelet_Asy_Inc2}. Also, clearly
\begin{equation}
\label{eq:mu_2}
\begin{split}
\mu_{\mathbf{N},\mathbf{M}}(1,j) &= \sqrt{\mu(P_{N_1}^{N_0} U 
P_{M_j}^{M_{j-1}})\cdot \mu(P_{N_1}^{N_0} U)} \leq B_{\Phi,\Psi} \sqrt{\mu(P_{N_1}^{N_0} U P_{M_j}^{M_{j-1}})},
\end{split}
\end{equation}
for $j \in \{1,\hdots, r\}$.
Thus, for $k \geq 2$, it follows that 
$\mu_{\mathbf{N},\mathbf{M}}(k,k)\leq \mu(P_{N_{k-1}}^\perp U)\leq 
B_{\Phi,\Psi} \frac{1}{N_{k-1}},
$
yielding the last part of (\ref{eq:estimate_mu_wavelet}). Also, the last part of 
(\ref{eq:estimate_mu_wavelet2}) is clear from (\ref{eq:mu_2}). 

As for the middle 
part of (\ref{eq:estimate_mu_wavelet}), note that for $k \geq 2$,
and with $j\leq k-1$, we may use (iii) of Theorem \ref{t:Wavelet_Asy_Inc2} to 
obtain
$$
\sqrt{\mu(P_{N_k}^{N_{k-1}} U P_{M_j}^{M_{j-1}})}\leq 
\sqrt{\mu(P_{N_{k-1}}^\perp U P_{M_j})}\leq B_{\Phi,\Psi}\cdot
 \frac{1}{\sqrt{N_{k-1}}}\left(\frac{2^{R_{j -1}}}{\omega 
 N_{k-1}}\right)^{\alpha -1/2},
$$
and thus, in combination with (\ref{eq:mu_1}), we obtain the $j\leq k-1$ part 
of (\ref{eq:estimate_mu_wavelet}). Observe that if $k \in \{1,\hdots, r\}$ and
 $j\geq k+1$, then by applying (iv) of Theorem \ref{t:Wavelet_Asy_Inc2}, we 
obtain
\begin{equation}\label{est_1}
\sqrt{\mu(P_{N_k}^{N_{k-1}} U P_{M_j}^{M_{j-1}})} \leq \sqrt{\mu(P_{N_k} U 
P_{M_{j-1}}^\perp)} \leq B_{\Phi,\Psi}\cdot
\frac{ \sqrt{\omega}}{\sqrt{2^{R_{j-1}}}}\cdot  \left(\frac{\omega 
N_{k}}{2^{R_{j-1}}}\right)^{v}.
\end{equation}
Thus, by combining (\ref{est_1}) with (\ref{eq:mu_1}), we obtain the $j\geq k+1$ part of 
(\ref{eq:estimate_mu_wavelet}). Also, by combining (\ref{est_1}) with (\ref{eq:mu_1}) we get the $j \geq 2$ part of 
(\ref{eq:estimate_mu_wavelet2}).
Finally, recall that
\bes{
\mu_{\mathbf{N},\mathbf{M}}(k,\infty) = \sqrt{\mu(P_{N_k}^{N_{k-1}} U 
P_{M_{r-1}}^\perp)\cdot \mu(P_{N_{k-1}}^\perp U)}
}
and similarly to the above, (\ref{eq:estimate_mu_inf}) and (\ref{eq:estimate_mu_inf2}) are direct consequences 
of parts (ii) and (iv) of Theorem \ref{t:Wavelet_Asy_Inc2}.
\end{proof}

The following lemmas inform us of the range of Fourier samples required for 
accurate reconstruction of wavelet coefficients. Specifically, Lemma \ref{cor:l_inf_normbounds} will provide a quantitative understanding of the balancing property, whilst Lemma \ref{lem:op_wave_bound} and Lemma \ref{lem:nm_bound} will be used in bounding the relative sparsity terms.

\begin{lemma}[{\cite[Corollary 5.4]{gs_l1}}] \label{cor:l_inf_normbounds}
Consider the setup in \S \ref{Setup}. Let the sampling density $\omega$ be such that $\omega^{-1}\in\bbN$ and suppose that there exists $C_\Phi,C_\Psi>0$ and $\alpha \geq 1.5$ such that 
\bes{
\abs{\hat{\Phi}^{(k)}(\xi)}\leq \frac{C_\Phi}{(1+\abs{\xi})^\alpha}, \quad 
\abs{\hat{\Psi}^{(k)}(\xi)}\leq \frac{C_\Psi}{(1+\abs{\xi})^\alpha}, \quad 
\xi\in\mathbb{R},\qquad k=0,1,2.
}
Then given $\gamma\in(0,1)$, we have that
$
\nm{ P_M U^* P_N U P_M - P_M}_{l^\infty \rightarrow l^\infty} \leq \gamma
$
wherever $N \geq C \gamma^{-1/(2\alpha -1)} M$ 
and
$
\nm{ P_M^\perp U^* P_N U P_M}_{l^\infty \rightarrow l^\infty} \leq \gamma
$
wherever $N \geq C \gamma^{-1/(\alpha -1)} M$ where
 $C$ is some constant independent of $N$ but dependent on $C_\Phi$, $C_\Psi$ and $\omega$.
\end{lemma}

\begin{lemma}[{\cite[Lemma 5.1]{gs_l1}}]\label{lem:op_wave_bound}
Let $\varphi_k$ denote the $k^{th}$ wavelet via the ordering in (\ref{ordering}). 
Let $R \in \mathbb{N}$ and $M\leq W_R$ be such that
$
\{\varphi_{j}: j \leq M \} \subset\Lambda_{R,a},
$
where $W_R$ and $\Lambda_{R,a}$ are defined in (\ref{eq:W_R}) and (\ref{eq:omega_r}) respectively.
Also, let the sampling density $\omega$ be such that $\omega^{-1}\in\bbN$.
Then for any $\gamma\in (0,1)$, we have that
$
\nm{P_N^\perp U P_M} \leq \gamma,
$
whenever $N$ is such that
$$
N \geq \omega^{-1} \left( \frac{4 C_\Phi^2}{(2\pi)^{2\alpha} \cdot (2\alpha -1)}\right)^{\frac{1}{2\alpha -1}}\cdot 2^{R+1}\cdot \gamma^{-\frac{2}{2\alpha-1}}
$$ and $ C_\Phi$ is a constant depending on $\Phi$.
\end{lemma}

\begin{lemma} \label{lem:nm_bound}

Let $\varphi_k$ denote the $k^{th}$ wavelet the ordering in (\ref{ordering}). 
Let $R_1,R_2\in\bbN$ with $R_2>R_1$, and $M_1, M_2 \in \mathbb{N}$ with $M_2 > M_1$ be 
such that
$$
\{\varphi_{j}: M_2\geq j>M_1 \} \subset \Lambda_{R_2,a} \setminus 
\Lambda_{R_1,a},
$$
where $\Lambda_{R_i,a}$ is defined in (\ref{eq:omega_r}).
Then for any $\gamma\in(0,1)$
$$
\nm{P_N U P_{M_2}^{M_1}} \leq   \frac{\pi^2}{4}  \|\theta_\Psi\|_{L^\infty} 
\cdot (2\pi\gamma)^{v} \cdot\sqrt{\frac{1-2^{2v(R_1-R_2)}}{1-2^{-2v}}} 
$$
whenever $N$ is such that
$
N \leq \gamma \omega^{-1} 2^{R_1}$.
\end{lemma}

\begin{proof}
Let $\eta \in l^2(\bbN)$ be such that $\nm{\eta}=1.$  Note that, by the definition of $U$ in (\ref{inf_U}), it follows that 
\begin{equation*}
\begin{split}
\| P_N U P_{M_2}^{M_1} \eta\|^2 &
\leq \sum_{\abs{k}\leq N/2} 
\abs{\ip{\psi_k}{\sum_{j = M_1+1}^{M_2} \eta_j \varphi_j}}^2 \leq  \sum_{\abs{k}\leq N/2} 
\abs{\ip{\psi_k}{\sum_{l=R_1}^{R_2-1}\sum_{j\in\Delta_l} \eta_{\rho(l,j)} \Psi_{l,j}}}^2,
\end{split}
\end{equation*}
where we have defined
$$
\Delta_l = \{j \in \mathbb{Z}: \Psi_{l,j} \in  \Lambda_{l+1,a} \setminus 
\Lambda_{l,a}\}, \quad \rho: \{(l,\Delta_l)\}_{l \in \mathbb{N}} \rightarrow \mathbb{N} \setminus \{1, \hdots, |\Lambda_{1,a}|\}
$$ to be the bijection such that $\varphi_{\rho(l,j)} = \Psi_{l,j}$. Now, observe that we may argue as in the proof of Theorem \ref{t:Wavelet_Asy_Inc2} and use (\ref{Psi_estimate}) to deduce that 
given $l\in\bbN$, $-\lceil a\rceil < j < 2^l\lceil a\rceil$ and $k\in\bbZ$, we have that 
$\ip{\Psi_{l,j}}{\psi_k} = 
\sqrt{ \frac{\omega}{2^l}} \hat \Psi\left( -\frac{2\pi \omega k}{ 2^l}\right) 
e^{2\pi i \omega jk}$.
Hence, it follows that 
$$
\sum_{\abs{k}\leq N/2} 
\abs{\ip{\psi_k}{\sum_{l=R_1}^{R_2-1}\sum_{j\in\Delta_l} \eta_{\rho(l,j)} \Psi_{l,j}}}^2 = \sum_{\abs{k}\leq N/2} \abs{ 
\sum_{l=R_1}^{R_2-1}\frac{\sqrt{\omega}}{\sqrt{2^l}}\sum_{j\in\Delta_l}  \eta_{\rho(l,j)} 
\hat\Psi\left(-\frac{2\pi \omega k}{2^l}\right) e^{2\pi i \omega j k /2^l}}^2,
$$
which again gives us that 
\begin{equation}\label{key1}
\begin{split}
\| P_N U P_{M_2}^{M_1} \eta\|^2 &\leq \sum_{\abs{k}\leq N/2} \abs{ 
\sum_{l=R_1}^{R_2-1}\frac{\sqrt{\omega}}{\sqrt{2^l}} 
\hat\Psi\left(-\frac{2\pi\omega k}{2^l}\right) f^{[l]} \left(\frac{\omega 
k}{2^l}\right)}^2\\
&\leq \sum_{\abs{k}\leq N/2}  
\sum_{l=R_1}^{R_2-1} 
\abs{\hat\Psi\left(-\frac{2\pi\omega k}{2^l}\right)}^2 \cdot  \sum_{l=R_1}^{R_2-1}  \abs{\frac{\sqrt{\omega}}{\sqrt{2^l}} f^{[l]} \left(\frac{\omega 
k}{2^l}\right)}^2\\
&\leq \sum_{l=R_1}^{R_2-1} \max_{\abs{k} \leq N/2}\abs{\hat 
\Psi\left(-\frac{2\pi \omega k}{2^l}\right)}^2 \cdot \sum_{l=R_1}^{R_2-1} 
\sum_{\abs{k}\leq N/2} \frac{\omega}{2^l} \abs{f^{[l]} \left(\frac{\omega 
k}{2^l}\right)}^2,
\end{split}
\end{equation}
where $f^{[l]}(z) = \sum_{j\in\Delta_l}  \eta_{\rho(l,j)} e^{2\pi i zj}$.
Let $H=\chi_{[0,1)}$ and, for $l\in\bbN$,  
$-\lceil a\rceil < j < 2^j\lceil a\rceil$, define 
$
H_{l,j} = 2^{\frac{l}{2}}H(2^l \cdot - j).
$
By the choice of $j$, we have that $H_{l,j}$ is supported on $[-T_1,T_2]$.  Also,  since by (\ref{eq:sampling_den_assump}) we have $\omega\in (0,1/(T_1+T_2)),$ we may argue as in 
(\ref{Psi_estimate}) and find that 
$
\ip{H_{l,j}}{\psi_k} = 
\sqrt{\frac{\omega}{2^l}}\hat H\left(\frac{-2\pi 
k\omega}{ 2^l}\right)e^{2\pi i \omega kj/2^l}.
$
Thus, 
\begin{equation}\label{almost_f}
\ip{\sum_{j\in\Delta_l} \eta_{\rho(l,j)}  H_{l,j}}{\psi_k}  = \sqrt{\frac{\omega}{2^l}} \sum_{j\in\Delta_l} \eta_{\rho(l,j)} \hat H\left(\frac{-2\pi 
k\omega}{ 2^l}\right)e^{2\pi i \omega kj/2^l}.
\end{equation}
It is straightforward to show that  $\inf_{\abs{x} \leq \pi} 
\abs{\hat H(x)} \geq 2/\pi$, and since $N \leq 2^{R_1}/\omega$, for each 
$l\geq R_1$, it follows directly from (\ref{almost_f}) and the definition of $f^{[l]}$ that
\eas{
\sum_{\abs{k}\leq N/2} \frac{\omega}{2^l} \abs{f^{[l]} \left(\frac{\omega 
k}{2^l}\right)}^2 
&\leq \left(\inf_{\abs{x} \leq \pi} \abs{\hat H(x)}^2\right)^{-1} 
\sum_{\abs{k}\leq N/2} \abs{\ip{\sum_{j\in\Delta_l} \eta_{\rho(l,j)}  H_{l,j}}{\psi_k}}^2\\
&\leq \frac{\pi^2}{4} \nm{\sum_{j\in\Delta_l} \eta_{\rho(l,j)}  H_{l,j}}^2 \leq   
\frac{\pi^2}{4}\nm{P_{\Delta_l}\eta}^2.
}
Hence, we immediately get that 
\begin{equation}\label{key3}
\sum_{l=R_1}^{R_2-1} \sum_{\abs{k}\leq N/2} \frac{\omega}{2^l} \abs{f^{[l]} 
\left(\frac{\omega k}{2^l}\right)}^2 \leq \frac{\pi^2}{4}\sum_{l=R_1}^{R_2-1} 
\nm{P_{\Delta_l}\eta}^2 \leq \frac{\pi^2}{4}\nm{\eta}^2 \leq \frac{\pi^2}{4}.
\end{equation}
Also, since $\Psi$ has $v$ vanishing moments, we have that $\hat \Psi (z) = 
(-i z)^v \theta_{\Psi}(z)$ for some bounded $L^{\infty}$ function $\theta_{\Psi}$. 
Thus, since $N \leq \gamma \cdot 2^{R_1}/\omega$, we have
\eas{
\sum_{l=R_1}^{R_2-1} \max_{\abs{k}\leq N/2}\abs{\hat \Psi \left(\frac{2\pi 
\omega k}{2^l}\right)}^2
&\leq \frac{\pi^2}{4} \|\theta_\Psi\|^2_{L^\infty} \sum_{l=R_1}^{R_2-1} 
\left(2\pi \gamma 2^{R_1 -l}\right)^{2v} \\
&\leq \frac{\pi^2}{4}  (2\pi\gamma)^{2v} 
\|\theta_\Psi\|^2_{L^\infty}\frac{1-2^{2v(R_1-R_2)}}{1-2^{-2v}}.
}
Thus, by applying (\ref{key1}), (\ref{almost_f}) and (\ref{key3}), it follows that 
\bes{
\begin{split}
\| P_{N} U P_{M_2}^{M_1} \eta\|^2
 \leq  \frac{\pi^2}{4} \|\theta_\Psi\|^2_{L^\infty} \cdot (2\pi \gamma)^{2v} 
 \frac{1-2^{2v(R_1-R_2)}}{1-2^{-2v}},
\end{split}
}
and we have proved the desired estimate.
\end{proof}

\subsubsection{The proof}
\begin{proof}[Proof of Theorem \ref{Four_to_wave}]
In this proof, we will let $B_{\Phi,\Psi}$ be some constant which depends only 
on $\Phi$ and $\Psi$, although its value may change from instance to instance. 
The assertions of the theorem will follow if we can show that the conditions 
in Theorem  \ref{main_full_inf_noise2} are satisfied. We will begin with 
condition (i). 
First observe that since $U$ is an isometry we have that $\|P_{M}U^* P_NUP_{M} -P_{M}\|_{l^{\infty}} = \|P_{M}U^* P_N^{\perp}UP_{M}\|_{l^\infty \rightarrow l^\infty} \leq \sqrt{M} \nm{P_N^\perp U P_M}$ and $\|P_M^{\perp}U^*P_NUP_M\|_{l^\infty \rightarrow l^\infty} = \|P_M^{\perp}U^*P_N^{\perp}UP_M\|_{l^\infty \rightarrow l^\infty} \leq \sqrt{M} \nm{P_N^\perp U P_M}$. So 
$N$, $K$ satisfy the strong balancing property with respect to $U$, $M$ and 
$s$ if
$$
\nm{P_N^\perp U P_M} \leq\frac{1}{8}\left( M \log_2(4KM\sqrt{s})\right)^{-1/2}.
$$

In the case  of $\alpha\geq 1$, by applying Lemma \ref{lem:op_wave_bound} with $\gamma = \frac{1}{8}\left( M \log_2(4KM\sqrt{s})\right)^{-1/2}$, it follows that $N$, $K$ satisfy the 
strong balancing property with respect to $U$, $M$, $s$ whenever
$$
N \geq C_{\omega, \Phi} \cdot 2^{R+1}\cdot \left( \frac{1}{8}\left( M \log_2(4KM\sqrt{s})\right)^{-1/2}\right)^{-\frac{2}{2\alpha-1}},
$$
where $R$ is the smallest integer such that $M\leq W_R$ (where $W_R$ is defined in (\ref{eq:W_R})) and $C_{\omega, \Phi}$ is a constant which depends only on the Fourier decay of $\Phi$ and $\omega$. By the choice of $R$, we have that $M=\ord{2^R}$ since $W_R = \ord{2^R}$ by (\ref{eq:W_R}). Thus, the strong balancing property holds provided that
$$N \gtrsim  M^{1+1/(2\alpha -1)} \cdot 
\left(\log_2(4MK\sqrt{s})\right)^{1/(2\alpha -1)}
$$
where the constant involved depends only on $\omega$ and the Fourier decay of $\Phi$. 
Furthermore, if (\ref{eq:cond2_fdecay_mthm}) holds, then a direct application of Lemma \ref{cor:l_inf_normbounds} gives that $N$, $K$ satisfy the 
strong balancing property with respect to $U$, $M$, $s$ whenever $N \gtrsim M \cdot 
\left(\log_2(4K M \sqrt{s})\right)^{1/(4\alpha-2)}$. So, condition 
(i) of Theorem \ref{Four_to_wave} implies condition (i) of Theorem \ref{main_full_inf_noise2}.

To show that (ii) in Theorem \ref{main_full_inf_noise2} is satisfied, we need 
to demonstrate that 
\be{
\label{conditions331}
1 \gtrsim \frac{N_k-N_{k-1}}{m_k} \cdot \log(\epsilon^{-1})  \cdot \left(
\sum_{l=1}^r \mu_{\mathbf{N},\mathbf{M}}(k,l) \cdot s_l\right) \cdot 
\log\left(K \tilde M \sqrt{s}\right),
 }
 (with $\mu_{\mathbf{N},\mathbf{M}}(k,r)$ replaced by 
 $\mu_{\mathbf{N},\mathbf{M}}(k,\infty)$, and also recall that $N_0 = 0$)
 and 
\be{\label{conditions_on_hatm331}
\begin{split}
& \qquad m_k \gtrsim \hat m_k \cdot  \log(\epsilon^{-1}) \cdot \log\left(K 
\tilde M \sqrt{s}\right),\\
1 \gtrsim \sum_{k=1}^r & \left(\frac{N_k-N_{k-1}}{\hat m_k} - 1\right) \cdot 
\mu_{\mathbf{N},\mathbf{M}}(k,l)\cdot \tilde s_k, \qquad \forall \, l = 1, 
\hdots, r,
 \end{split}
 }
 where 
 \begin{equation}\label{MM}
\tilde{M} = \min\{i\in\bbN: \max_{k\geq i}\| P_N U e_k \| \leq 
1/(32K\sqrt{s})\}.
\end{equation}
We will first consider (\ref{conditions331}). By applying the bounds (\ref{eq:estimate_mu_wavelet}) and (\ref{eq:estimate_mu_inf}) on the 
local coherences derived in Corollary \ref{lem:inc_bd}, we have that 
(\ref{conditions331}) is implied by
\be{\label{imply1}
\begin{split}
&\frac{m_k}{(N_k - N_{k-1})} \gtrsim  B_{\Phi,\Psi}\cdot \Bigg(
\sum_{j=1}^{k-1}  \frac{s_j}{ N_{k-1}} \left(\frac{2^{R_{j -1}}}{\omega 
N_{k-1}}\right)^{\alpha -1/2}
+  \frac{s_k   }{N_{k-1}}\\
& \qquad + \sum_{j=k+1}^r \frac{ s_j \cdot  
\sqrt{\omega}}{\sqrt{N_{k-1}2^{R_{j-1}}}} \cdot \left(\frac{\omega 
N_{k}}{2^{R_{j-1}}}\right)^{v} \Bigg) \cdot \log(\epsilon^{-1}) \cdot \log\left(K 
\tilde M \sqrt{s}\right), \qquad k = 2,\hdots,r
\end{split}
} 
\begin{equation}\label{imply11}
\begin{split}
&\frac{m_1}{N_1} \gtrsim  B_{\Phi,\Psi}\cdot \Bigg(s_1 + \sum_{j=2}^r \frac{ s_j \cdot  
\sqrt{\omega}}{\sqrt{2^{R_{j-1}}}} \cdot \left(\frac{\omega 
N_{1}}{2^{R_{j-1}}}\right)^{v} \Bigg) \cdot \log(\epsilon^{-1}) \cdot \log\left(K 
\tilde M \sqrt{s}\right).
\end{split}
\end{equation}
To obtain a bound on the value of $\tilde{M}$ in (\ref{MM}), observe that by Lemma 
\ref{lem:nm_bound},
$
\nm{ P_N  U P_{\{j\}}} \leq  1/(32K\sqrt{s})
$ 
whenever $j=2^J$ such that $2^J \geq (32K\sqrt{s})^{1/v}\cdot N \cdot \omega$. 
Thus, $\tilde M \leq \lceil (32K\sqrt{s})^{1/v}\cdot N \cdot \omega \rceil$,
and by recalling that $N_k = 2^{R_k} \omega^{-1}$, we have that 
(\ref{imply1}) is implied by
\begin{equation}\label{implied1}
\begin{split}
\frac{m_k \cdot N_{k-1}}{N_k - N_{k-1}} 
&\gtrsim   B_{\Phi,\Psi}\cdot \log(\epsilon^{-1})  \cdot 
\log\left((K\sqrt{s})^{1+ 1/v} N\right)\\
& \quad\cdot \Bigg(\sum_{j=1}^{k-1} s_j\cdot 
\left( 2^{\alpha-1/2}\right)^{-(R_{k-1}-R_{j-1})} + s_k
 + s_{k+1} \cdot 2^{-(R_{k}-R_{k-1})/2} \\
 & \qquad + \sum_{j=k+2}^r s_j\cdot 
 2^{-(R_{j-1}-R_{k-1})/2} \cdot 2^{-v(R_{j-1}-R_{k})} \Bigg), \quad k \geq 2,
 \end{split}
 \end{equation}
 and when $k=1$, (\ref{imply11}) is implied by
 \begin{equation}\label{implied11}
\begin{split}
&\frac{m_1}{N_1} 
\gtrsim   B_{\Phi,\Psi}\cdot \log(\epsilon^{-1})  \cdot 
\log\left((K\sqrt{s})^{1+ 1/v} N\right)\\
&\cdot \Bigg(s_1 + s_{2} \cdot 2^{-R_{1}/2}  
+ \sum_{j=k+2}^r s_j\cdot 
 2^{-(R_{j-1}-R_{k-1})/2} \cdot  2^{-v(R_{j-1}-R_{k})} \Bigg).
 \end{split}
 \end{equation}
 However, the condition (\ref{eq:cond_wav1}) obviously implies (\ref{implied1}) and (\ref{imply11}), hence we have established that 
 condition (\ref{eq:cond_wav1}) implies (\ref{conditions331}).
As for condition (\ref{conditions_on_hatm331}), we will first derive upper 
bounds for the $\tilde s_k$ values.
Recall that according to Theorem \ref{main_full_inf_noise2} we have 
$$
\tilde s_k \leq S_k(\mathbf{N},\mathbf{M},\mathbf{s}) =  \max\{\|P_{N_k}^{N_{k-1}}U\eta\|^2: \|\eta\|_{l^{\infty}} \leq 1, 
|\mathrm{supp}(P_{M_l}^{M_{l-1}}\eta)| = s_l, \, l=1,\hdots, r\},
$$
where $N_0 = M_0 = 0$. Thus, we will concentrate on bounding $S_k$.
First note that by a direct rearrangement of terms in Lemma \ref{lem:op_wave_bound},   for any $\gamma\in(0,1)$ and $R\in\bbN$ such that $M\leq W_R$, we have that
  $
  \nm{P_N^\perp U P_M} \leq \gamma
  $
  whenever $N$ is such that
  $$
  \gamma \geq \left(\frac{2^R}{\omega N}\right)^{\frac{2\alpha-1}{ 2}} \cdot \sqrt{\frac{2}{2\alpha-1}}\cdot \frac{C_\Phi}{\pi^\alpha}.
  $$ 
So for any $L>0$, by letting 
$
\gamma =  \sqrt{\frac{2}{2\alpha-1}}\cdot \frac{C_\Phi}{\pi^\alpha} \cdot L^{-\frac{2\alpha-1}{2}},
$
if $\gamma\in (0,1)$, then $\nm{P_N^\perp U P_M}\leq \gamma$ provided that $N\geq \omega^{-1} \cdot L\cdot 2^R$. Also, if $\gamma>1$, then $\nm{P_N^\perp U P_M}\leq \gamma$ is trivially true since $\nm{U}= 1$.
Therefore, for $k \geq 2$ we have that 
$$
\|  
P_{N_{k-1}}^\perp U P_{M_{l}}\| <
  \sqrt{\frac{2}{2\alpha-1}} \cdot \frac{ C_\Phi}{\pi^\alpha}\cdot 
  \left(\frac{2^{R_l}}{2^{R_{k-1}}}\right)^{\alpha - 1/2}, \qquad 
  l\leq k-1.$$
Also, by Lemma \ref{lem:nm_bound}, it follows that 
$$  \|  P_{N_{k}} U  P_{M_l}^{M_{l-1}}\eta \|<   (2\pi)^v 
\cdot \nm{\theta_\Psi}_{L^\infty}\cdot \left(\frac{2^{R_k}}{2^{R_{l-1} 
}}\right)^v, \qquad l\geq k+1.$$  Consequently, for $k=3,\ldots,r$
\bes{
\begin{split}
\sqrt{\tilde s_k }&\leq \sqrt{S_{k}} = \max_{\eta\in\Theta} \| 
P_{N_{k}}^{N_{k-1}} U \eta \| 
\leq  \sum^{r}_{l=1}\| P^{N_{k-1}}_{N_k} U P^{M_{l-1}}_{M_l} \| \sqrt{s_l} \\
&\leq B_{\Phi,\Psi} \Bigg(\sum_{l=1}^{k-2}  \sqrt{s_l}  \cdot   
\left(\frac{2^{R_l}}{2^{R_{k-1}}}\right)^{\alpha - 1/2} +\sqrt{ s_{k-1}} + 
\sqrt{s_k} + \sqrt{s_{k+1}} + \sum_{l=k+2}^r \sqrt{s_l}\cdot  
\left(\frac{2^{R_k}}{2^{R_{l-1 }}}\right)^v  \Bigg),
\end{split}
}
where 
\bes{
\Theta = \{\eta : \|\eta\|_{l^{\infty}} \leq 1, 
|\mathrm{supp}(P_{M_l}^{M_{l-1}}\eta)| = s_l, \, l=1,\hdots, r\},
}
and for $k = 1,2$ we have 
$$
\sqrt{\tilde s_k } \leq B_{\Phi,\Psi} \left(\sqrt{ s_{k-1}} + 
\sqrt{s_k} + \sqrt{s_{k+1}} + \sum_{l=k+2}^r \sqrt{s_l}\cdot  
\left(\frac{2^{R_k}}{2^{R_{l-1 }}}\right)^v\right),
$$
where we let $s_0 = 0$.
Hence, for  $k=3,\ldots,r$, 
$
A_\alpha = 2^{\alpha-1/2}$ and $ A_v = 2^v$ 
$$
\tilde s_k \leq  B_{\Phi,\Psi}\Bigg( \sqrt{\hat s_k} + \sum_{l=1}^{k-2}  \sqrt{s_l}  \cdot   A_\alpha^{-(R_{k-1} - 
R_{l}) } +   \sum_{l=k+2}^r \sqrt{s_l}\cdot  A_v^{-(R_{l-1}-R_k)} \Bigg)^2,
$$
where $\hat s_k = \max\{s_{k-1},s_k,s_{k+1}\}$. So, by using the Cauchy-Schwarz inequality, we obtain
\begin{equation*}
\begin{split}
\tilde s_k & \leq B_{\Phi,\Psi} \left( 1 +  \sum_{l=1}^{k-2}  A_\alpha^{-(R_{k-1} - 
R_{l}) } +   \sum_{l=k+2}^r   A_v^{-(R_{l-1}-R_k)}\right) \\
&\qquad\qquad\qquad\qquad \cdot \left(\hat s_k +  \sum_{l=1}^{k-2}  s_l \cdot   
A_\alpha^{-(R_{k-1} - R_{l}) } + \sum_{l=k+2}^r s_l \cdot  
A_v^{-(R_{l-1}-R_k)} \right)\\
&\leq B_{\Phi,\Psi}\Bigg(\hat s_k 
+\sum_{l=1}^{k-2}  s_l \cdot   A_\alpha^{-(R_{k-1} - R_{l}) } +   
\sum_{l=k+2}^r s_l \cdot  A_v^{-(R_{l-1}-R_k)} \Bigg),
\end{split}
\end{equation*}
and similarly, for $k = 1,2$, it follows that $\tilde s_k \leq B_{\Phi,\Psi}(\hat s_k +   
\sum_{l=k+2}^r s_l \cdot  A_v^{-(R_{l-1}-R_k)})$. 
Finally, we will use the above results to show that condition (\ref{eq:cond_wav1}) implies 
 (\ref{conditions_on_hatm331}): By our coherence estimates in 
 (\ref{eq:estimate_mu_wavelet}), (\ref{eq:estimate_mu_wavelet2}), (\ref{eq:estimate_mu_inf}) and (\ref{eq:estimate_mu_inf2}), we see that 
(\ref{conditions_on_hatm331})
 holds if  $m_k \gtrsim \hat m_k \cdot 
 (\log(\epsilon^{-1}) + 1)  \cdot \log\left((K\sqrt{s})^{1+ 1/v} N\right) $ 
 and  for each $l=2,\ldots, r$,
\be{\label{eq:cond_2_interm}
\begin{split}
& 1\gtrsim B_{\Phi,\Psi}\Bigg(\left(\frac{N_1}{\hat m_1} -1\right)\cdot \tilde s_1 \cdot 
\sqrt{\frac{\omega}{2^{R_{l-1}}}} \cdot  \left( \frac{ \omega 
N_1}{2^{R_{l-1}}}\right)^v\\ 
&+ \sum_{k=2}^{l-1} \left(\frac{N_k - 
N_{k-1}}{\hat m_k} -1\right)\cdot \tilde s_k \cdot 
\sqrt{\frac{\omega}{N_{k-1}2^{R_{l-1}}}} \cdot  \left( \frac{ \omega 
N_k}{2^{R_{l-1}}}\right)^v \\
&+ \left(\frac{N_l - N_{l-1}}{\hat m_l} -1\right)\cdot \tilde s_l \cdot 
\frac{1 }{N_{l-1}} + \sum_{k=l+1}^r  \left(\frac{N_k - N_{k-1}}{\hat m_k} -1\right) \cdot \tilde 
s_k \cdot \frac{1}{ N_{k-1}} \left( \frac{2^{R_{l-1}}}{\omega 
N_{k-1}}\right)^{\alpha - 1/2} \Bigg),
\end{split}
}
(where we with slight abuse of notation define $\sum_{k=2}^{l-1} (\frac{N_k - 
N_{k-1}}{\hat m_k} -1)\tilde s_k 
\sqrt{\frac{\omega}{N_{k-1}2^{R_{l-1}}}}( \frac{ \omega 
N_k}{2^{R_{l-1}}})^v = 0$ when $l=2$),
and for $l = 1$
\be{\label{eq:cond_2_interm2}
\begin{split}
1\gtrsim B_{\Phi,\Psi}\left(\left(\frac{N_1}{\hat m_1} -1\right)\cdot \tilde s_1 + \sum_{k=2}^r  \left(\frac{N_k - N_{k-1}}{\hat m_k} -1\right) \cdot \tilde 
s_k \cdot \frac{1}{ N_{k-1}} \left( \frac{1}{\omega 
N_{k-1}}\right)^{\alpha - 1/2} \right).
\end{split}
}
Recalling that $N_k = \omega^{-1} 2^{R_k}$, (\ref{eq:cond_2_interm}) becomes, for $l = 2,\hdots, r$,
\eas{
&1\gtrsim B_{\Phi,\Psi}\cdot \Bigg(\left(\frac{N_1}{\hat m_1} -1\right)\cdot \frac{\tilde s_k}{N_{k-1}} \cdot     
2^{-v(R_{l-1}-R_k)} 
+  \sum_{k=1}^{l-1} \left(\frac{N_k - 
N_{k-1}}{\hat m_k} -1\right)\cdot \frac{\tilde s_k}{N_{k-1}} \cdot     
2^{-v(R_{l-1}-R_k)}\\
&+ \left(\frac{N_l - N_{l-1}}{\hat m_l} -1\right)\cdot \frac{\tilde 
s_l}{N_{l-1}} 
+ \sum_{k=l+1}^r  \left(\frac{N_k - N_{k-1}}{\hat m_k} -1\right) \cdot 
\frac{\tilde s_k}{N_{k-1}} \cdot  \left( 2^{\alpha - 
1/2}\right)^{-(R_{k-1}-R_{l-1})} \Bigg),
}
and (\ref{eq:cond_2_interm2}) becomes
\eas{
&1\gtrsim  B_{\Phi,\Psi}\cdot \Bigg(\left(\frac{N_1}{\hat m_1} -1\right)\cdot \tilde s_1
+ \sum_{k=l+1}^r  \left(\frac{N_k - N_{k-1}}{\hat m_k} -1\right) \cdot 
\frac{\tilde s_k}{N_{k-1}} \cdot  \left( 2^{\alpha - 
1/2}\right)^{-R_{k-1}} \Bigg).
}
Observe that for $l = 2,\hdots, r$
\eas{
1+ \sum_{k=1}^{l-1}   2^{-v(R_{l-1}-R_k)}  + \sum_{k=l+1}^r  
\left( 2^{\alpha - 1/2}\right)^{-(R_{k-1}-R_{l-1})} \leq B_{\Phi,\Psi},
}
and that $1+  \sum_{k=l+1}^r  
\left( 2^{\alpha - 1/2}\right)^{-(R_{k-1})} \leq B_{\Phi,\Psi}$.
Thus, (\ref{conditions_on_hatm331}) holds provided that for each $ k=2,\ldots, r$,
$$
 \hat m_k \geq B_{\Phi,\Psi} \cdot \frac{N_k - N_{k-1}}{N_{k-1}}\cdot \tilde 
 s_k, \qquad \hat m_1 \geq B_{\Phi,\Psi} \cdot N_1\cdot \tilde 
 s_1, 
$$
 and combining with our estimates of $\tilde{s}_k$, we may deduce that 
 (\ref{eq:cond_wav1}) implies (\ref{conditions_on_hatm331}).
\end{proof}

\section*{Acknowledgements}

The authors would like to thank Akram Aldroubi, Emmanuel Cand\`es, Massimo
Fornasier, Karlheinz Gr\"ochenig, Felix Krahmer, Gitta Kutyniok, Thomas
Strohmer, Gerd Teschke, Michael Unser, Martin Vetterli and Rachel Ward for
useful discussions and comments. The authors also thank Stuart Marcelle and
Homerton College, University of Cambridge for the provision of computing 
hardware used in some of the experiments.  BA acknowledges support from the NSF DMS grant 1318894.  ACH acknowledges support from a Royal Society University Research Fellowship as well as the UK Engineering and Physical Sciences Research Council (EPSRC) grant EP/L003457/1.  CP acknowledges support from the EPSRC grant EP/H023348/1 for the University of Cambridge Centre for Doctoral Training, the Cambridge Centre for Analysis.

\bibliographystyle{abbrv}
\small
\bibliography{BreakingCoherenceRefs}

\end{document}